\documentclass[twoside]{article}
\usepackage[accepted]{aistats2024}

\usepackage{graphicx} 
\usepackage[round]{natbib}

\usepackage{microtype}
\usepackage{graphicx}
\usepackage{booktabs} 
\usepackage{amsmath, amssymb, amsthm}
\usepackage{bm}
\usepackage{mathtools}

\newcommand{\mat}[1]{\boldsymbol {#1}}

\renewcommand{\vec}[1]{\boldsymbol {#1}}

\newcommand{\rvar}[1]{\mathrm {#1}}
\newcommand{\rvec}[1]{\boldsymbol{\mathrm{#1}}}

\newcommand{\set}[1]{\mathbb {#1}}

\newcommand{\R}[0]{\set{R}}
\newcommand{\norm}[1]{\left\lVert#1\right\rVert}

\newcommand{\T}[0]{{^\top}}

\renewcommand{\exp}[1]{\mathrm{exp}\left(#1\right)}

\DeclareMathOperator*{\argmin}{arg\,min}

\newcommand{\1}[1]{\mathbb{I}_{#1}}

\renewcommand{\Pr}[1]{\mathbb{P}\left[#1\right]}

\newcommand{\Manifold}{\mathcal{M}}
\newcommand{\Sphere}[1]{\mathcal{S}^{#1}}

\newcommand{\TPManifold}[1]{T_{#1}\Manifold}
\newcommand{\expmap}{\text{exp}}
\newcommand{\Prob}[1]{\mathcal{P}({#1})}

\usepackage{ulem}
\usepackage{enumitem}
\usepackage{multirow}
\usepackage{amsfonts}       
\usepackage{nicefrac} 
\usepackage[utf8]{inputenc} 
\usepackage[T1]{fontenc} 
\usepackage{overpic}
\usepackage{wrapfig}
\usepackage{caption}
\usepackage{subcaption}
\usepackage{xcolor}         
\usepackage{pgfplots}
\usepackage{pgf}
\usepackage{array}
\usepackage{hyperref}
\usepackage{url}
\usepackage[capitalize,noabbrev]{cleveref}

\theoremstyle{plain}

\theoremstyle{definition}

\theoremstyle{remark}

\begin{document}

\runningauthor{ Pegoraro,  Vedula, Rosenberg, Tallini, Rodolà, Bronstein}

\twocolumn[

\aistatstitle{Vector Quantile Regression on Manifolds}

\aistatsauthor{  Marco Pegoraro \And Sanketh Vedula \And 
Aviv A. Rosenberg }

\aistatsaddress{ Sapienza, University of Rome \\ Technion 
\And  Technion\\ Sibylla 
\And  Technion \\Sibylla } 

\aistatsauthor{  Irene Tallini  
\And Emanuele Rodolà  \And Alex M. Bronstein }

\aistatsaddress{  Sapienza, University of Rome \\  Technion 
\And Sapienza, University of Rome  
\And  Technion \\Sibylla }

]

\begin{abstract}
Quantile regression (QR) is a statistical tool for distribution-free estimation of conditional quantiles of a target variable given explanatory features.
QR is limited by the assumption that the target distribution is univariate and defined on an Euclidean domain.
Although the notion of quantiles was recently extended to multi-variate distributions,
QR for multi-variate distributions on manifolds remains underexplored, even though 
many important applications inherently involve data distributed on, e.g., spheres (climate and geological phenomena), and tori (dihedral angles in proteins).
By leveraging optimal transport theory and $c$-concave functions, we meaningfully define conditional vector quantile functions of high-dimensional variables on manifolds (M-CVQFs).
Our approach allows for quantile estimation, regression, and computation of conditional confidence sets and likelihoods.
We demonstrate the approach's efficacy and provide insights regarding the meaning of non-Euclidean quantiles through synthetic and real data experiments.
\end{abstract}

\section{\uppercase{Introduction}}

\begin{figure*}[t]
\centering
\begin{subfigure}[b]{0.31\linewidth}
\centering
\begin{overpic}
[trim=2.9cm 2.7cm 3.35cm 2.9cm,clip,width=0.48\linewidth, grid=false]{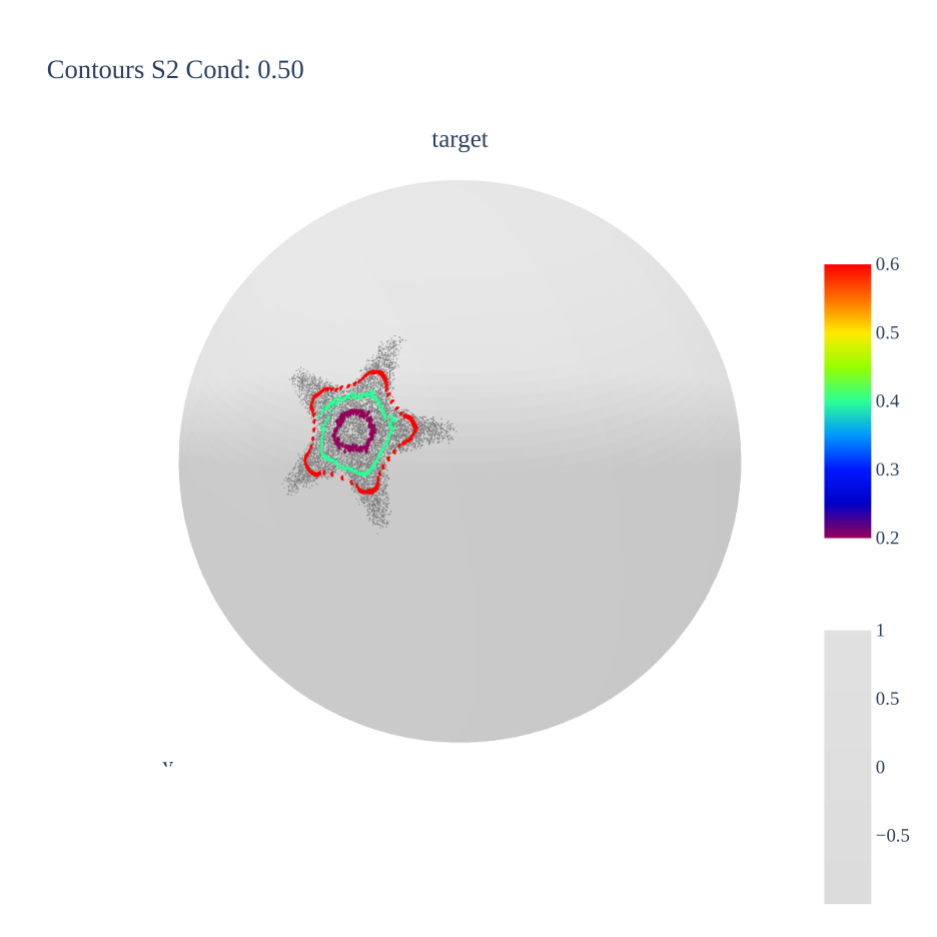}
\put(25,-6){\footnotesize $x=0.5$}
\end{overpic}
\begin{overpic}
[trim=2.9cm 2.7cm 3.35cm 2.9cm,clip,width=0.48\linewidth, grid=false]{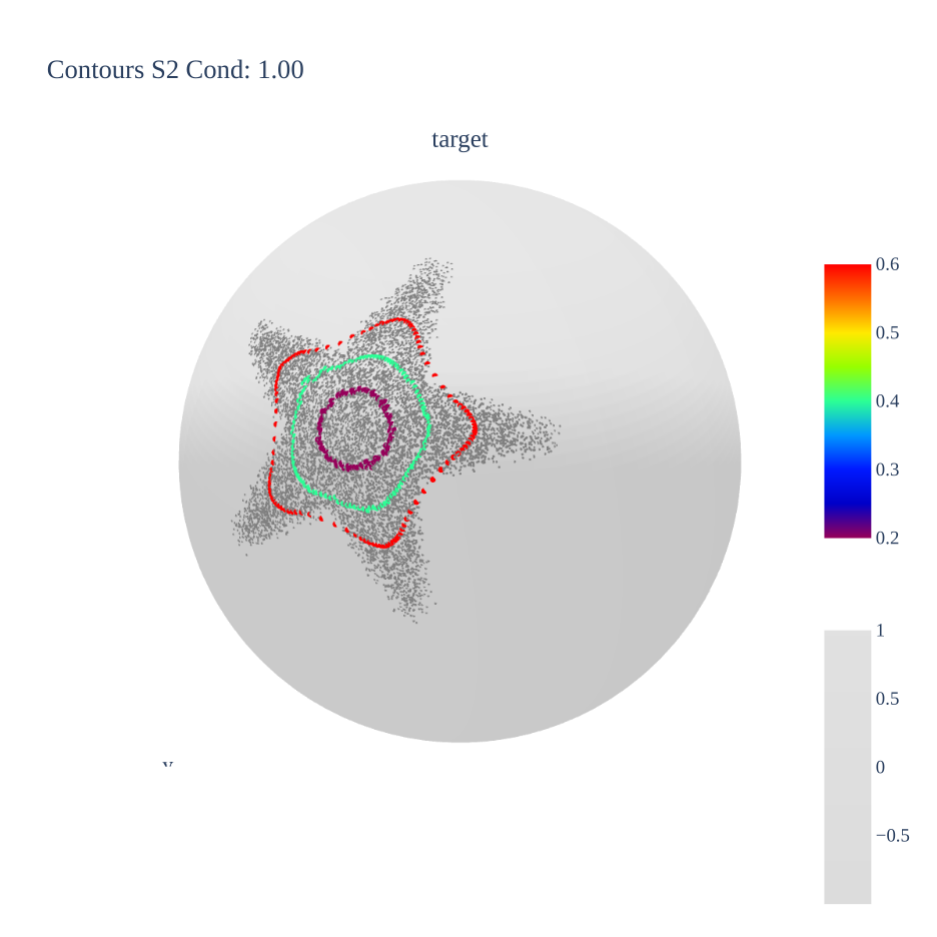}
\put(28,-6){\footnotesize $x=1$}
\end{overpic}
\end{subfigure} 
\hfill
\begin{subfigure}[b]{0.31\linewidth}
\centering
\begin{overpic}
[trim=2.9cm 2.7cm 3.35cm 2.9cm,clip,width=0.48\linewidth, grid=false]{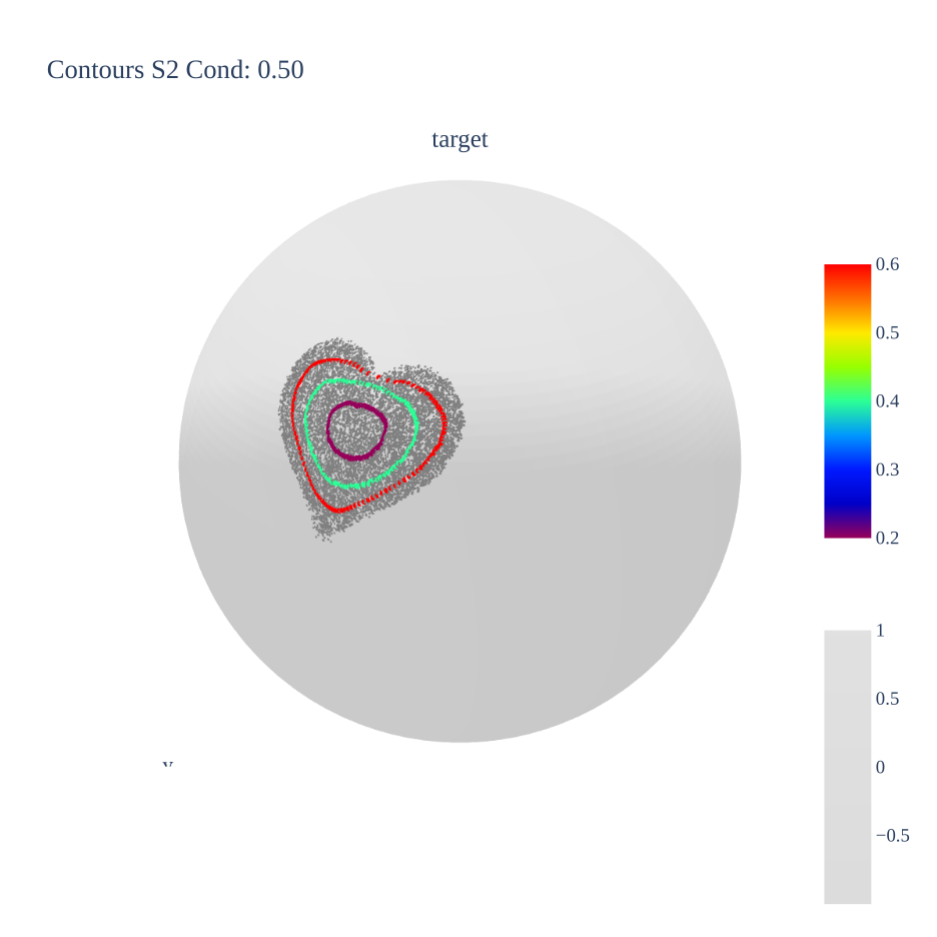}
\put(25,-6){\footnotesize $x=0.5$}
\end{overpic}
\begin{overpic}
[trim=2.9cm 2.7cm 3.35cm 2.9cm,clip,width=0.48\linewidth, grid=false]{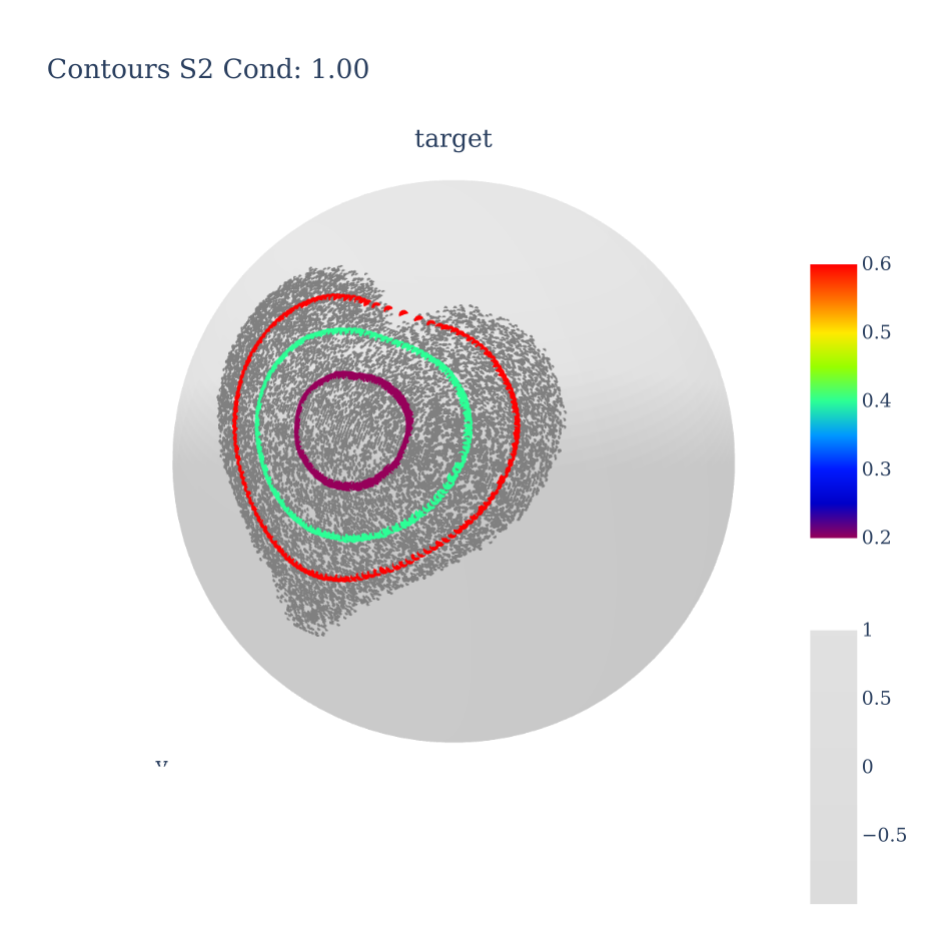}
\put(28,-6){\footnotesize $x=1$}
\end{overpic}
\end{subfigure} 
\hfill
\begin{subfigure}[b]{0.31\linewidth}
\centering
\begin{overpic}
[trim=3.25cm 3cm 3.46cm 5.18cm,clip,width=0.48\linewidth, grid=false]{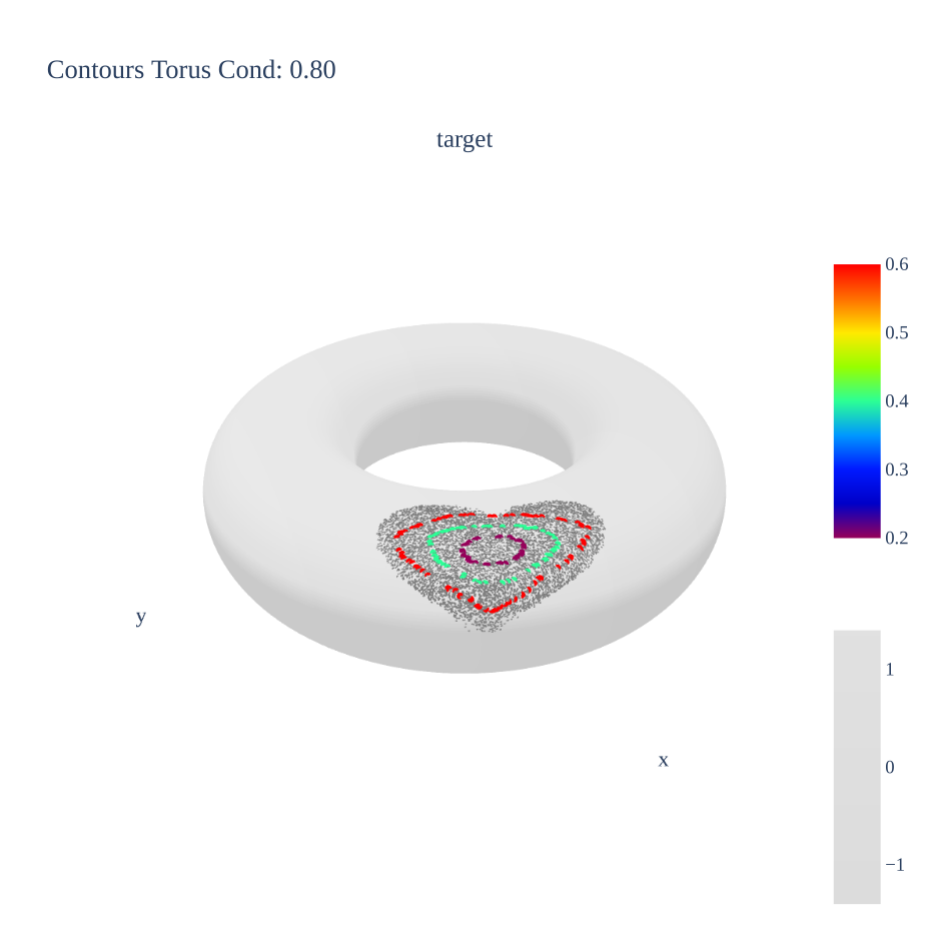}
\put(25,-6){\footnotesize $x=0.8$}
\end{overpic}
\begin{overpic}
[trim=3.25cm 3cm 3.46cm 5.18cm,clip,width=0.48\linewidth, grid=false]{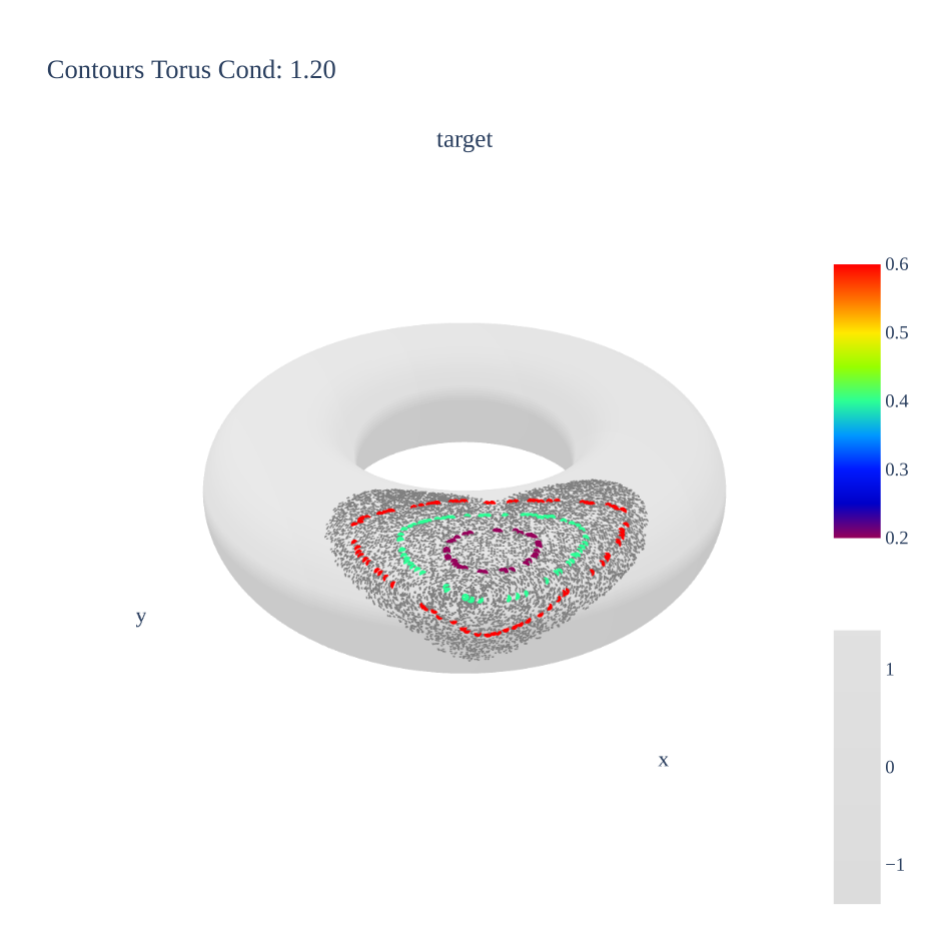}
\put(28,-6){\footnotesize $x=1.2$}
\end{overpic}
\end{subfigure}
\begin{subfigure}[b]{0.03\linewidth}
\centering
\begin{overpic}
[trim=4.52cm -4cm 3.45cm 2.79cm,clip,width=1\linewidth, grid=false]{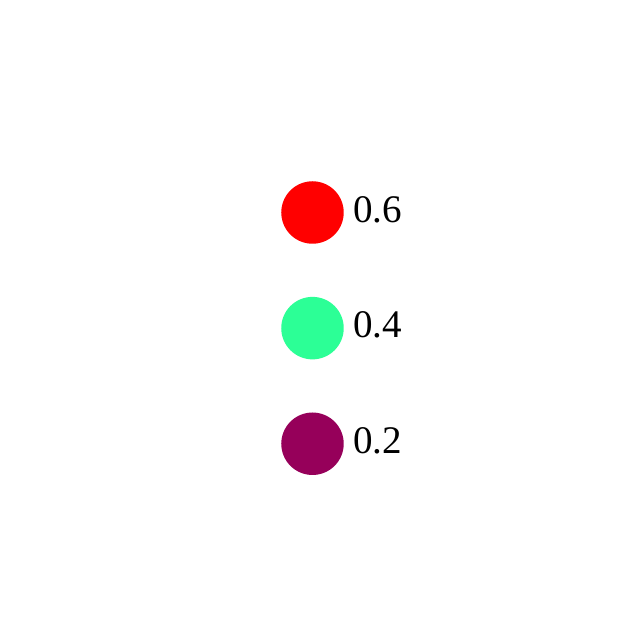}
\put(4,105){\footnotesize $\tau$}
\end{overpic}
\end{subfigure}
\caption{\label{fig:vqr_cont}
\textbf{Sampling and confidence sets for the `Scaled Heart' and  `Scaled Star' distributions $\rvec{Y}|\rvec{X}$ on $\Sphere{2}$ and $\mathcal{T}_2$, under different conditioning values.}
The conditioning variable $x$ controls the scale of the distribution. $\tau$-contours shown as colored lines. The probability of $\rvec{Y}|\rvec{X}$ falling inside a $\tau$-contour is $\tau$.}

\end{figure*}

Quantile regression (QR) \citep{Koenker1978} is a powerful statistical tool that estimates \textit{conditional quantiles} of a target variable $\rvar{Y}$, given covariates $\rvec{X}$. 
QR is usually formulated as a regression problem minimizing the \textit{pinball loss}, the definition of which inherently assumes both scalar and Euclidean data.
QR has been extensively employed in applications with scalar target variables in the Euclidean domain, where the notion of a quantile is both well-defined and widely understood.
However, there exist many real-world applications in which the target data are distributed on a \textit{manifold}, such as a sphere or a cylinder in case of climate measurements \citep{lagona2018correlated, scher2020spherical}, a torus in case of protein dihedral angles \citep{rosenberg2022codon}, or more general manifolds for medical images \citep{pennec2019riemannian}.
These datasets are better represented as points lying on manifolds, which are geometric spaces with nontrivial topological and geometric properties.
A meaningful extension of QR to manifolds would, therefore, unlock the power of this statistical tool for a broader range of applications.

Recently, new perspectives on quantile functions have emerged, allowing for their extension into high-dimensional data.
Notably, \citet{Carlier2016} proposed \textit{vector quantile regression} (VQR), defining the \textit{vector quantile function} (VQF) as a co-monotonic map---which can be obtained by solving a suitable optimal transport (OT) problem---between a multivariate uniform distribution and the target distribution of interest.
They further showed that quantile regression can be framed as a \textit{conditional} OT problem where a family of OT maps, parametrized by covariates $\rvec{X}$, are solved for simultaneously. The resulting maps are the \textit{conditional vector quantile functions} (CVQFs) of the multivariate target variable $\rvec{Y}|\rvec{X}$.
This approach is, however, limited by the assumption of linearity and the use of the primal OT formulation, which is hard to scale.
More recently, \citet{rosenberg2022fastnlvqr} extended VQR by proposing a non-linear extension and introducing a more scalable solver which exploits the entropic-regularized dual of the conditional OT problem. Although this work introduced a learned embedding, to incorporate inductive bias through the structure of $\rvec{X}$, the approach is unable to exploit any intrinsic structure present in $\rvec{Y}$.

Leveraging the OT-based formulation of quantile functions, \citet{hallin2022nonparametric} proposed quantile estimation for spherical data, by solving OT between the base and target distributions on the $n$-sphere.
However, their approach has the following limitations: (i) it is limited to quantile \textit{estimation}, as opposed to regression (only supports \textit{unconditional} quantile functions); (ii) it is defined only for the sphere, and (iii) it requires solving the primal OT formulation, resulting in a large-scale linear program which becomes impractical even for moderately-sized problems.
Despite these limitations, \citeauthor{hallin2022nonparametric} provide a motivating example for the potential of OT methods for estimating distributions on manifolds, and is, to our knowledge, the only approach proposed so far for defining non-Euclidean quantiles.

\paragraph{Contributions.} We propose a novel, \textit{scalable} approach for estimating \textit{multi-dimensional conditional} quantile functions on \textit{manifolds}.
We build upon a dual formulation of the Riemannian conditional OT problem to solve for a family of OT maps parametrized in $\rvec{X}$.
We model the conditional OT maps as gradients of $c$-convex potentials, which are represented by partially input $c$-convex neural networks.
Thus, we address the major limitations of both previous works \citep{hallin2022nonparametric,rosenberg2022fastnlvqr}, in a holistic framework supporting multivariate distributions on any manifold for which the exponential map is known.
To the best of our knowledge, this is the first work to estimate \textit{conditional} VQFs on manifolds, and, more generally, to address the problem of conditional OT on manifolds.
Furthermore, our method extends beyond the capabilities of existing approaches by not only computing quantiles but also addressing additional aspects, including sampling and likelihood computation. This ability distinguishes our work and aligns with recent advancements in generative modeling on Riemannian geometries \citep{brehmer2020flows,chen2023riemannian}.
We demonstrate the effectiveness of our approach through extensive experiments on synthetic and real datasets on the sphere and torus.
The code is available at  \url{https://github.com/Marco-Peg/mvqr}.

\section{\label{app:background} \uppercase{Background}}
In the upcoming section, we offer a concise introduction to fundamental concepts: quantile regression, $c$-convexity, and Riemannian Optimal Transport (OT). 
For a more extensive overview of Riemannian manifolds and all the pertinent formulas employed throughout this paper, we refer to Section \ref{app:diff_geom} of the Appendix and \cite{do1992riemannian}.

\paragraph{Notation.}
Throughout, $\rvar{Y}$, $\rvec{X}$ denote random variables and vectors, respectively; deterministic scalars, vectors and matrices are denoted as $y$, $\vec{x}$, and $\mat{X}$.
$p_{\rvec{Y}}$ denotes the density of the random vector $\rvec{Y}$ and $p_{(\rvec{Y},\rvec{X})}$ denotes the joint density of $\rvec{X}$ and $\rvec{Y}$.
$\rvec{1}_N$ denotes an $N$-dimensional vector of ones, $\odot$ denotes the elementwise (Hadamard) product, and $\1{A}$ is the indicator function of a set $A$.
$Q_{\rvec{Y}|\rvec{X}}(\vec{u};\vec{x})$ is the \textit{manifold conditional vector quantile function} M-CVQF (defined in the sequel) of the  variable $\rvec{Y}|\rvec{X}$, evaluated at the vector quantile level $\vec{u}$, for $\rvec{X}=\vec{x}$. $\mathcal{P}(\Manifold)$ denotes the set of probability measures over $\Manifold$.

\paragraph{Quantile functions.}
The \textit{quantile function} $Q_{\rvar{Y}}$ of a scalar-valued random variable $\rvar{Y} \in \R$ is commonly defined as $Q_{\rvar{Y}}(u) = \inf\{y \in \R: \Pr{\rvar{Y} \leq y} \geq u\}$. Of note are two important facts concerning $Q_{\rvar{Y}}$: (i) it is a unique monotonic map that maps a uniform random variable $\rvar{U} \sim [0, 1]$ to $\rvar{Y}$; and (ii) it is the optimal transport map between $\rvar{U}$ and $\rvar{Y}$ where the ground cost is the negative inner product.

\paragraph{Quantile regression.} Given a scalar-valued response $\rvar{Y}$ and covariates $\rvec{X}$, \textit{quantile regression} (QR) aims to estimate the quantile of a variable $\rvar{Y}|\rvec{X}$. 
The standard approach for solving QR is via the minimization of the \textit{pinball loss}.
\cite{Carlier2016} showed that QR can be equivalently written as an optimal transport problem between $\rvar{U}$ and $\rvar{Y}$, with the ground cost as the negative inner product, and subject to additional mean-independence constraint on the transport plan which depends on $\rvec{X}$. While the pinball loss, being a pointwise loss, is not well-defined for vector-valued targets, the optimal transport formulation of QR can be generalized to vector-valued targets by simply modifying the ground cost. This gave rise to \textit{vector quantile regression} (VQR), a multivariate analogue of QR. The goal of the current work is to further generalize VQR to densities defined on manifolds.

\paragraph{$c$-convexity.} In this paper, we consider Riemannian $d$-dimensional manifolds $(\Manifold, g)$ with the Riemannian metric $g$, embedded in $\R^D$. A function $\varphi: \Manifold \rightarrow \R \cup \{ + \infty\}$ is $c$-\textit{convex} with respect to cost $c$ if and only if it is not identically $+\infty$ and there exists an ${\alpha}: \Manifold \rightarrow \mathbb{R} \cup \{ \pm \infty\}$ such that,
\begin{equation*}
\label{eq:cconvex}
\varphi(\vec{y}) = \sup_{\vec{z} \in \Manifold} \{ -c(\vec{z}, \vec{y}) + {\alpha}(\vec{z})\}.
\end{equation*}
Similarly, a $c$-concave function $\tilde{\varphi}: \Manifold \rightarrow \R \cup \{-\infty\}$ is defined as, 
\begin{equation*}
\label{eq:cconcave}
\tilde{\varphi}(\vec{y}) = \inf_{\vec{z} \in \Manifold} \{ c(\vec{z}, \vec{y}) + {\alpha}(\vec{z})\}.
\end{equation*} 
Furthermore, if $\varphi$ is $c$-convex, then $- \varphi$ is $c$-concave, and
a convex-combination of $c$-convex functions is $c$-convex~\citep{villani2021topics}.
The $c$-transform of a function $\varphi : \Manifold \rightarrow \mathbb{R}$ is defined via infimal convolution, 
\begin{equation*} \label{eq:ctransform}
\varphi^c(\vec{u}) = \inf_{\vec{z} \in \Manifold} \{ c(\vec{z}, \vec{u}) - \varphi(\vec{z})\}.
\end{equation*}
Lastly, a function $\varphi$ is $c$-convex if and only if it satisfies the involution property: $\varphi^{cc} = \varphi$. In the case of squared Euclidean cost, $c$-convexity reduces to standard convexity, and $c$-transform becomes the well-known \textit{Legendre-Fenchel transform}.

\paragraph{Riemannian optimal transport.} Given two probability distributions and a cost function, the optimal transport (OT) problem consists of finding the mapping that pushes one distribution into the other while minimizing the overall cost. Here we are interested in distributions defined on $\Manifold$, and a cost function given by $c(\vec{y},\vec{z}) = \frac{1}{2} d_{\Manifold}(\vec{y},\vec{z})^2$, where $d_{\Manifold}$ is a geodesic distance on ${\Manifold}$. We refer to the associated optimal transport problem as \textit{Riemannian OT}.
Formally, we define $\mathcal{S}(\mu,\nu)$ as the set of maps from $\Manifold$ to $\Manifold$, pushing a base probability measure $\mu \in \Prob{\Manifold}$ to a target measure $\nu \in \Prob{\Manifold}$, i.e. $\mathcal{S}(\mu,\nu) = \{\vec{s}: \Manifold \rightarrow \Manifold | \vec{s}_{\#}\mu = \nu$\}. The Monge problem on $\Manifold$ with cost $c$ consists of finding
\begin{equation} \label{eq:Monge_cost}
   \inf_{\vec{s} \in \mathcal{S}(\mu,\nu)} \int_{\Manifold} c(\vec{u},\vec{s}(\vec{u})) d\mu (\vec{u}).
\end{equation}
The optimal $\vec{s}(\vec{u})$ in \eqref{eq:Monge_cost} is the optimal transport map.
To avoid the non-convex optimization in Monge's formulation, we consider the Kantorovich relaxation, in which we search for a joint distribution of the measures instead of a map between them. Denoting by $\Gamma(\mu,\nu)$ the set of joint distributions on $\Manifold \times \Manifold$ that admit $\mu$ and $\nu$ as marginals, the Kantorovich formulation is given by
\begin{equation} \label{eq:Kantorovich_cost}
   \inf_{\gamma \in \Gamma(\mu,\nu)} \int_{\Manifold \times \Manifold} c(\vec{u},\vec{y}) d\gamma (\vec{u},\vec{y}),
\end{equation}
and its dual formulation is given by,
\begin{equation}
    \label{eq:dual_kantorovich}
    \begin{split}
    \sup_{\varphi,\psi} \int_{\Manifold} \varphi(\vec{u}) d\mu(\vec{u}) + \int_{\Manifold} \psi(\vec{y}) d\nu(\vec{y}) \\
    \text{s.t.} \quad \varphi(\vec{u}) + \psi(\vec{y}) \leq c(\vec{u},\vec{y}),
    \end{split}
\end{equation}
where $\varphi,\psi : \Manifold \rightarrow \mathbb{R}$ are bounded and continuous $c$-concave functions, referred to as the \textit{potential functions}.
As shown in Theorem 9 of \citet{mccann2001polar}, the optimal transport map $\vec{s}: \Manifold \rightarrow \Manifold$ is a unique minimizer of the Monge problem (\ref{eq:Monge_cost}) and can be obtained from the potential $\varphi$ as,
\begin{equation} \label{eq:transport_plan}
   \vec{s}(\vec{u}) = \expmap_{\vec{u}}[-\nabla_{\vec{u}}
 {\varphi}(\vec{u})]
\end{equation}
where $\nabla$ is the intrinsic gradient on $\Manifold$, and $\expmap$ is the exponential map.
The optimal dual potentials, referred to as $\varphi^*$ and $\psi^*$, are each other's $c$-transform:
\begin{align*}
& \varphi^*(\vec{y}) = \inf_{\vec{z} \in \Manifold} \{ c(\vec{z}, \vec{y}) - {\psi^*}(\vec{z})\} \\
& \psi^*(\vec{y}) = \inf_{\vec{z} \in \Manifold} \{ c(\vec{z}, \vec{y}) - {\varphi^*}(\vec{z})\}
\end{align*}
and, therefore, both $c$-concave.
For a more extensive overview of Optimal Transport and $c$-convexity, we refer to \cite{villani2009optimal} and \cite{ santambrogio2015optimal}.

\section{\uppercase{Quantile Regression on Manifolds}}

\subsection{Vector quantile functions on manifolds}
Let $\rvec{Y}$ be a random variable supported on the manifold $\Manifold$, with distribution $\nu$. We define the \textit{manifold uniform distribution} $\mathcal{U}_{\mathcal{M}}$ on $A \subseteq \Manifold$ as the distribution having the density
$p_{\rvec{Y}}(\vec{y})=\1{A}(\vec{y})/{V(A)}$,
where $V(A)$ denotes the volume of $A$.
The manifold vector quantile function (M-VQF) of $\rvec{Y}$, $Q_{\rvec{Y}}: \Manifold \rightarrow \Manifold$, is defined by the OT map obtained as a solution to the Riemannian OT problem \eqref{eq:Monge_cost}, where the base distribution is $\mu = \mathcal{U_{\Manifold}}$ and the target distribution $\nu$ is that of $\rvec{Y}$~\citep{hallin2022nonparametric}. We refer to this problem as manifold vector quantile estimation (M-VQE). The resulting M-VQF can therefore be written as,
$$
Q_{\rvec{Y}} (\vec{u}) = \text{exp}_{\vec{u}} \left[ - \nabla_{\vec{u}} \varphi(\vec{u}) \right].
$$
In the Euclidean case, the exponential map is identity and $c$-convexity is simply convexity, so the M-VQF reduces to the Euclidean vector quantile function as defined in \cite{Carlier2016, chernozhukov2017monge, rosenberg2022fastnlvqr}.
Moreover, when the random variable is Euclidean and scalar-valued, this definition naturally recovers the one-dimensional quantile function.

\subsection{Extension to quantile regression}
In the regression case, we estimate the conditional quantile function of $\rvec{Y} | \rvec{X} = \vec{x}$, denoted by $Q_{\rvec{Y}|\rvec{X}}$.
This requires solving a family of OT problems parameterized by $\vec{x}$.
Following the approach of \cite{Carlier2016} for the Euclidean domain, we average the OT losses over $\vec{x}$.
Denoting by $\xi$ the distribution of the joint variable $(\rvec{X}, \rvec{Y})$ and by $\mu = \mathcal{U_{\Manifold}}$, the uniform base distribution, the quantile regression problem on manifolds becomes:
\begin{equation}
    \label{eq:dual_OT}
    \small
    \begin{split}
    \sup_{\varphi,\psi} 
    \int\limits_{(\mathcal{X} \times \Manifold ) \times \Manifold} \left( \varphi(\vec{u}; \vec{x}) + \psi(\vec{y}; \vec{x}) \right)d\xi(\vec{x}, \vec{y})d\mu(\vec{u})  \\
    \text{s.t. } \forall\; \vec{x}, \vec{y}, \vec{u} \quad \varphi(\vec{u}; \vec{x}) + \psi(\vec{y}; \vec{x}) \leq c(\vec{u},\vec{y}) 
    \end{split}
\end{equation}
where $\varphi,\psi : \Manifold \times \mathcal{X} \rightarrow \mathbb{R}$ are bounded and continuous $c$-concave functions in $\mathcal{\mathbf{u}}$.
The manifold conditional vector quantile function (M-CVQF) will thus be a map $\Manifold \times \mathcal{X} \rightarrow \Manifold$ such that:
\begin{equation*}
    Q_{\rvec{Y}|\rvec{X}}(\vec{u}, \vec{x}) = \expmap_{\vec{u}}[-\nabla_{\vec{u}} {\varphi}(\vec{u}, \vec{x})].
\end{equation*}

\paragraph{Discretization.}
We can dicretize Equation \ref{eq:dual_OT} by sampling $\left\{ \vec{u}_i \right\}_{i=1}^T  \sim \mathcal{U}_\Manifold$ from $\mu$ and $\left\{ (\vec{y}_j, \vec{x}_j)\right\}_{j=1}^N$  from $\xi$, getting:
\begin{equation*} \label{eq:OT_discr_vqr}
    \begin{split}
     \max_{\varphi,{\psi}} 
      \sum_{i=1}^T {\mu_i}  \sum_{j=1}^N {\xi_j} \varphi(\vec{u}_i;\vec{x}_j)  +
     \sum_{j=1}^N {\xi_j} {\psi}(\vec{y}_j;\vec{x}_j)  \\
    \text{s.t.}\,\, \forall i,j: ~ \varphi(\vec{u}_i;\vec{x}_j) +{\psi}(\vec{y}_j;\vec{x}_j) \leq c(\vec{u}_i,\vec{y}_j).
    \end{split}
\end{equation*}

By writing one of the potentials as the $c$-transform of the other we obtain the following optimization problem,
\begin{align}\label{eq:vqr_loss}
    \max_{\varphi}
  & \sum_{i=1}^T {\mu_i}  \sum_{j=1}^N {\xi_j} \varphi(\vec{u}_i;\vec{x}_j) \\
\nonumber + & \sum_{j=1}^N \xi_j \min_{\vec{u} \in \Manifold} \left\{ c(\vec{u},\vec{y}_j) -  \varphi(\vec{u};\vec{x}_j) \right\},
\end{align}
with $\vec{\mu }= \frac{1}{T} \mat{1}_T$, $\vec{\xi} = \frac{1}{N} \mat{1}_N$ and where  $\varphi(\vec{u};\vec{x}): \Manifold \times \mathcal{X} \rightarrow \R$ is a partial $c$-concave function in $\vec{u}$. We refer to this problem as manifold vector quantile regression (M-VQR).
The other potential can be retrieved via the $c$-transform,
\begin{equation}\label{eq:vqr_dual}
    {\psi}(\vec{y};\vec{x}_j) = \min_{\vec{u} \in \Manifold} \left\{ c(\vec{u},\vec{y}) - \varphi(\vec{u};\vec{x}_j) \right\}.
\end{equation}

\subsection{Confidence sets} \label{sec:confidence_sets}
With the estimated quantile function $Q_{\rvec{Y}|\rvec{X}}$, we can compute \textit{confidence sets} on the target distribution.
To define a set of nested regions with a $\mu$-probability content of $\tau \in [0,1]$ on $\Manifold$, we must first choose a central point, or \textit{pole}, $\vec{\omega} \in \Manifold$. This point will play the role of the median for $\nu$, around which the contours will be nested. 
We opted to compute the pole using the \textit{Fr\'echet mean} of $\nu$, defined as
\begin{equation}
    \label{eq:frechetmean}
    \vec{\omega} = \argmin_{\vec{y} \in \Manifold} \mathbb{E}_{\rvec{y}\sim\nu}[ c( \rvec{y}, \vec{y})];
\end{equation}
other choices are also possible. Under the base distribution, the $\tau$-contour, containing $\mu$-probability of $\tau$ and centered at $\vec{\omega}$, can be defined as
\begin{equation}
    \label{eq:contour_base}
    \mathcal{C}_{\tau}^{\rvec{U}} = \left\{\vec{u} \in \Manifold : 
    C^*_{\vec{\omega}}( \vec{u}) = \tau \right\}
\end{equation}

where $C^*_{\vec{\omega}}$ is a function mapping the geodesic distance of $\vec{u}$ from the pole $\vec{\omega}$ to the probability $\tau$. A more detailed explanation on how we define $C^*_{\vec{\omega}}$ is provided in Section \ref{app:conf_sets} of the Appendix.
The $\tau$ conditional quantile contour, under the target distribution of $\rvec{Y}|\rvec{X}$, can then be obtained via the image $\mathcal{C}_{\tau}^{\rvec{Y}|\rvec{X}}:=Q_{\rvec{Y}|\rvec{X}}(\mathcal{C}_{\tau}^{\rvec{U}}; \vec{x})$.
\Cref{fig:vqr_cont} shows $\mathcal{C}_{\tau}^{\rvec{Y}|\rvec{X}}$ for multiple distributions and values of $\tau$.

\subsection{\label{sec:lh}Likelihood} 
Once the conditional quantile function has been estimated, we can use it to also obtain the conditional likelihood function $p_{\rvec{Y}|\rvec{X}}(\vec{y};\vec{x} )$:
\begin{align*}
    & p_{\rvec{Y} | \rvec{X}}(\vec{y}; \vec{x}) = 
    \label{eq_lh}
     \frac{1}{V(\mathcal{M})} \cdot
    \left|\mat{\nabla}_{\vec{y}}Q^{-1}_{\rvec{Y}|\rvec{X}}(\vec{y}; \vec{x})\right| 
\end{align*}
where $\mat{\nabla}$ denotes the Jacobian on $\Manifold$ and $\rvec{U} \sim \mathcal{U}_{\Manifold}$ (with a slight abuse of notation).

\section{\label{sec:impl}\uppercase{Implementation Details}}
Below we discuss details related to the implementation of $c$-concave functions, partial input $c$-concave neural networks, and regularization techniques employed.

\paragraph{Discrete $c$-concave functions.} The fundamental building block for our input c-concave potentials are $c$-concave functions. Following \cite{cohen2021rcpm}, we parameterize each $c$-concave function $\varphi$ with $\{(\vec{z}_i, {\alpha_i})\}_{i=1}^M \subset \Manifold \times \R$, where ${\alpha_i}=\alpha(\vec{z}_i)$ are the (learned) values of an implicitly defined function on $\Manifold$, and $\vec{z}_i$ are the (learned) points on which it is sampled. A $c$-concave function is thus obtained by applying the $c$-transform to the implicit function $\alpha(z)$:
\begin{equation}
\label{eq:cconvex_params}
{\varphi}(\vec{u}) = \min_{i=1,\dots,M} \left\{ c(\vec{z}_i, \vec{u}) + {\alpha}(\vec{z}_i)\right\}.
\end{equation}
This formulation results in a piecewise smooth \textit{approximation} of the $c$-concave potential. \cite{cohen2021rcpm} prove that this discrete approximation of $c$-concave potential has the expressive power to represent arbitrary $c$-concave potentials on compact manifolds.

\paragraph{Input $c$-concave networks.}
We use two properties of $c$-concave functions to build input $c$-concave neural networks: (i) a convex combination of $c$-concave functions is $c$-concave, and (ii) applying a concave and monotone function to a $c$-concave function retains its $c$-concavity.
This is similar in spirit to input-convex neural networks~\citep{amos2017input}.
An $L$-layered $c$-concave function $\beta$ can be obtained by combining $L+1$ $c$-concave functions $\varphi_i$, as follows:
\begin{equation}\label{eq:cconv_impl}
\begin{split}
   \beta_0(\vec{u}) = &\, \varphi_0(\vec{u}) \\
   \beta_l(\vec{u}) = &\, (1-\omega_{l}) \varphi_{l}(\vec{u}) + \omega_{l} \sigma( \beta_{l-1}(\vec{u})),
\end{split}
\end{equation}
where $l\in\{1,\dots,L\}$, $\omega_l\in[0,1]$ are learnable weights, and $\sigma(s)=\min\{0,s\}$ is a concave monotone function.

\paragraph{Partial input $c$-concave networks.} 
In the regression setting, we model the potentials as a partially input $c$-convex neural networks: $c$-convex in $\vec{u}$ but not in $\vec{x}$.
A similar approach was also used in \cite{bunne2022supervised} where they model conditional OT maps on Euclidean domain using partially input convex neural networks.
Assuming $\mathcal{X} = \mathbb{R}^k$, we implement a partial $c$-concave potential $\varphi(\vec{u};\vec{x})$ as a non-negative sum of $n+1$ functions which are $c$-concave in $\vec{u}$, and apply non-linear trainable transformations on $\vec{x}$:
\begin{equation}\label{eq:part_cconv}
\begin{split}
   \varphi_0(\vec{u};\vec{x}) = & \beta_0(\vec{u}),  \hspace{1cm} \beta_0: \Manifold \rightarrow \R \\
   \varphi_i(\vec{u};\vec{x}) = &  \varphi_{i-1}(\vec{u};\vec{x}) + \vec{\beta}_i(\vec{u})\T \vec{g}_i\circ...\circ\vec{g}_1(\vec{x}), \\
   \varphi(\vec{u};\vec{x}) = & \varphi_n(\vec{u};\vec{x}) 
\end{split}
\end{equation}
with $i \in \{1, \dots, n\}$, $\vec{\beta}_i: \Manifold \rightarrow \R^{k_i}$ and $\vec{g}_i(\vec{x}): \R^{k_{i-1}} \rightarrow \R_{+}^{k_i}$ ($k_0 = k$) parametrized using neural networks.
 In particular, each $\vec{\beta}_i$ is implemented as a $k_i$-stack of $c$-concave functions (\ref{eq:cconvex_params}):
$\vec{\beta}_i = \left( \beta_{i1},\dots,  \beta_{ik_i} \right)$ where $\beta_{ij}: \Manifold \rightarrow \R $ with $j \in \{1, \dots, {k_i}\}$.
Conversely, the $\vec{g}_i$'s can be arbitrary learned embedding functions. For example if $\rvec{X}$ are images, a CNN-based $\vec{g}(\vec{x})$ can be used to leverage translation equivariance and the hierarchical nature of image features.
In our experiments, we implement the $\vec{g}_i$'s as a multilayer perceptron (MLP) with rectified linear unit (ReLU) activation as the last layer to ensure non-negative values. 

\begin{figure}[!t]
    \centering
    \begin{overpic}
    [trim=1.2cm 1cm 2cm 1.6cm,clip,width=0.7\linewidth, grid=false]{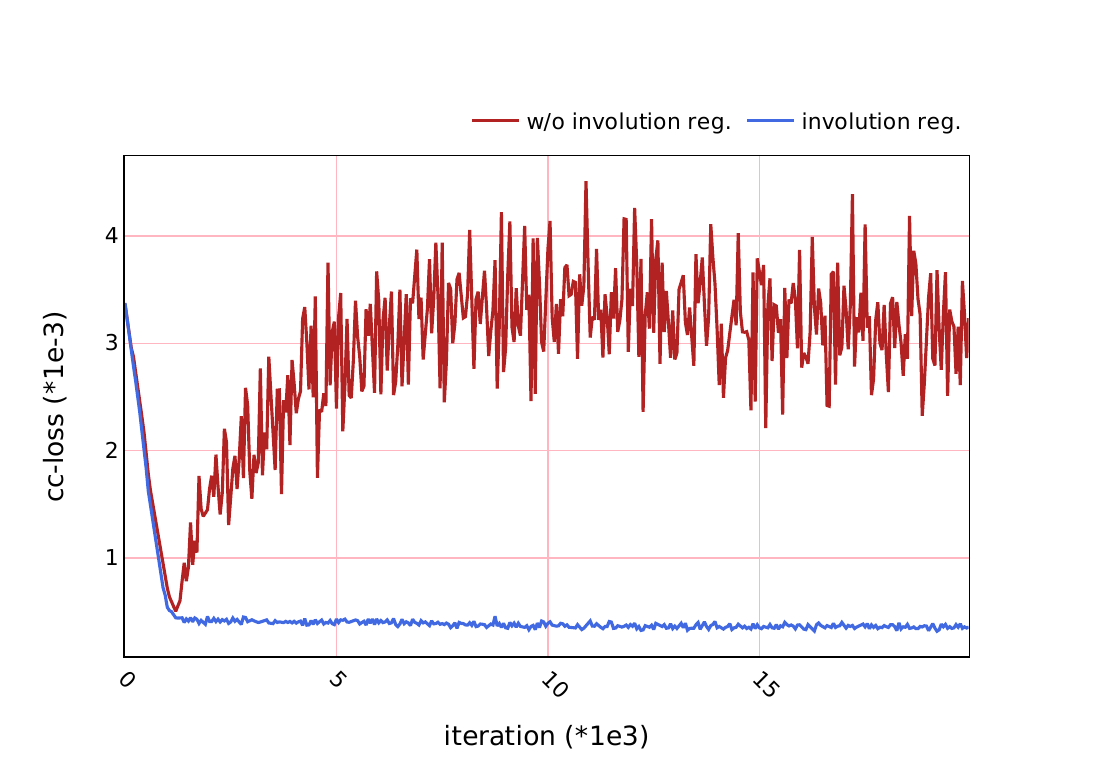}
    \put(39,-1){\footnotesize iteration ($\times 10^{3}$)}
    \put(-3,35){\footnotesize \rotatebox[origin=c]{90}{involution error ($\times 10^{-3}$)}}
    \end{overpic}
    \caption{\textbf{Impact of involution regularization.} The results are from a M-VQE trained on $\mathcal{S}^2$, where the target distribution is a von-Mises distribution and the c-concave potential consists of 3 layers and $\gamma=0.1$. Involution error is dramatically reduced when training with the involution regularization.}
    \label{fig:ccloss}
\end{figure}

\paragraph{Computing convex conjugates.} Given a $c$-concave potential $\varphi(\vec{u}, \vec{x})$, its convex conjugate is defined by the $c$-transform,
$$
\varphi^c(\vec{y}; \vec{x}) = \inf_{y \in \Manifold} \left\{c(\vec{u}, \vec{y}) - \varphi(\vec{u}; \vec{x}) \right\}.
$$
In practice, we compute this conjugate numerically by sampling several points on $\Manifold$, and evaluating the conjugate explicitly, i.e.,
\begin{equation}
\label{eq:conv_conjugate}
\varphi^c(\vec{y}; \vec{x}) = \min_{i=1 \ldots T} \left\{ c(\vec{u}_i, \vec{y}) - \varphi(\vec{u}_i; \vec{x}) \right\},
\end{equation}
where $\vec{u}_1 \ldots \vec{u}_T \sim \mathcal{U}_{\Manifold}$. 

Both in Equation \ref{eq:cconvex_params} and \ref{eq:conv_conjugate}, we replace the minimum with a soft-minimum to maintain differentiability: 
$$
    \min_\gamma(a_1,\cdots, a_n) = - \gamma \log \sum^n_{i=1} \exp{- \frac{a_i}{\gamma}}
$$
with $\min_\gamma \rightarrow \min$ as $ \gamma \rightarrow 0$. Notice that using the soft-minimum is {equivalent to} performing {entropic OT} (see Appendix A.4, A.5 in \cite{rosenberg2022fastnlvqr}).

\begin{figure*}[t]
\centering
\begin{subfigure}[b]{0.49\linewidth}
\centering
\begin{overpic}
[trim=2cm 5.5cm 3cm 6cm,clip,width=\linewidth, grid=false]{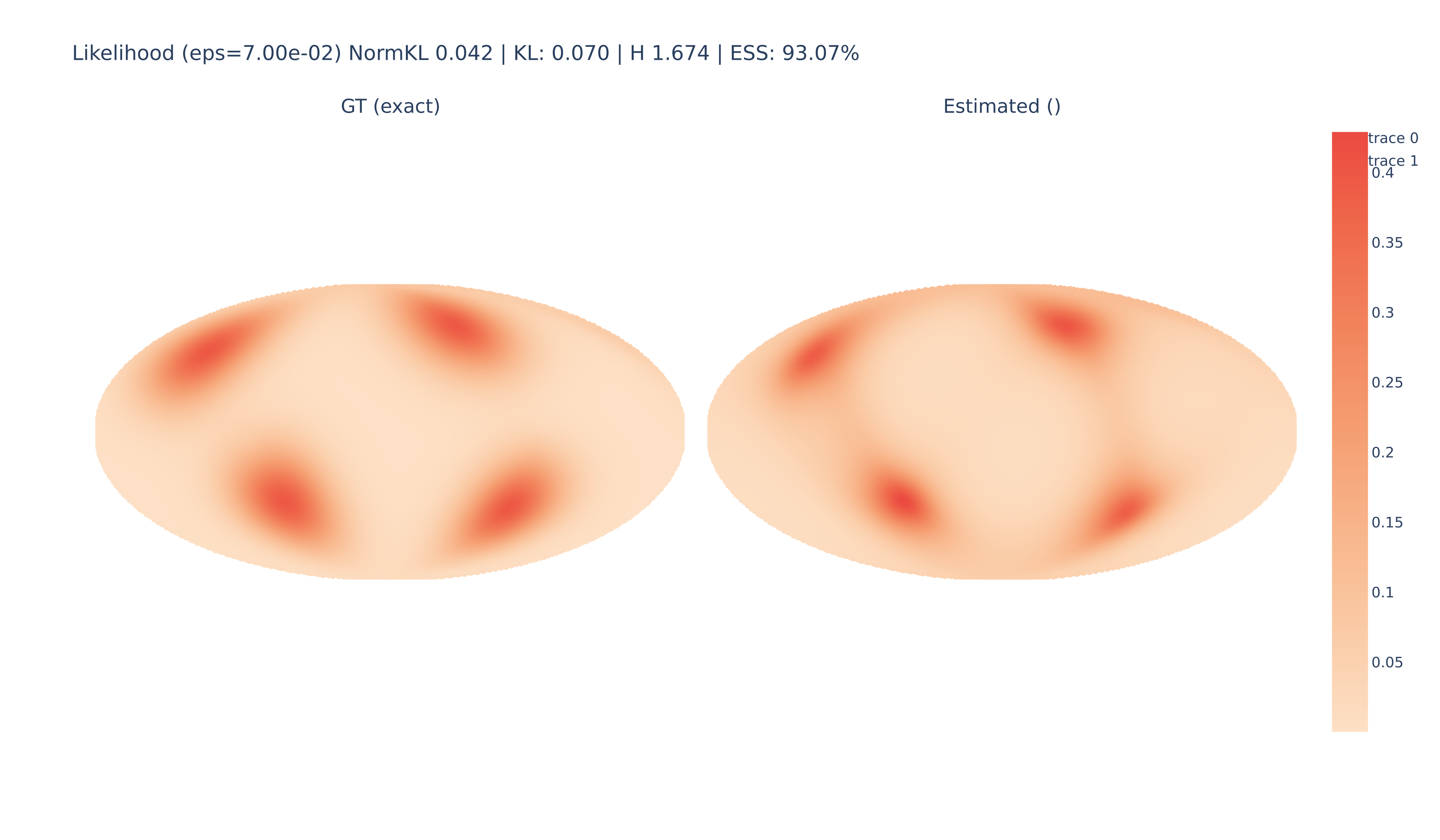}
\put(11,26){Ground Truth}
\put(66,26){Estimated}
\put(54,-1.5){\tiny  $ESS_\% = 93\%$} 
\end{overpic}
\caption{\label{fig:vqe_lh} Likelihood}
\end{subfigure} 
\hfill
\begin{subfigure}[b]{0.29\linewidth}
\centering
\begin{overpic}
[trim=1.9cm 4.7cm 2.85cm 5.2cm,clip,width=0.88\linewidth, grid=false]{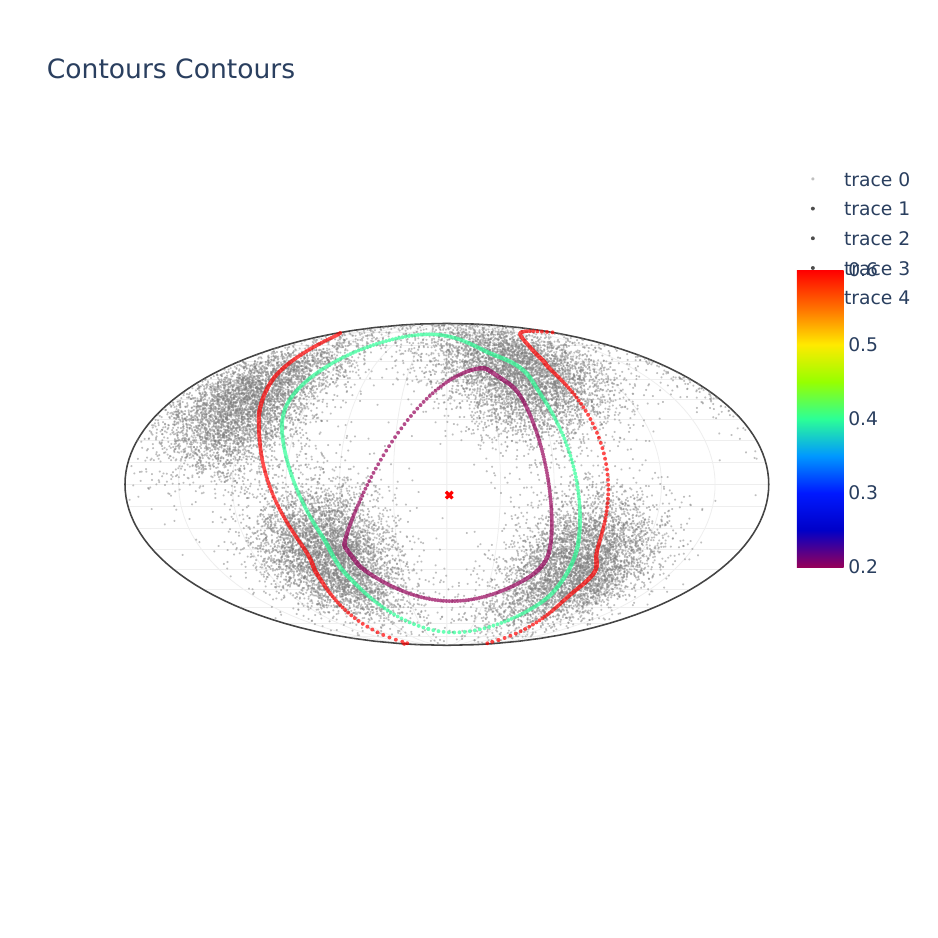}
\end{overpic}
\begin{overpic}
[trim=4.72cm -2cm 3.45cm 2.79cm,clip,width=0.1\linewidth, grid=false]{./figs/synth/legend_synth}
\put(10,105){\footnotesize $\tau$}
\end{overpic}
\caption{\label{fig:vqe_cont} $\tau$-contours}
\end{subfigure} 
\hfill
\begin{subfigure}[b]{0.2\linewidth}
\centering
\begin{overpic}
[trim=0cm 0cm 1cm 1.5cm,clip,width=\linewidth, grid=false]{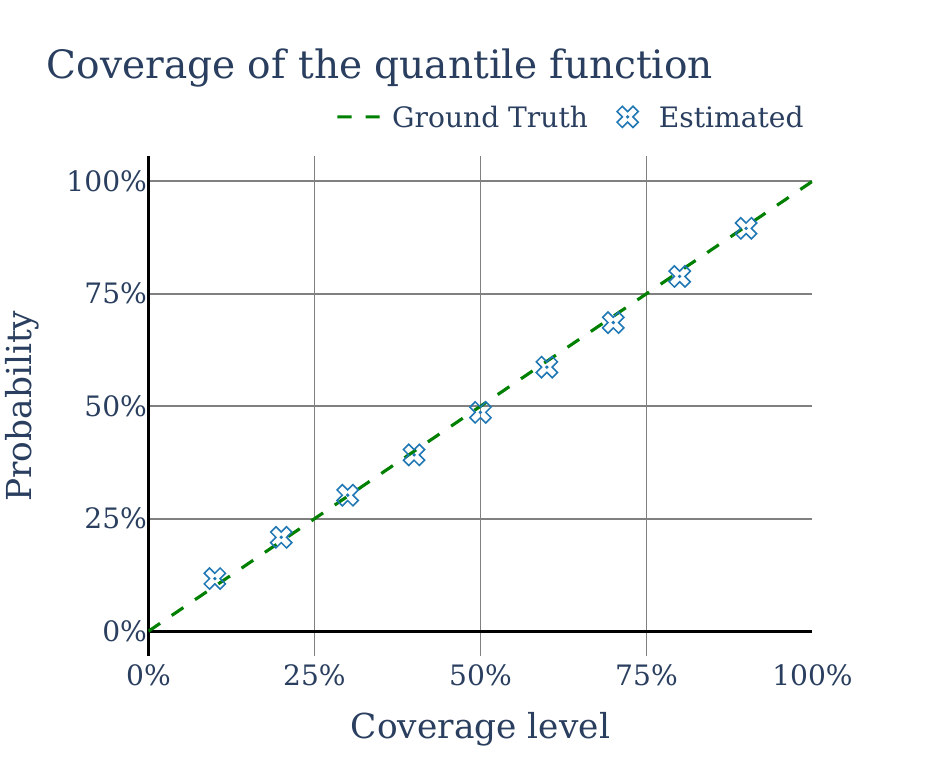}
\end{overpic}
\caption{\label{fig:vqe_cov} Coverage}
\end{subfigure} 
\caption{\label{fig:vqe}
\textbf{M-VQE approximation of the quantile function $Q_{\rvec{Y}}$ of the `Multimodal von-Mises' distribution produces nested, smooth, and valid contours and correctly estimates the likelihood function.} Subfigure (a) shows ground truth and estimated likelihood functions. We use the Mollweide projection to plot the whole sphere surface. Subfigure (b) shows $\tau$-contours overlayed on the ground truth samples, for different values of $\tau$. Graph (c) plots the requested coverage level on the horizontal axis and the coverage achieved by the model on the vertical axis.}
\end{figure*}

\paragraph{Involution regularization.}
The use of the soft-minimum, while improving gradient computation over the $c$-concave function parameters, may compromise the $c$-concavity of the network. This compromise can lead to a violation of the involution property during training, as already noted in \cite{cohen2021rcpm} and demonstrated in Figure \ref{fig:ccloss}. To address this issue, we introduce an extra term in the VQR loss \eqref{eq:vqr_loss} which penalizes deviations the learned $c$-concave potentials may have from the involution property.
Given $\left\{ \vec{u}_i \right\}_{i=1}^T  \sim \mathcal{U}_\Manifold$ and $\left\{ \vec{y}_j, \vec{x}_j\right\}_{i=1}^N  \sim P_{(\rvec{Y},\rvec{X})}$, we compute the \textit{involution regularization} as: 
\begin{equation}
    \label{eq:inv_reg}
    \sum_{j=1}^{N} \sum_{i=1}^{T} || \varphi^{cc} (\vec{u}_i; \vec{x}_j) - \varphi(\vec{u}_i; \vec{x}_j)||,
\end{equation}
where $\varphi^{cc}$ is obtained by computing the numerical conjugate via equation \eqref{eq:conv_conjugate} twice on $\varphi$. This regularization is conceptually similar to the cycle-consistency regularization proposed by \cite{korotin2019wasserstein}. However, our approaches differ in two ways: (i) cycle consistency implies involution but only up to a constant, and (ii) we train a single c-convex potential and involution regularizer is used to ensure the c-convexity of the potential, whereas  \cite{korotin2019wasserstein} train two potentials and cycle consistency is employed for promoting convex-conjugacy between them. \Cref{fig:ccloss} demonstrates that this regularizer effectively reduces the involution error, and thus strongly promotes $c$-concavity of the learned potentials.

\paragraph{Identity initialization.} Similarly to \cite{korotin2019wasserstein} and \cite{cohen2021rcpm}, we pre-train our model to represent the identity map.  In practice, we observe that this serves as a good initialization for training.

\begin{table*}[!t]
\centering
\caption{\textbf{M-VQR approximation of $Q_{\rvec{Y}|\rvec{X}}$ allows building confidence sets with good marginal coverage, to perform good quality conditional sampling, and to approximate the likelihood accurately.} The table reports on the left marginal coverage values for different confidence sets built using the estimated quantile function $Q_{\rvec{Y}|\rvec{X}}$, averaged over 50 values of the continuous conditioning $\rvec{X}$, with relative standard deviation. KDE$-L_1$ values in the are also averages over 50 values of $\rvec{X}$. The last column reports the {ESS}$_\%$ computed over samples drawn from the whole distribution ${\rvec{Y}|\rvec{X}}$.}
    \label{tab:coverage_synth}
\begin{tabular}{cccccccc}
        & & \multicolumn{4}{c}{Coverage (\%)} & KDE-$L_1$ & \multirow{2}{*}{$ESS_\%$} \\ \cmidrule{3-6}
    $\Manifold$  & ${\rvec{Y}|\rvec{X}}$  & 20 & 30 &  60 & Mean error & ($\times 10^{-4}$)  \\ \toprule
\multicolumn{1}{l|}{\multirow{3}{*}{ $\Sphere{2}$ }}     & Cond. Multimodal  &  $20.22 \pm 0.44$ & $40.28 \pm 0.56$ & $60.22 \pm 0.49$ &  $  0.44\pm 0.34 $ & $ 14.2\pm  2.09$ & 94.93\%\\
 \multicolumn{1}{l|}{}     & Scaled Star  &  $19.50 \pm 0.51$ & $38.40 \pm 0.63$ & $59.35 \pm 0.70$ &  $  0.86\pm 0.64 $ & $ 2.25\pm  0.99$ & 89.30\%\\
 \multicolumn{1}{l|}{}     & Scaled Heart &  $20.57 \pm 0.46$ & $40.82 \pm 0.60$ & $60.02 \pm 0.53$ &  $  0.60\pm 0.42 $ & $ 4.46 \pm 1.18 $  & 89.84\%\\ \midrule
 \multicolumn{1}{l|}{\multirow{3}{*}{ $\mathcal{T}^2$ }} & Cond. Multimodal  &  $20.06 \pm 0.43$ & $40.68 \pm 0.61$ & $60.46 \pm 0.68$ &  $  0.54\pm 0.43 $ & $ 17.5\pm 2.92 $ & 96.82\%\\
\multicolumn{1}{l|}{} & Scaled Star  & $19.81 \pm 0.50$ & $39.86 \pm 0.55$& $60.39 \pm 0.61$  &  $  0.47\pm0.35  $ & $ 5.85 \pm 1.92 $ & 84.79\% \\
\multicolumn{1}{l|}{} & Scaled Heart  & $19.95 \pm 0.47$ & $40.50 \pm 0.65$& $61.18 \pm 0.60$  &  $ 0.61 \pm 0.49 $ & $7.93 \pm 2.14$ & 88.05\%\\
    \end{tabular}

\end{table*}
\begin{figure*}[t]
\centering
    \begin{subfigure}[b]{0.4\linewidth}
        \begin{subfigure}[b]{\linewidth}
        \centering
        \begin{overpic}
        [trim=1.9cm 4.3cm 2.8cm 4.5cm,clip,width=\linewidth, grid=False]{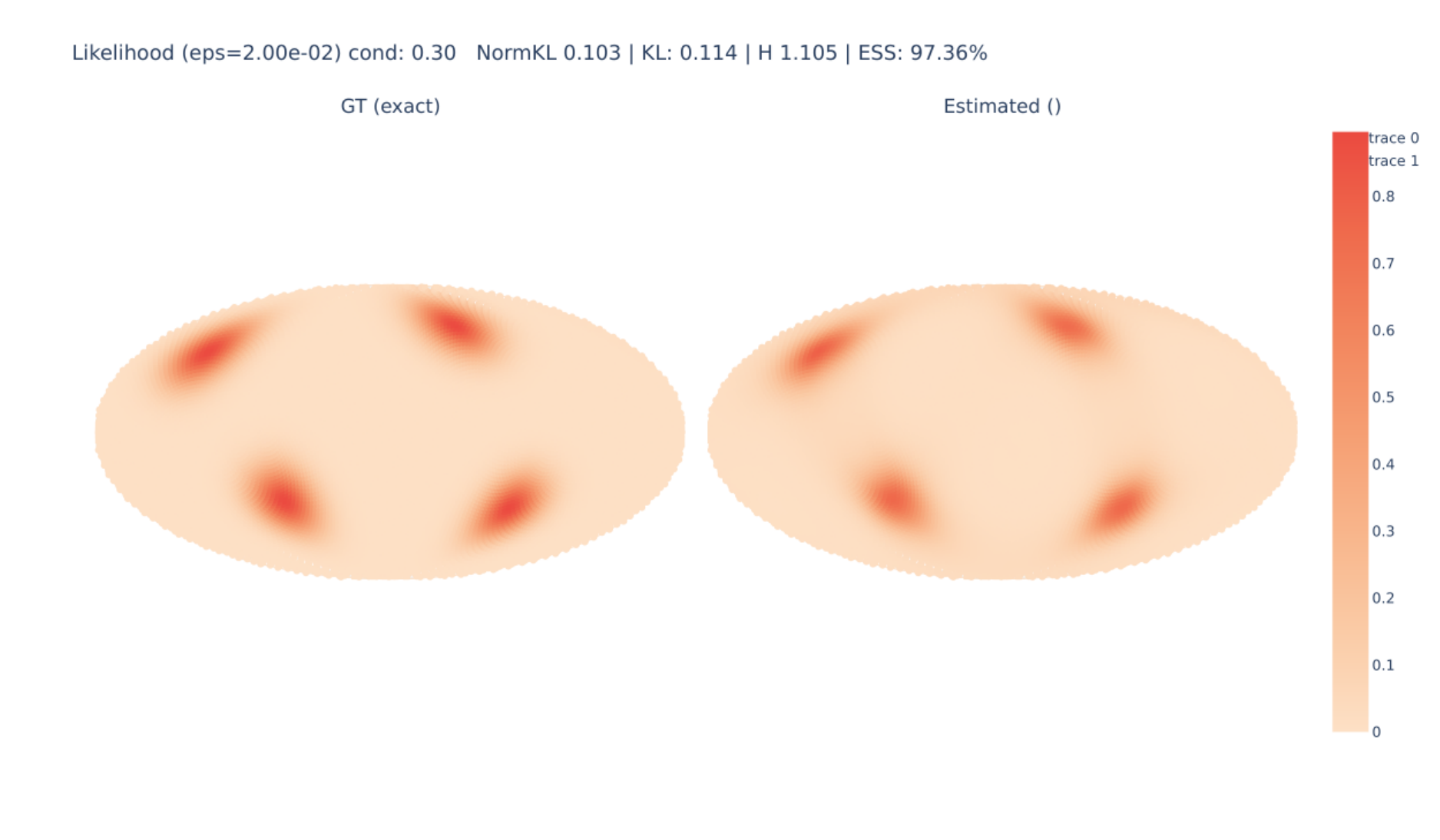}
        \put(9,31){Ground Truth}
        \put(64,31){Estimated}
        \put(53,2){\tiny $ESS_\% = 97\%$} 
        \put(-4,15){\rotatebox[origin=c]{90}{$x=0.3$}}
        \end{overpic}
        \end{subfigure} 
        \begin{subfigure}[b]{\linewidth}
        \centering
        \begin{overpic}
        [trim=1.9cm 4.3cm 2.8cm 4.5cm,clip,width=\linewidth, grid=false]{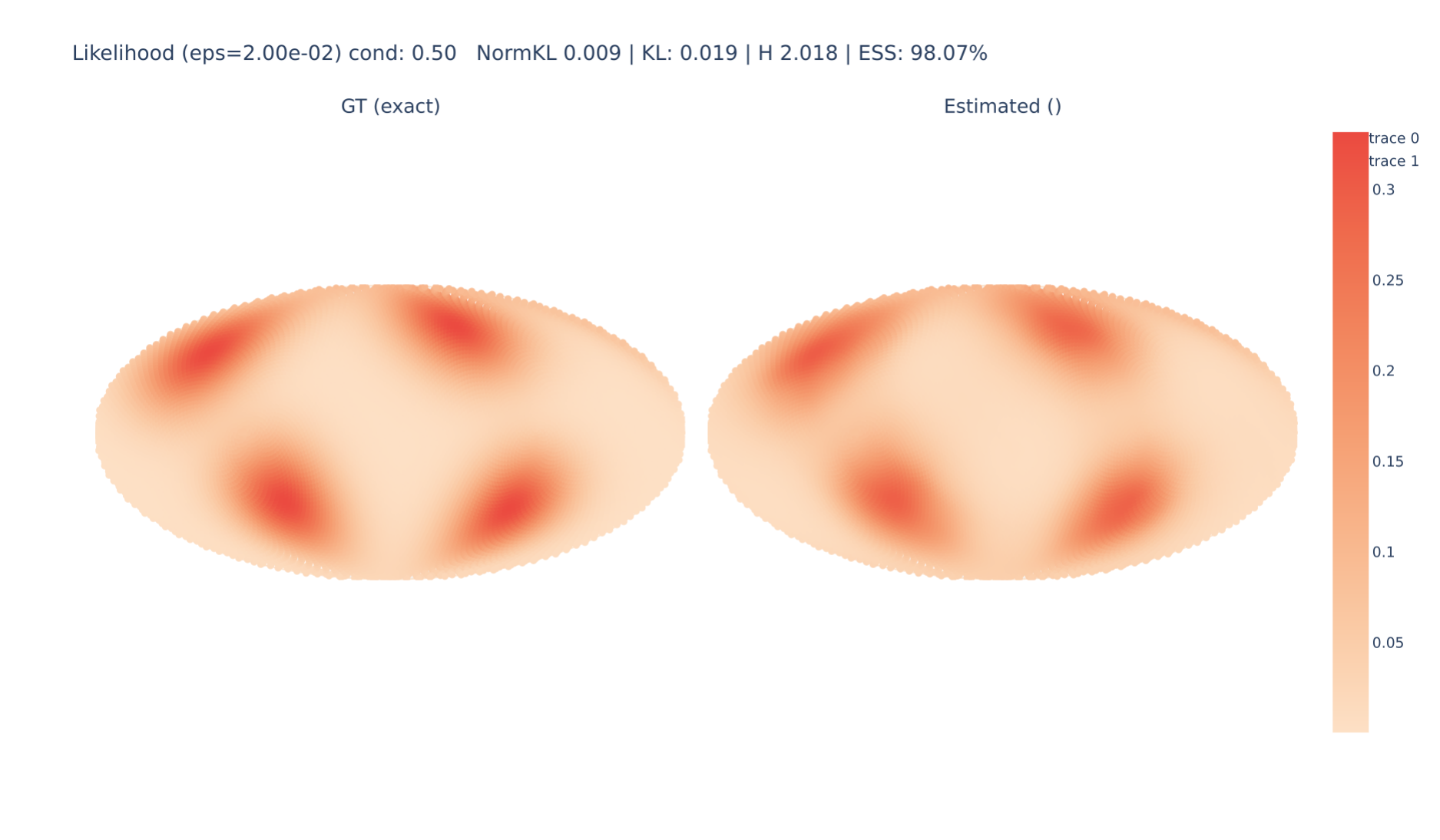}
        \put(53,2){\tiny $ESS_\% = 98\%$} 
        \put(-4,15){\rotatebox[origin=c]{90}{$x=0.5$}}
        \end{overpic}
        \end{subfigure} 
        \caption{\label{fig:s2_lh} $\Sphere{2}$ }
    \end{subfigure}
    \hspace{0.5cm}
    \begin{subfigure}[b]{0.4\linewidth}
        \begin{subfigure}[b]{\linewidth}
        \centering
        \vspace{0.3cm}
        \begin{overpic}
        [trim=1.7cm 6.8cm 3cm 10cm,clip,width=0.8\linewidth, grid=false]{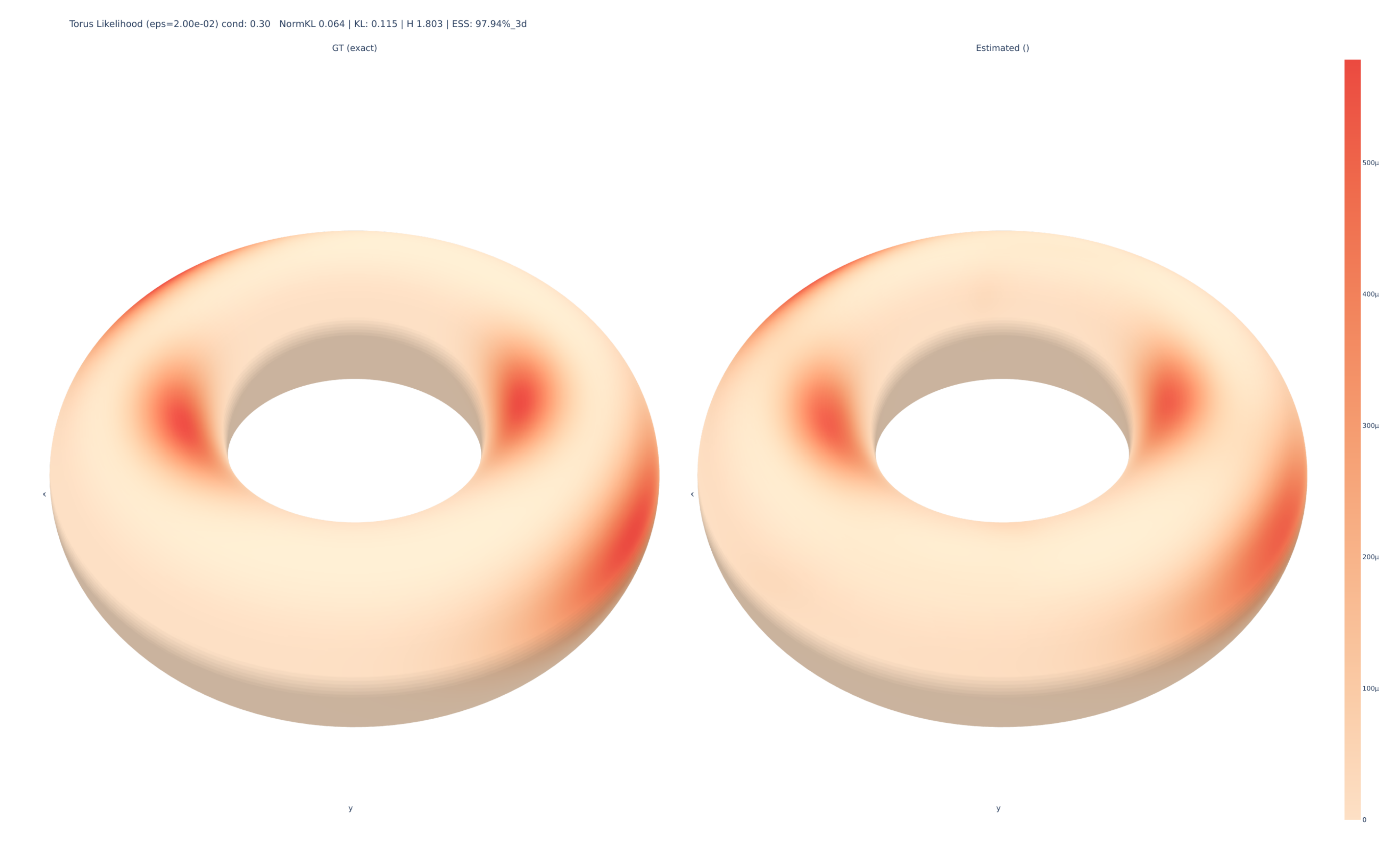}
        \put(4,40){Ground Truth}
        \put(60,40){Estimated}
        \put(85,0){\tiny $ESS_\% = 97\%$} 
        \put(-4,17){\rotatebox[origin=c]{90}{$x=0.3$}}
        \end{overpic}
        \end{subfigure} 
        \begin{subfigure}[b]{\linewidth}
        \centering
        \vspace{0.3cm}
        \begin{overpic}
        [trim=1.7cm 6.8cm 3cm 10cm,clip,width=0.8\linewidth, grid=false]{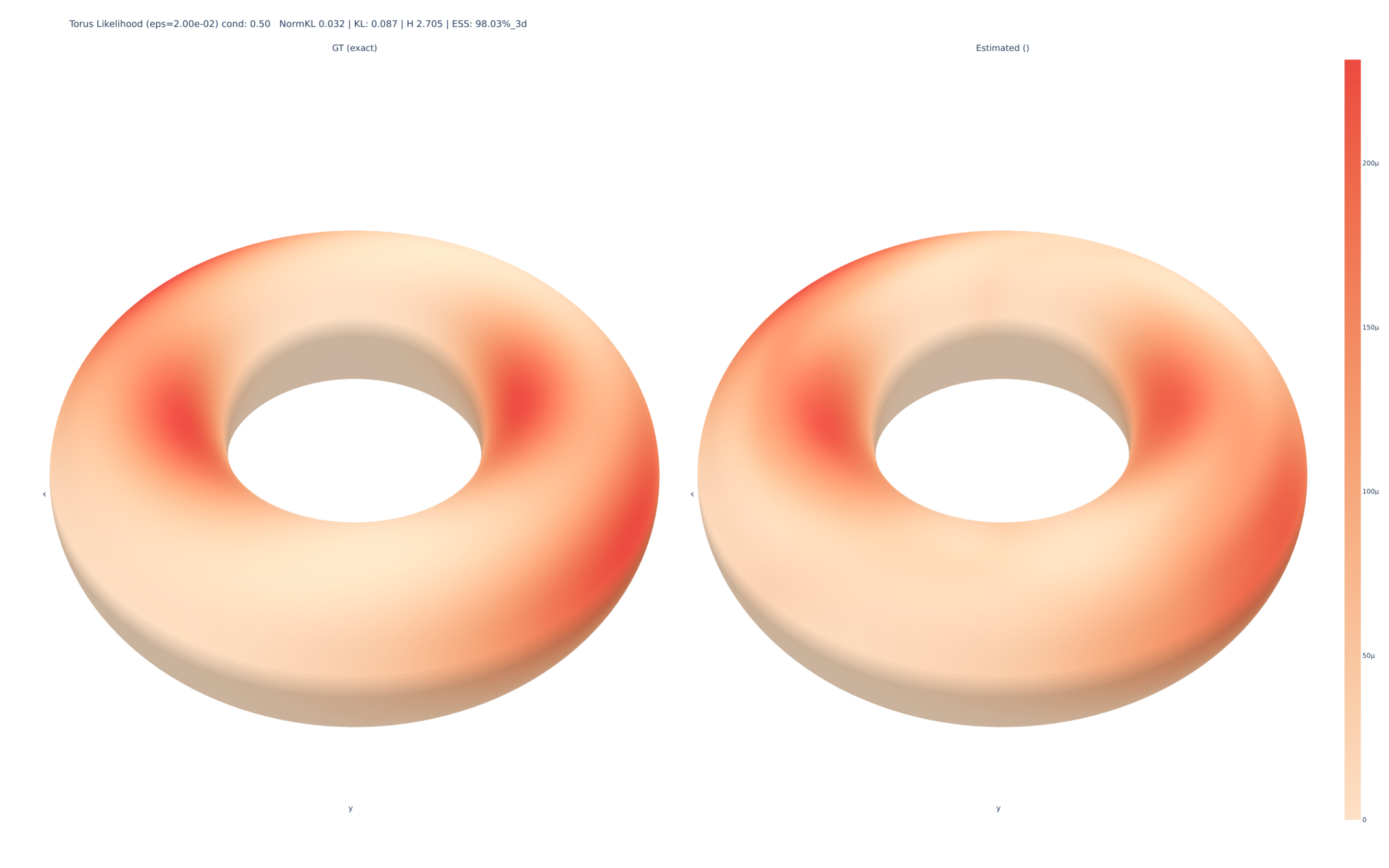}
        \put(85,0){\tiny $ESS_\% = 98\%$} 
         \put(-4,17){\rotatebox[origin=c]{90}{$x=0.5$}}
        \end{overpic}
        \end{subfigure} 
\caption{\label{fig:t2_lh} $\mathcal{T}^2$}
\end{subfigure}
\caption{\label{fig:vqr_lh}
\textbf{Likelihood function $p_{\rvec{Y}|\rvec{X}}$ for the `Conditional Multimodal' distribution.}
The covariate $\rvar{X}$ controls the scale of the distribution. $ESS_{\%}$ values are also reported. We use the mollweide projection to plot the whole sphere surface.}

\end{figure*}

\section{\uppercase{Experiments}}\label{sec:results}
We validate our method on synthetic and real-world datasets, and on two different manifolds: the sphere $\Sphere{2}$ and the 2-dimensional torus $\mathcal{T}^2 = \Sphere{1} \times \Sphere{1}$.
Both manifolds have a closed-form expression for distances, exp/log-maps, and the ground cost is the squared geodesic distance. We report all the formulas in Section \ref{app:diff_geom} of the Appendix.
For the synthetic datasets, where we have access to the data generating process, we measure the quality of sampling, likelihood estimation, and confidence sets constructed from the estimated M-CVQFs.
In real-world scenarios, where ground truth likelihood are unavailable, we quantify sampling quality and confidence set validity, and visually present the estimated likelihoods.

\subsection{Evaluation Metrics}\label{sec:evalm}
Below we describe the quantitative metrics used to evaluate $\hat{Q}_{\rvec{Y}|\rvec{X}}$, the estimated M-CVQF.

\paragraph{Sampling.}
We assess sampling quality by computing $L_1$ distance between kernel density estimates (KDE) obtained using samples drawn from the groundtruth and estimated distributions. 
We employ manifold-specific kernels for the KDEs. The KDE-$L_1$ distance is measured as:
\begin{equation*}
    \text{KDE-}L_1(\mathcal{Y}, \mathcal{Y}^{(gt)}) = \left| p^{\text{\tiny{KDE}}}_{\mathcal{Y}} - p^{\text{\tiny{KDE}}}_{\mathcal{Y}^{\text{gt}}} \right|,
\end{equation*}
where $\mathcal{Y}, \mathcal{Y}^{\text{gt}}$ are equally-sized sets of samples drawn from the estimated and true distribution, respectively.
In the conditional setting, we first sample $\{\vec{x}_i\}_{i=1}^M \sim \rvec{X}$. For each $\vec{x}_i$, we sample a set of points $\mathcal{Y}_{{x}_i}$ from $\hat{Q}_{\rvec{Y}|\rvec{X}}$, a set of points $\mathcal{Y}_{\vec{x}_i}^{(gt)}$ from the true $p_{\rvec{Y}|\rvec{X}=\vec{x}_i}$, and report the mean KDE-$L_1$ distance over all $\vec{x}_i$. 

\paragraph{Likelihood.}
When groundtruth likelihood is available, we assess the quality of the model's likelihood using estimated sample size (ESS), originally proposed by \cite{KishSampling} as also used in \cite{cohen2021rcpm}. It is computed as follows:
\begin{align*}
 ESS_{\%} = 100 \times \frac{(\sum^N_j w_j)^2}{N \cdot \sum^N_j w_j^2} \quad \text{with} \quad w_j = \frac{ p^{gt}_{\rvec{Y} | \rvec{X}}(\vec{y}_j, \vec{x}_j)}{p_{\rvec{Y} | \rvec{X}}(\vec{y}_j, \vec{x}_j)},
\end{align*}
{where $\left\{ \vec{y}_j, \vec{x}_j\right\}_{j=1}^N  \sim p_{(\rvec{Y},\rvec{X})}$, $p^{gt}_{\rvec{Y} | \rvec{X}}$ and $p_{\rvec{Y} | \rvec{X}}$ are the groundtruth and estimated conditional likelihoods, respectively.}

\paragraph{Confidence sets.}
Estimating the validity of a confidence set $\mathcal{C}_{\tau}^{\rvec{Y}|\rvec{X}}$ requires computing $\Pr{\vec{y} \in \mathcal{C}_{\tau}^{\rvec{Y}|\rvar{X}}}$, which is non-trivial because confidence sets on the target distribution assume an arbitrary shape on the manifold. However, checking whether a point is inside a $\tau$ confidence set for $\rvec{U}$, i.e., $\mathcal{C}_{\tau}^{\rvec{U}}$,
is straightforward: it can be verified by checking that $C^*_{\vec{\omega}}( \vec{u}) \leq \tau$.
To exploit this property we use the inverse M-CVQF, which maps $\rvec{Y}|\rvec{X}$ back to $\rvec{U} \sim \mathcal{U}_{\Manifold}$, computed as $Q_{\rvec{Y}|\rvec{X}}^{-1} (\vec{y}; \vec{x}) = \expmap_{\vec{y}}[-\nabla_{\vec{y}} {\psi}(\vec{y}; \vec{x})]$. 
The coverage of the confidence set   $\mathcal{C}_{\tau}^{\rvec{Y}|\rvec{X}}$ is then estimated by:
\begin{align*}
\small
    &\Pr{\vec{y} \in \mathcal{C}_{\tau}^{\rvec{Y}|\rvec{X}}} = 
    \Pr{Q_{\rvec{Y}|\rvec{X}}^{-1} (\vec{y}; \vec{x}) \in \mathcal{C}_{\tau}^{\rvec{U}}} \\
    & = \mathbb{E}_{\rvec{U}}[\1{C^*_{\vec{\omega}}( \rvec{U}) \leq \tau}] 
    \approx \frac{1}{N}\sum_{i=1}^N \1{C^*_{\vec{\omega}}( \vec{u}_i) \leq \tau}
\end{align*}
where $u_i = Q^{-1}_{Y|X}(y_i;x_i)$.

\begin{figure*}[h!]
\centering
\begin{subfigure}[b]{0.3\linewidth}
\centering
\begin{overpic}
[trim=2.07cm 1.9cm 3cm 2.5cm,clip,width=\linewidth, grid=False]{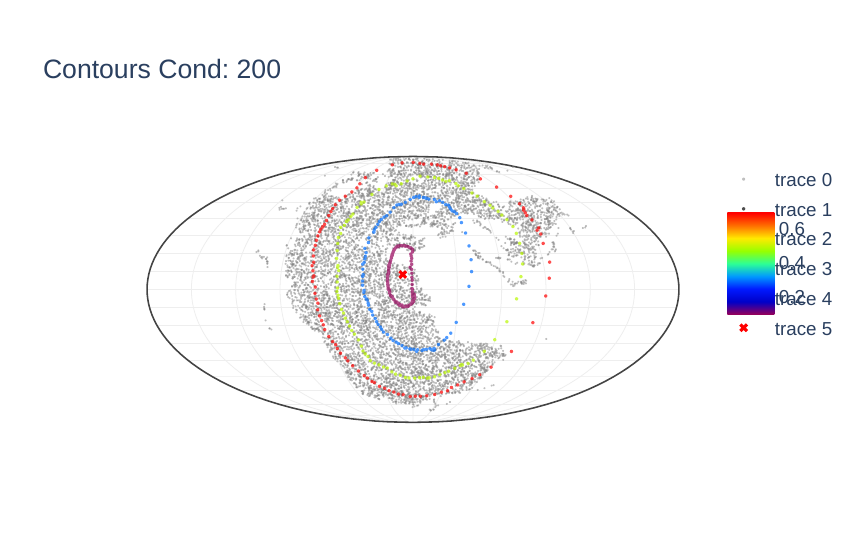}
\end{overpic}
\caption{\label{fig:cont_duec} \rvec{Y}|\rvec{X}=200}
\end{subfigure} 
\hspace{0.2cm}
\begin{subfigure}[b]{0.3\linewidth}
\centering
\begin{overpic}
[trim=2.07cm 1.9cm 3cm 2.5cm,clip,width=\linewidth, grid=false]{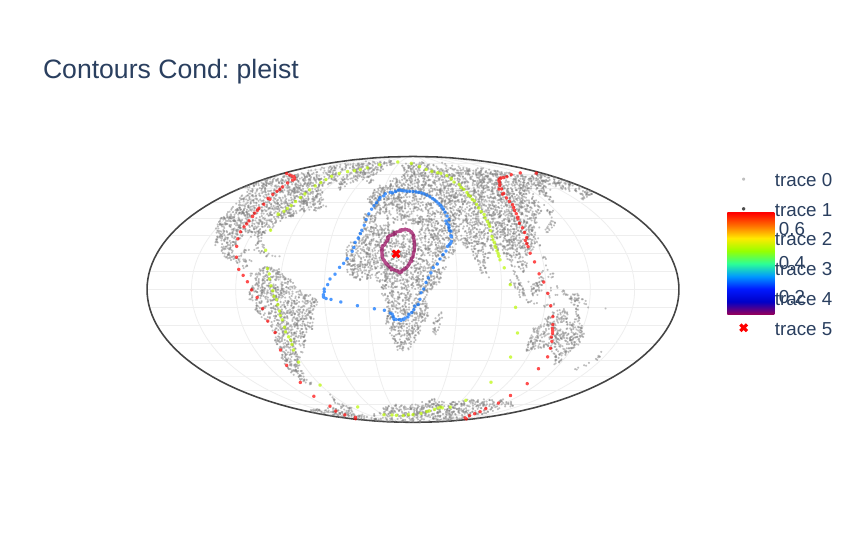}
\end{overpic}
\caption{\label{fig:cont_pleist} \rvec{Y}|\rvec{X}=pleist}
\end{subfigure} 
\begin{subfigure}[b]{0.04\linewidth}
\centering
\begin{overpic}
[trim=4.52cm -3.5cm 3.45cm 2.79cm,clip,width=\linewidth, grid=false]{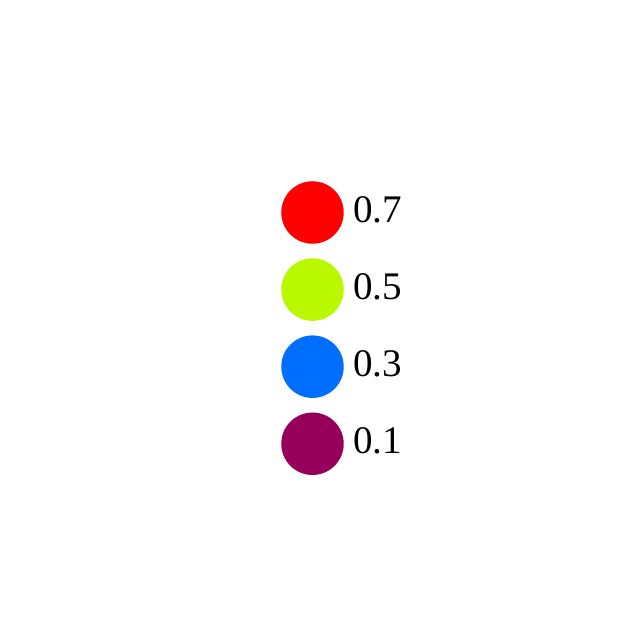}
\put(10,100){\footnotesize $\tau$}
\end{overpic}
\end{subfigure} 
\hfill
\begin{subfigure}[b]{0.3\linewidth}
\centering
\begin{overpic}
[trim=0cm 0.5cm 0cm 2cm,clip,width=\linewidth, grid=false]{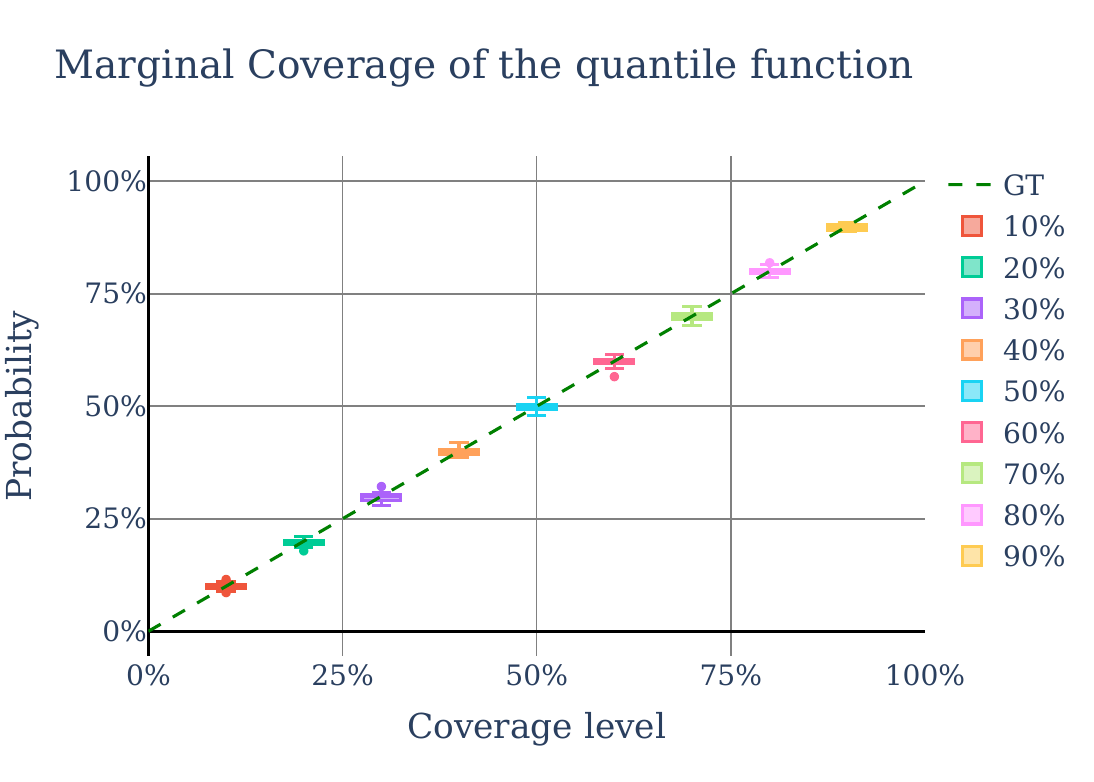}
\end{overpic}
\caption{\label{fig:cov_contdrift} Marginal Coverage}
\end{subfigure} 
\caption{\label{fig:cont_contdrift}
\textbf{$\tau$-confidence sets constructed with M-VQR on the `Continental Drift' dataset are smooth, nested, and valid.} Subfigures (a) and (b) report $\tau$-contours overlayed on the ground truth samples, for different values of $\tau$; 
each subfigure represents conditioning on a different era.
Mollweide projection is used to visualize the whole sphere.
Graph (c) shows the coverage achieved by the model as a function of the requested coverage level, averaged over the different conditionings with relative confidence bars.
}
\end{figure*}

\begin{figure*}[h]
\centering
\begin{subfigure}[b]{0.2\linewidth}
\centering
\begin{overpic}
[trim=0.4cm 1cm 2cm 2cm,clip,width=\linewidth, grid=False]{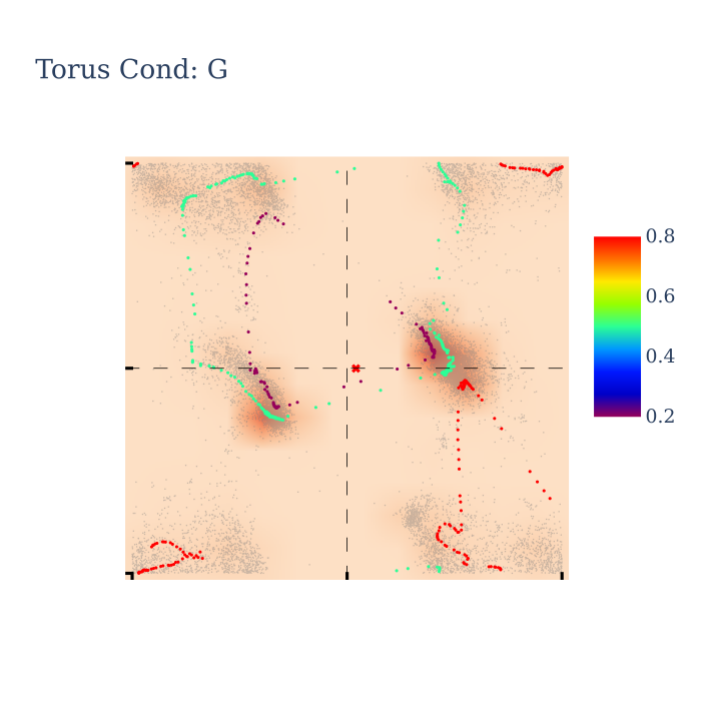}
\put(6,84){\tiny 180}
\put(14,47.5){\tiny 0}
\put(4,12){\tiny -180}
\put(0,50){\tiny \rotatebox[origin=c]{90}{$\psi (^{\circ})$}}
\put(89,7){\tiny 180}
\put(55.5,7){\tiny 0}
\put(12,7){\tiny -180}
\put(50,0){\tiny {$\phi (^{\circ})$}}
\end{overpic}
\caption{\label{fig:cont_G} \rvec{Y}|\rvec{X}=G}
\end{subfigure} 
\hfill
\begin{subfigure}[b]{0.2\linewidth}
\centering
\begin{overpic}
[trim=0.4cm 1cm 2cm 2cm,clip,width=\linewidth, grid=false]{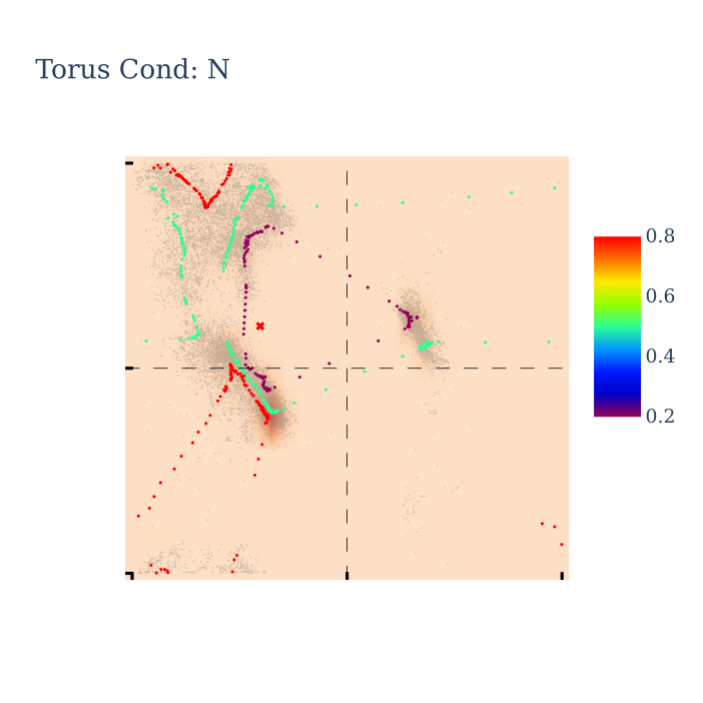}
\put(6,84){\tiny 180}
\put(14,47.5){\tiny 0}
\put(4,12){\tiny -180}
\put(0,50){\tiny \rotatebox[origin=c]{90}{$\psi (^{\circ})$}}
\put(89,7){\tiny 180}
\put(55.5,7){\tiny 0}
\put(12,7){\tiny -180}
\put(50,0){\tiny {$\phi (^{\circ})$}}
\end{overpic}
\caption{\label{fig:cont_N} \rvec{Y}|\rvec{X}=N}
\end{subfigure} 
\hfill
\begin{subfigure}[b]{0.2\linewidth}
\centering
\begin{overpic}
[trim=0.4cm 1cm 2cm 2cm,clip,width=\linewidth, grid=false]{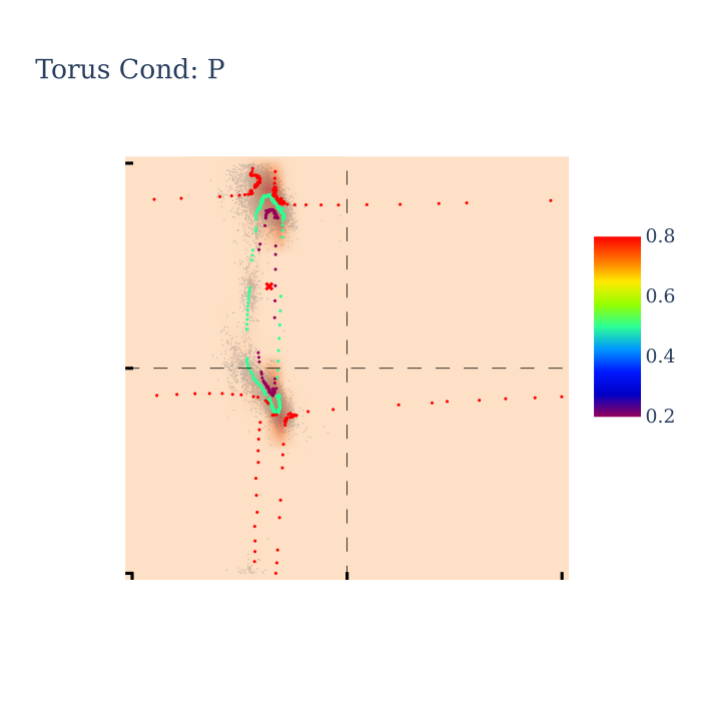}
\put(6,84){\tiny 180}
\put(14,47.5){\tiny 0}
\put(4,12){\tiny -180}
\put(0,50){\tiny \rotatebox[origin=c]{90}{$\psi (^{\circ})$}}
\put(89,7){\tiny 180}
\put(55.5,7){\tiny 0}
\put(12,7){\tiny -180}
\put(50,0){\tiny {$\phi (^{\circ})$}}
\end{overpic}
\caption{\label{fig:cont_P} {\rvec{Y}|\rvec{X}=\text{P}}}
\end{subfigure} 
\begin{subfigure}[b]{0.03\linewidth}
\centering
\begin{overpic}
[trim=4.52cm -9cm 3.45cm 2.79cm,clip,width=\linewidth, grid=false]{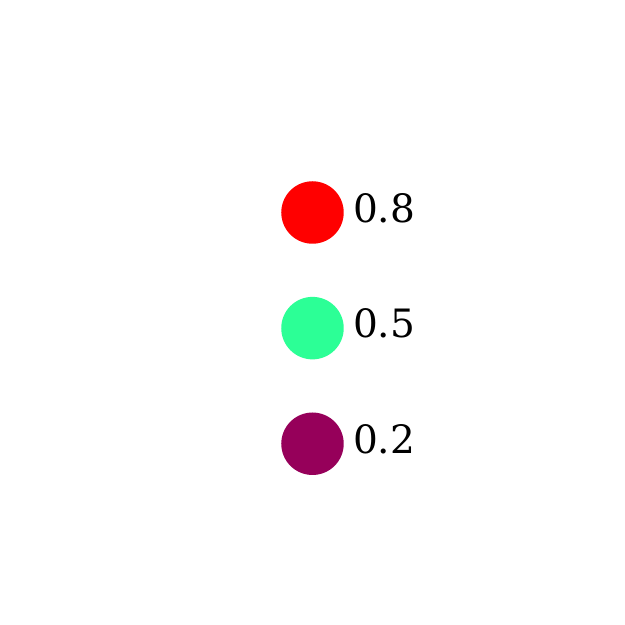}
\put(6,101){\footnotesize $\tau$}
\end{overpic}
\end{subfigure} 
\hfill
\begin{subfigure}[b]{0.3\linewidth}
\centering
\begin{overpic}
[trim=0cm 0.5cm 0cm 2cm,clip,width=\linewidth, grid=false]{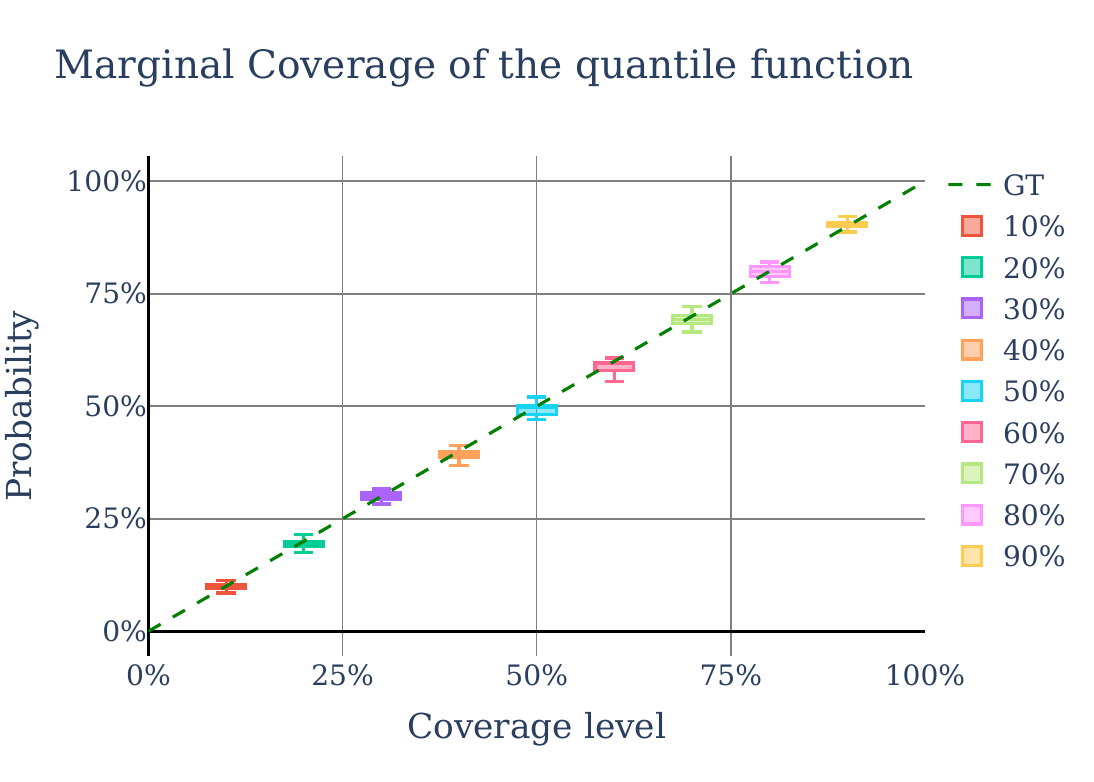}
\end{overpic}
\caption{\label{fig:cov_codon} Marginal Coverage}
\end{subfigure} 
\caption{\label{fig:cont_codon}
\textbf{$\tau$-contours constructed with M-VQR on the `Dihedral Angles' dataset are smooth, nested, and valid.}
Subfigures (a), (b) and (c) report $\tau$-contours overlayed on the ground truth samples, for different values of $\tau$;
each subfigure depicts conditioning by a different amino acid.
The dihedral angles are plotted on a ``flat torus'', with the axes representing the dihedral angles in degrees.
The wrap-around, due to the toroidal domain, is clearly visible in the contours.
Graph (d) shows the the coverage achieved by the model as a function of the requested coverage level, averaged over the different conditionings with relative confidence bars.
}

\end{figure*}

\subsection{\label{sec:datasets} Datasets}
\paragraph{Synthetic datasets.}
For synthetic data experiments, we defined three data generating processes. 
The `Conditional Multimodal' distribution is a mixture of four von-Mises distributions on $\Manifold$, conditioned on $\rvar{X}\in\R$ which controls the covariance of the distributions.
`Scaled Heart' and `Scaled Star' are uniform distributions on a subset $A \subset \Manifold$ which is heart-shaped and star-shaped, respectively. The conditioning variable $\rvar{X}\in\R$ controls the scale of $A$. 
All aforementioned distributions are defined both on $\Sphere{2}$ and $\mathcal{T}^2$.

\paragraph{Real-world datasets.}
As a candidate for distributions defined on $\Sphere{2}$, we consider the `Continental drift' dataset, consisting of continental drift measured over 26 eras. We consider the era as the conditioning variable, and continental drift as the target variable.
The source maps are taken from the 2020 Colorado Plateau Geosystems \footnote{\url{https://deeptimemaps.com/}}. We convert each Mollweide earth image into a spherical point cloud and into Euclidean coordinates.  

As a real-world use case on the torus $\mathcal{T}=\Sphere{1} \times \Sphere{1}$, we consider a dataset of local dihedral angles ($\varphi, \psi \in [0,2\pi]$) measured in the backbone of folded protein structures.
Pairs of angles can be naturally represented as points of on a torus, because each angle is represented by a point on $\Sphere{1}$. 
Protein structures are composed of chains of common amino acids (AAs), with each AA in the chain assuming different dihedral angles depending on its location, chemistry, context, and other biological factors.
We use the AA identity as the conditioning variable, and the corresponding dihedral angles as the target.
Note that here $\rvar{X}$ is categorical, with 20 possible values indicating the AA identity.
We use the dihedral angle dataset curated by \citet{rosenberg2022codon} and group the dihedral angles by their AA identity.

\subsection{Synthetic data experiments}
Below we report results both for unconditional estimation (approximating $Q_{\rvec{Y}}$, without covariates) and for regression (approximating $Q_{\rvec{Y}|\rvec{X}}$). 

\paragraph{Quantile estimation.}
Though it is not the primary focus of this paper, evaluating the performance of our approach in the simpler quantile estimation scenario is worthwhile, because until now the only existing approach for this problem was that of \cite{hallin2022nonparametric}, which is limited to the sphere and neither scalable nor differentiable.
Figure \ref{fig:vqe} visually and quantitatively demonstrates that the estimated M-VQF results in smooth, nested, and valid contours. Both the likelihood visualization and the ESS value (93\%) indicate a close approximation to the ground truth likelihood. We also report a KDE-$L_1$ distance of $3.9\times 10^{-3}$; compared to an upper bound of $0.7\times 10^{-3}$ when measured between two independent pairs of groundtruth samples. This indicates good sampling quality of the M-VQF.

With regard to scalability, in Section \ref{app:hallin} of the Appendix, we demonstrate that our approach, based on the relaxed dual OT problem, is substantially more scalable than the previous approach of \cite{hallin2022nonparametric}, which involves solving a discrete OT problem as a linear-program.

\paragraph{Quantile regression.}
Quantitative results for coverage, sampling, and likelihood in the regression setting are reported in Table \ref{tab:coverage_synth}. 
We observe that the confidence sets constructed by estimated M-CVQFs result in a maximum coverage violation of 1.5\% with respect to the nominal coverage. KDE-$L_1$ shows an order of magnitude of $10^{-3}$ at maximum against a magnitude of $10^{-4}$ for  KDE-$L_1$ between ground truth samples, indicating good sampling. ESS is between 84\%-96\%, demonstrating that the conditional likelihoods computed from the estimated M-CVQFs are reasonably accurate compared to the groundtruth.

Figure \ref{fig:vqr_cont} provides a visualization of conditional contours for the `Scaled Heart' and `Scaled Star' distributions.
Figures \ref{fig:s2_lh} and \ref{fig:t2_lh} depict the estimated conditional likelihoods for the `Conditional Multimodal' distribution dataset on $\Sphere{2}$ and $\mathcal{T}_2$, respectively. 
Additional visualizations are provided in Section \ref{app:add_exp} of the Appendix.

\subsection{Real data experiments}
Figure \ref{fig:cont_contdrift} presents results on the `Continental Drift' dataset, reporting the coverage and confidence sets visualizations for two different conditioning values.
Figure \ref{fig:cont_codon} reports the same results for the `Dihedral Angles' dataset. In both cases, the resulting contours are nested and smooth, as desired.
The coverage values are almost perfectly aligned with the ground truth, and demonstrate low variance over the conditioning variables.
In Figure \ref{fig:cont_codon}, the wrap-around of the confidence sets due to the manifold structure of the domain is clearly visible.
The mean KDE-$L_1$ for the `Continental Drift' dataset is $(1.72 \pm 0.26) \times 10^{-3}$ against $(9.45 \pm 1.05) \times 10^{-4}$ from the ground truth samples; while for the `Dihedral Angles' dataset we obtain a mean KDE-$L_1$ of $(2.61 \pm 0.56) \times 10^{-3}$ against $(7.34 \pm 2.88) \times 10^{-4}$.
The likelihood plots for both datasets are provided in Section \ref{app:add_exp} of the Appendix.

\section{\uppercase{Discussion and Conclusions}}
Our work provides the first formulation of conditional vector quantile functions on manifolds, by extending non-linear VQR to non-Euclidean domains, together with conditional OT and quantile regression on manifolds.
Our key contributions are the novel formulation of nonlinear VQR as a Riemannian OT problem, its parametrization with partial input $c$-concave neural networks, and the involution regularization approach for training.

One potential limitation of this work is that the estimated potential functions might not be $c$-concave, as this property is only promoted, but not enforced, through the regularized objective.
In future studies, we hope to analyze it theoretically and provide ways to overcome this limitation.
Another avenue for exploration is extending the proposed approach to domains possessing closed-form formulations for computing $c$-concave functions, such as Lie groups.

In summary, our approach enables the estimation of conditional quantiles and construction of confidence sets on general manifolds for which the exponential map is known by fitting data sampled directly from the joint distribution.
These capabilities may open the door to exciting new applications in diverse domains such as pose estimation, weather modeling, and protein structure prediction.
We believe that M-VQR is thus a powerful and useful new addition to the toolbox of directional statistics.

\subsubsection*{Acknowledgments}
S.V., A.A.R., and A.M.B. were partially supported by the European Research Council (ERC) under the European Union’s Horizon 2020 research and innovation programme (grant agreement No. 863839), by the Council For Higher Education - Planning \& Budgeting Committee, and by the Israeli Smart Transportation Research Center (ISTRC). I.T., M.P. and E.R. were partially supported by the ERC grant no.802554 (SPECGEO), PRIN 2020 project no.2020TA3K9N (LEGO.AI), and PNRR MUR project PE0000013-FAIR.

\bibliography{vqr}

\onecolumn
\section*{\uppercase{Appendix}}
\appendix
\section{\label{app:diff_geom} \uppercase{Differential geometry}}
In this Section, we provide an overview of the main concepts of differential geometry used in the main paper.

\paragraph{Manifolds.}
In this paper, we consider Riemannian $d$-dimensional manifolds $(\Manifold, g)$ with the Riemannian metric $g$, embedded in $\R^D$. Given a point $\vec{y} \in \Manifold$, the \textit{tangent space} $\TPManifold{\vec{y}}$ is defined as the linear subspace $\TPManifold{\vec{y}} = \left\{ {\vec{v} \in \R^D : \vec{v}\T \vec{y} = 0} \right\}$. 
The Riemannian metric defines an inner product $\langle \cdot, \cdot \rangle_g : \TPManifold{\vec{y}} \times \TPManifold{\vec{y}} \mapsto \mathbb{R}$ on the tangent space, which induces a geodesic distance $d_{\Manifold}(\vec{y}, \vec{z})$ between every pair of points $\vec{y}, \vec{z} \in \Manifold$, as the minimum length of a curve connecting the two points,
\begin{align*}
    & d_{\Manifold}(\vec{y}, \vec{z}) = \inf_{\vec{\gamma}}\int_0^1 \norm{\dot{\vec{\gamma}}(t)}_{g}\text{dt},
\end{align*}
where $\vec{{\gamma}}: [0,1] \rightarrow \Manifold$ and $\vec{{\gamma}}(0)=\vec{y}$, $\vec{{\gamma}}(1)=\vec{z}$.

The \textit{exponential map} of a manifold projects an infinitesimal displacement of $\vec{y}$ along the tangent vector $\vec{v}$ back to the manifold. Given a point $\vec{y} \in \Manifold$ with a tangent vector $\vec{v} \in \TPManifold{y}$, and given the unique geodesic $\vec{\gamma}:[0,1]\rightarrow\Manifold$ such that $\vec{\gamma}(0) = \vec{y}$ and $\dot{\vec{\gamma}}(0) = \vec{v}$, the exponential map at $\vec{y}$ is defined as $\expmap_{\vec{y}}(\vec{v}) = \vec{\gamma}(1)$. The tangent space inner product structure also allows one to define the intrinsic gradient and Jacobian over the manifold.

\paragraph{Manifold uniform distribution.}
Given the volume measure $d\Manifold(\vec{y})$ representing the infinitesimal volume element at each point $\vec{y}$ of the manifold, a random variable $\rvec{Y}$ follows a manifold uniform distribution $\mathcal{U}_A$ on the bounded subset $A \subseteq \Manifold$
if its probability density function (PDF) is constant within $A$:
\begin{equation}
p_{\rvec{Y}}(\vec{y})=\frac{\1{A}(\vec{y})}{V(A)}
\end{equation}
where $V(A)$ is the volume of the set $A$ with respect to the volume measure $d\Manifold$. The uniform distribution on the manifold, with respect to the volume measure, assigns probabilities to subsets of the manifold based on their intrinsic volumes.

\paragraph{Sphere.} On the $n$-sphere $\Sphere{n}$, the exponential map and the intrinsic distance are provided as closed-form expressions. If $\vec{y},\vec{u} \in \Sphere{n}$ and $\vec{v} \in T_{\vec{y}}\Sphere{n}$,
\begin{equation}
    \expmap^{\Sphere{n}}_{\vec{y}}(\vec{v}) = \vec{y} \cos(\norm{\vec{v}}) + \frac{\vec{v}}{\norm{\vec{v}}}\sin(\norm{\vec{v}}) \label{eq:expmap_sphere}
\end{equation}
\begin{equation}
    d_{\Sphere{n}}(\vec{y},\vec{u}) = \arccos(\vec{u}^T\vec{v})\label{eq:dist_sphere},
\end{equation}
where $\norm{\cdot}$ is the standard Euclidean norm.

\paragraph{Torus.} The torus $\mathcal{T}^2$ can be defines as a product manifold between two 1-sphere: $\mathcal{T}^2 = \Sphere{1} \times \Sphere{1}$. On general product manifolds of the form $\Manifold = \Manifold_1 \times\ldots\times \Manifold_l$, the squared intrinsic distance is simply
\begin{align}
    d^2_{\Manifold}(\vec{y},\vec{u}) =  d^2_{\Manifold_1}(\vec{y}_1,\vec{u}_1)+\ldots+d^2_{\Manifold_l}(\vec{y}_l,\vec{u}_l).\label{e:decomp_product_cost}
\end{align}
where  $\vec{y} = (\vec{y}_1,\ldots,\vec{y}_l)$, and $\vec{u}_j \in \Manifold_j,\ \ j\in [l]$ (and similarly for $y$). 
The exponential map on the product manifold is the cartesian product of exponential maps on the individual manifolds. 
Therefore, the  exponential map and intrinsic distance on the torus $\mathcal{T}^2$ is defined as:
\begin{align}
\expmap^{\mathcal{T}^2}_{\vec{y}}(\vec{v}) = \expmap^{\Sphere{1}}_{\vec{y}_1}(\vec{v}_1) \times \expmap^{\Sphere{1}}_{\vec{y}_2}(\vec{v}_2) \label{eq:expmap_torus}
\end{align}
\begin{align}
    d_{\mathcal{T}^2}(\vec{y},\vec{u}) =  \sqrt{d^2_{\Sphere{1}}(\vec{y}_1,\vec{u}_1)+d^2_{\Sphere{1}}(\vec{y}_2,\vec{u}_2)}.\label{eq:dist_torus}
\end{align}

\section{\uppercase{Derivation of manifold VQR loss function}}
The dual formulation of the Kantarovich problem is given by
\begin{align}
\label{eq:dual_kantarovich}
\begin{split}
    \sup_{\varphi,\psi} & \; \int_{\Manifold} \varphi(\vec{u}) d\mu(\vec{u}) + \int_{\Manifold} \psi(\vec{y}) d\nu(\vec{y})\\
    \text{s.t.} & \; \quad \varphi(\vec{u}) + \psi(\vec{y}) \leq c(\vec{u},\vec{y}),
\end{split}
\end{align}
due to the linearity of integration we have
\begin{align*}
\sup_{\varphi,\psi} & \;  \int_{\Manifold \times \Manifold} \left( \varphi(\vec{u}) + \psi(\vec{y})\right) d\mu(\vec{u}) d\nu(\vec{y})\\
\text{s.t.} & \; \quad \varphi(\vec{u}) + \psi(\vec{y}) \leq c(\vec{u},\vec{y}) 
\end{align*}
In the regression setting, $\varphi$ and $\psi$ are $c$-concave functions, parametric in $\vec{x}$ and concave in $\vec{u}$, and we denote the measure corresponding to the joint distribution of $\rvec{X}, \rvec{Y}$ as $\xi(\vec{x}, \vec{y})$.
The manifold vector quantile regression loss is simply obtained by taking an expectation of Equation \ref{eq:dual_kantarovich} with respect to $\vec{x}$, 
\begin{align*}
    \sup_{\varphi,\psi} & \; \int_{\mathcal{X} \times \Manifold \times \Manifold} \varphi(\vec{u}; \vec{x}) d\xi(\vec{x}, \vec{y}) d\mu(\vec{u}) + \int_{\mathcal{X} \times \Manifold} \psi(\vec{x}; \vec{y}) d\xi(\vec{x}, \vec{y})\\
    \text{s.t.} & \; \quad \quad \varphi(\vec{u}; \vec{x}) + \psi(\vec{y}; \vec{x}) \leq c(\vec{u},\vec{y}).
\end{align*}
In the finite sample setting where $\{ \vec{u} \}_{i=1}^T \sim \mathcal{U}_{\Manifold}$, $\left\{ \vec{x}_j, \vec{y}_j\right\} \sim p_{\{ \mathbf{X}, \mathbf{Y} \}}$, the above can be written as
\begin{equation}
    \label{eq:OT_discr_vqr}
    \begin{split}
      \max_{{\varphi},{\psi}}{ \sum_{j=1}^N {\xi_j} \sum_{i=1}^T {\mu_i} {\varphi}(\vec{u}_i; \vec{x}_j) }  + \sum_{j=1}^N  {{\xi}_j} {\psi}(\vec{y}_j; \vec{x}_j) \\
    \text{s.t.} \, \forall i,j: \quad   {\varphi}(\vec{u}_i; \vec{x}_j) + {\psi}(\vec{y}_j; \vec{x}_j) \leq c(\vec{u}_i,\vec{y}_j),
    \end{split}
\end{equation}
where $\vec{\mu }= \frac{1}{T} \mat{1}_T$, $\vec{\xi} = \frac{1}{N} \mat{1}_N$, are measures corresponding to the respective sample densities.

\textbf{Estimation.}
In the main paper, we evaluated the performance of example of manifold vector quantile estimation. For convenience, below we provide the finite-sample version of the loss function in the unconditional case.
Given samples $\left\{ \vec{u}_i \right\}_{i=1}^T  \sim \mathcal{U}_\Manifold$
 and $\left\{ \vec{y}_j \right\}_{j=1}^N  \sim \rvec{Y}$ from a random target variable $\rvec{Y}$, the dual formulation of the Kantorovich problem can be discretized as
\begin{equation}
    \label{eq:OT_discr_vqe}
    \begin{split}
      \max_{{\varphi},{\psi}}{ \sum_{i=1}^T {\mu_i} {\varphi}(\vec{u}_i) }  + \sum_{j=1}^N  {{\nu}_j} {\psi}(\vec{y}_j) \\
    \text{s.t.} \, \forall i,j: \quad   {\varphi}(\vec{u}_i) + {\psi}(\vec{y}_j) \leq c(\vec{u}_i,\vec{y}_j),
    \end{split}
\end{equation}
where $\vec{\mu }= \frac{1}{T} \mat{1}_T$, $\vec{\nu} = \frac{1}{N} \mat{1}_N$.

By writing one of the potentials in terms of the other leveraging the $c$-transform, we obtain the following max-min optimization problem,
\begin{equation}
    \label{eq:vqe_loss}
    \max_{\varphi} \sum_{i=1}^{T} {\mu_i}{\varphi}(\vec{u}_i)  + 
    \sum_{j=i}^N \nu_j \min_{\vec{u} \in \Manifold} \left\{ c(\vec{u},\vec{y}_j) - {\varphi}(\vec{u})\right\}.
\end{equation}

\begin{figure}[]
\centering
\begin{subfigure}[b]{0.48\linewidth}
\centering
\begin{overpic}
[trim=2cm 2cm 3.6cm 3cm,clip,width=1\linewidth, grid=false]{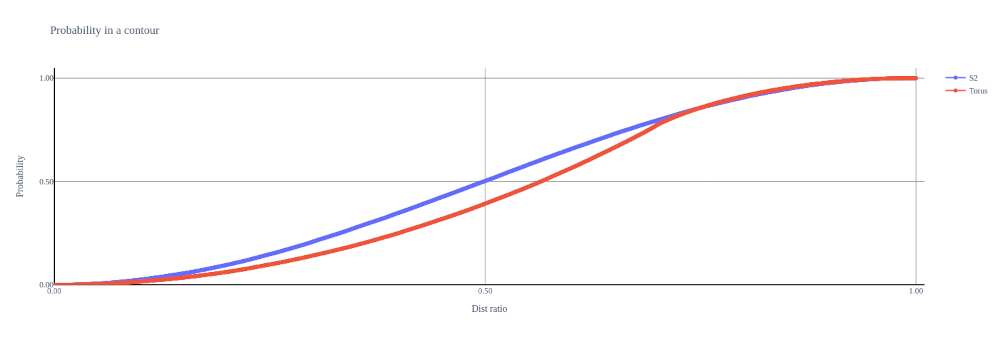}
\put(-4,9){ \rotatebox{90}{\tiny Probability}}
\put(48,-2){\tiny $d_\Manifold$}
\end{overpic}
\end{subfigure} 
\hfill
\begin{subfigure}[b]{0.48\linewidth}
\centering
\begin{overpic}
[trim=2cm 2cm 1.6cm 3cm,clip,width=1\linewidth, grid=false]{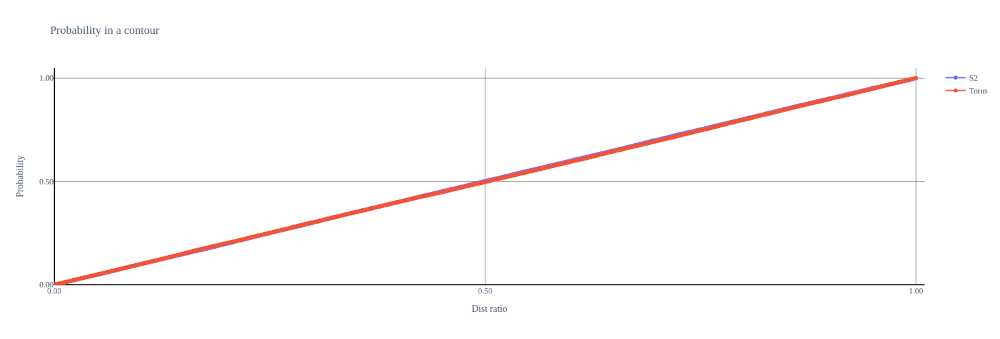}
\put(-4,9){ \rotatebox{90}{\tiny Probability}}
\put(46,-2){\tiny $C^*_{\vec{\omega}}$}
\put(100,23.6){\fontsize{2pt}{0.2pt}\selectfont \scalebox{0.5}{$\Sphere{2}$}}
\put(99,22.3){\fontsize{2pt}{0.2pt} \scalebox{0.5}{$\mathcal{T}^2$}}
\end{overpic}
\end{subfigure}
\caption{\label{fig:confset} \textbf{Probability contained in the contours computed using the inner distance $d_\Manifold(\omega, \vec{u})$ (left) and the mapping function $C^*_{\vec{\omega}}(\vec{u})$ (right).} The distances are normalized to be in the range $[0,1]$. The mapping function $C^*_{\vec{\omega}}$ creates a linear dependency between the distance of the points in a contour and the amount of probability contained. }
\end{figure}
\section{\uppercase{Comparison to prior works}}
In this section, we compare our method to two prior works, \cite{hallin2022nonparametric} and \cite{cohen2021rcpm}. We evaluate the scalability of our method in comparison to \cite{hallin2022nonparametric}.  We then study the impact of different training loss functions, to offer insights into how our approach compares to \cite{cohen2021rcpm}. 
Through this comparative exploration, we aim to provide a holistic view of our method's strengths and capabilities in relation to the existing state-of-the-art. It is important to note that these comparisons are carried out on manifold vector quantile estimation (M-VQE), as neither aforementioned works train conditional maps (M-VQR).

\subsection{\label{app:hallin} Comparison to \cite{hallin2022nonparametric}}
\cite{hallin2022nonparametric} propose solving the OT problem resulting from manifold vector quantile estimation (M-VQE) as a linear assignment problem with the ground cost set to be the squared geodesic distance. Given $\{ \vec{u}_i\}_{i=1}^{T} \sim \mathcal{U}_{\mathcal{M}}$ and $\{ \vec{y}_j \}_{j=1}^{N} \sim P_{\mathbf{Y}}$, the primal OT formulation proposed by  \cite{hallin2022nonparametric} solves the following optimization problem
\begin{align*}
\label{eq:hallin_primal}
    \min_{\mat{\Pi}\geq 0} \; &\sum_{j=1}^{N} \sum_{i=1}^{T} \pi_{ij} c(\vec{u}_i, \vec{y}_j)\\
    \text{s.t. } &\mat{\Pi} \vec{1}_N = \vec{1}_T,
    \mat{\Pi}\T \vec{1}_T = \vec{1}_N, 
\end{align*}
where $\mat{\Pi} \in \R^{T \times N}$ is the    assignment matrix, and $c(\vec{u}_i, \vec{y}_j) = d^2_{\mathcal{M}}(\vec{u}_i, \vec{y}_j)$. 
This approach has several limitations. First, the aforementioned problem solves a \textit{discrete} optimal transport problem, thus it recovers the transport map only at pre-specified points. This discrete representation of the map does not allow computation of the likelihood which is defined as the determinent of the Jacobian of the inverse map.
In contrast, we solve a \textit{continuous} optimal transport problem, we recover a continuous, differentiable, and invertible formulation of the transport map.
Second, the primal OT formulation is solved using a linear program solver. As a result, their approach scales poorly with the number of samples, both in the number of optimization variables and run-time. 
Instead our approach solves the relaxed formulation of the dual optimal transport problem which is amenable to gradient-based optimization and scales to large sample sizes.
For example, when $N, T = 10000$, our M-VQE solver converges in $8$ minutes, whereas the linear program solver, which estimates over $100$ million parameters, does not converge even in $3$ hours. Finally, \cite{hallin2022nonparametric} solve only the {estimation} problem, whereas our approach solves the more general {regression} problem.

\begin{figure}[h]
    \centering
    \begin{subfigure}[b]{\linewidth}
    \begin{subfigure}[b]{0.24\linewidth}
            \begin{overpic}
                [trim=1.9cm 5.5cm 16.2cm 6cm,clip,width=\linewidth, grid=False]{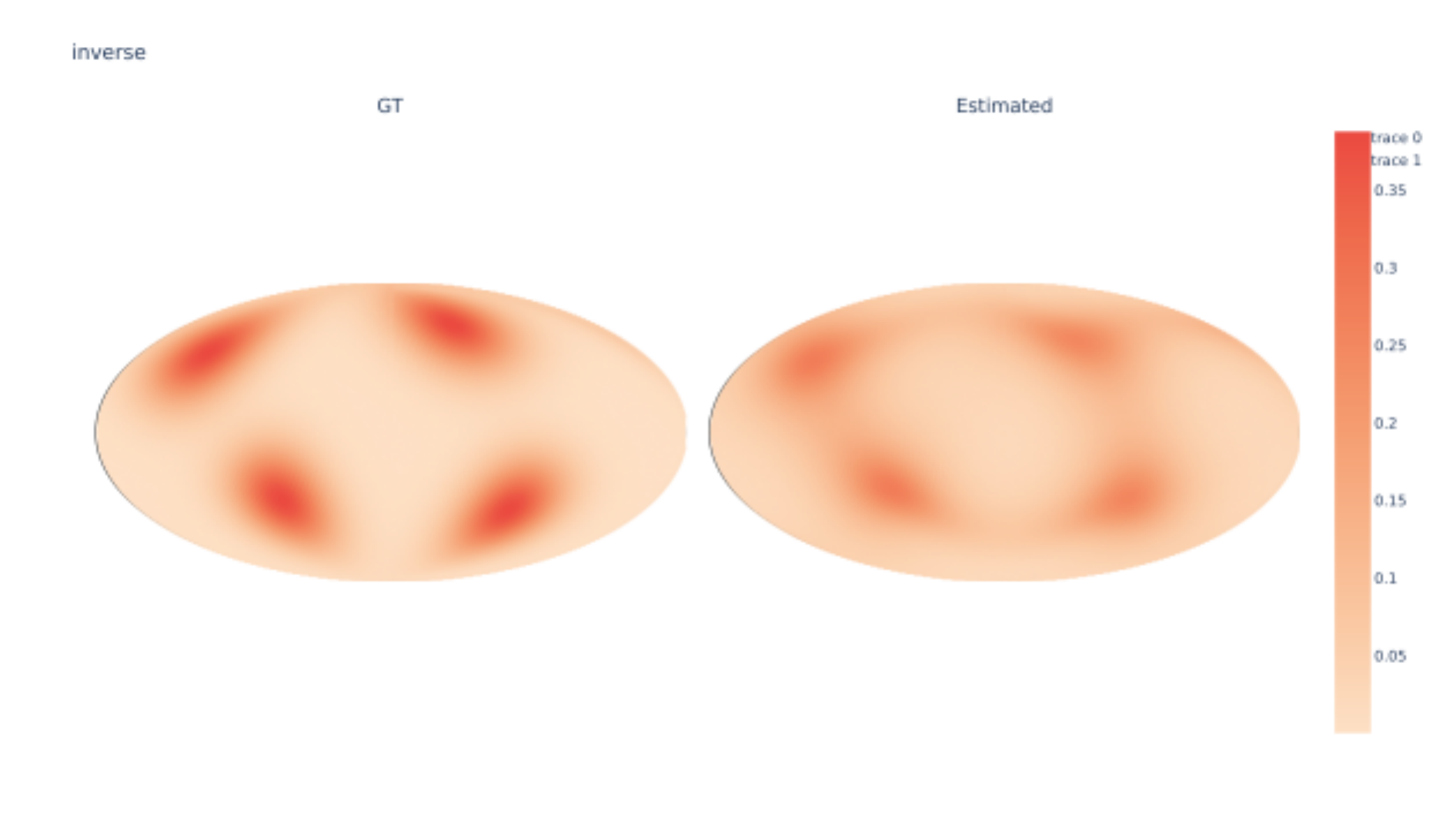}
            \end{overpic}
            \caption{\label{fig:rcpmvsvqe_gt_lh} Ground truth}
        \end{subfigure}
        \hfill
        \begin{subfigure}[b]{0.24\linewidth}
            \begin{overpic}
                [trim=15.35cm 5.5cm 3cm 6cm,clip,width=\linewidth, grid=False]{./figs/rezende/rcpm_lh.pdf}
                \put(79,-2){  $83\%$}
            \end{overpic}
            \caption{\label{fig:rcpmvsvqe_rcpm_lh} RCPM-KL}
        \end{subfigure}
        \hfill
        \begin{subfigure}[b]{0.24\linewidth}
            \begin{overpic}
                [trim=15.35cm 5.5cm 3cm 6cm,clip,width=\linewidth, grid=False]{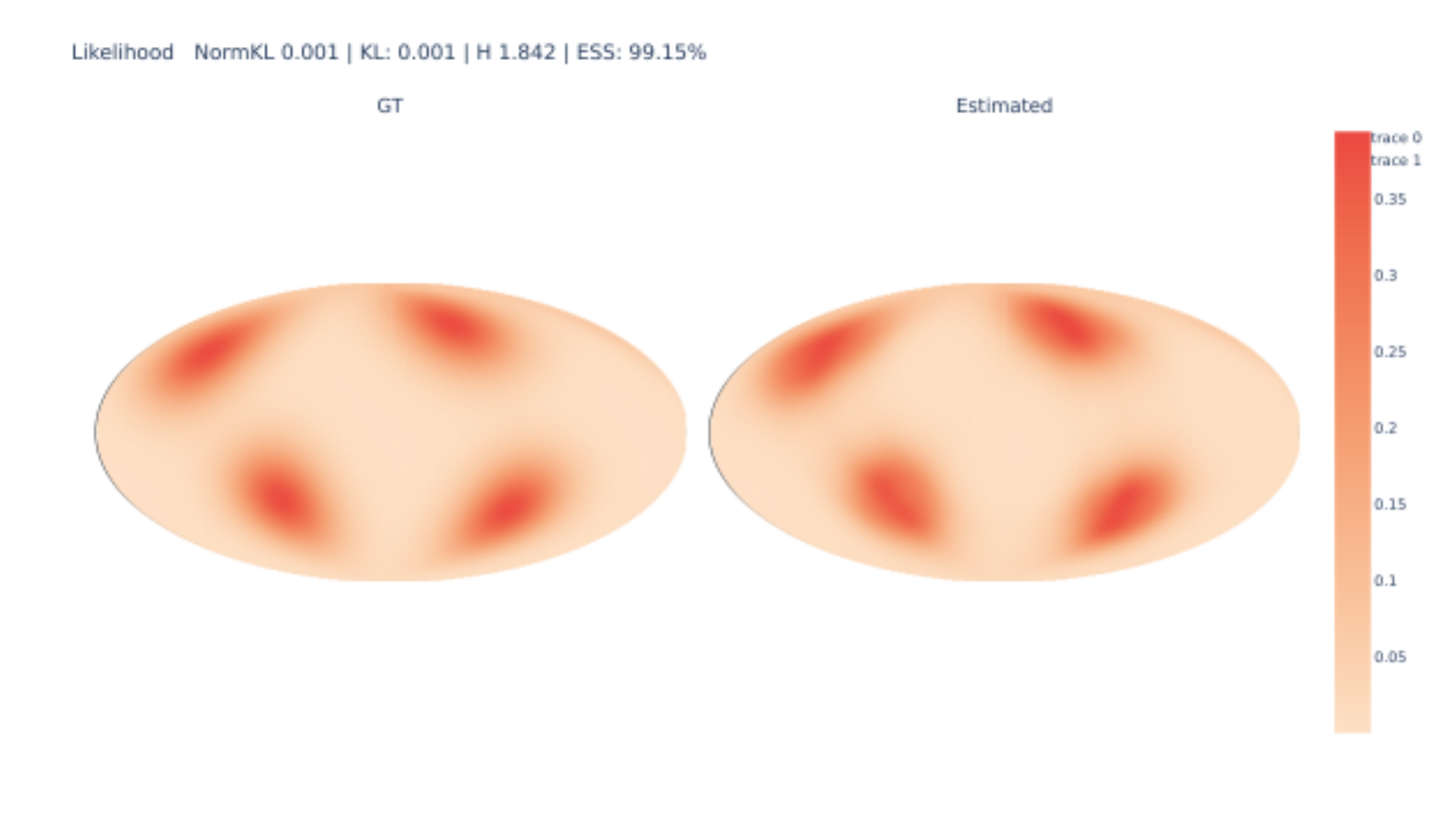}
                \put(79,-2){  $99\%$}
            \end{overpic}
            \caption{\label{fig:rcpmvsvqe_rcpmlh_lh} RCPM-LH}
        \end{subfigure}
        \hfill
        \begin{subfigure}[b]{0.24\linewidth}
            \begin{overpic}
                [trim=15.35cm 5.5cm 3cm 6cm,clip,width=\linewidth, grid=False]{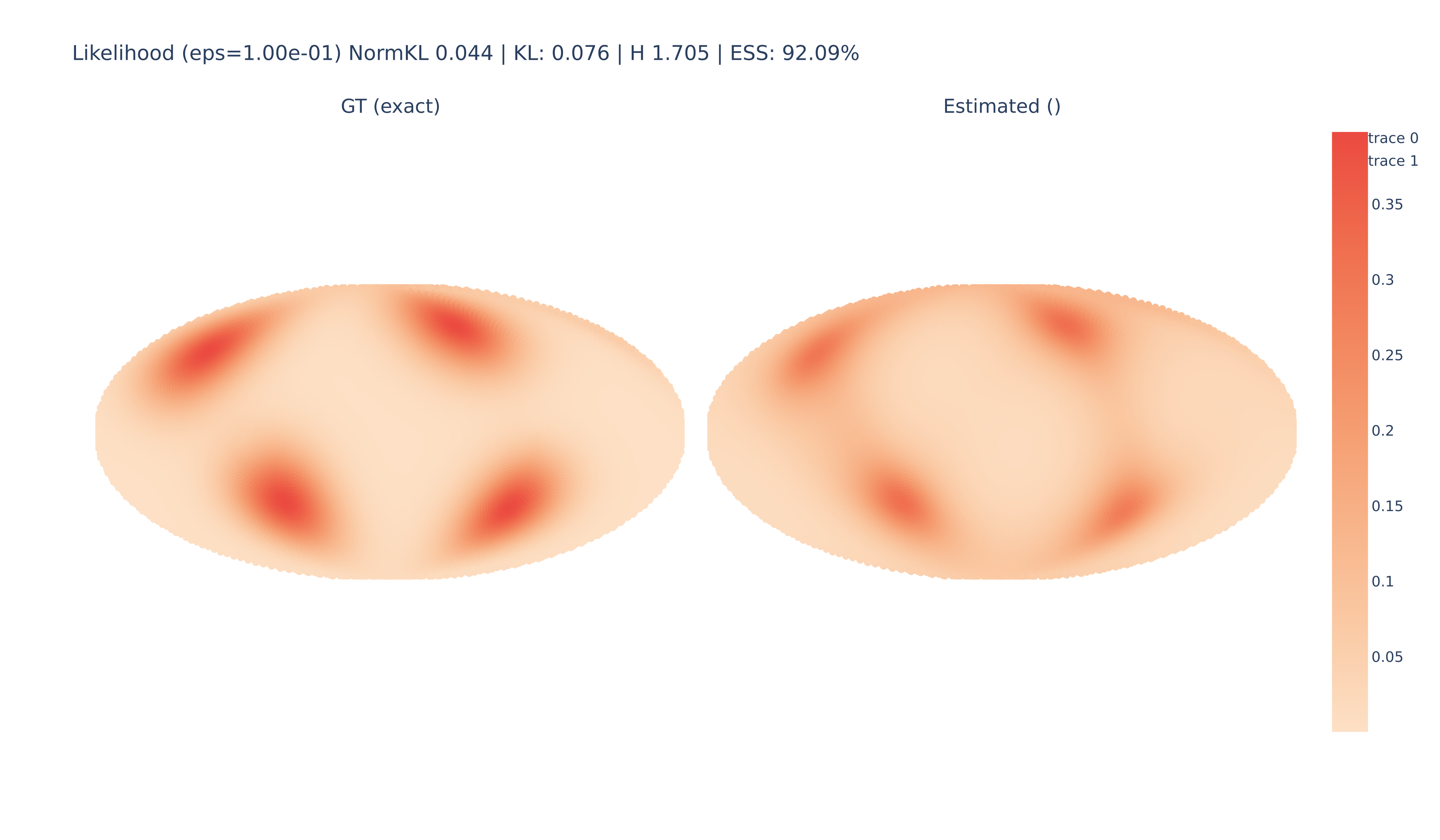}
                \put(79,-2){  $ 93\%$}
            \end{overpic}
            \caption{\label{fig:rcpmvsvqe_vqe_lh} M-VQE}
        \end{subfigure}
    \caption{\label{fig:rcpmvsvqe_vqe} Likelihood (ESS$_\%$)}
    \end{subfigure}

    \begin{subfigure}[b]{\linewidth}
        \begin{subfigure}[b]{0.24\linewidth}
            \begin{overpic}
                [trim=1.9cm 5.5cm 16.2cm 6cm,clip,width=\linewidth, grid=False]{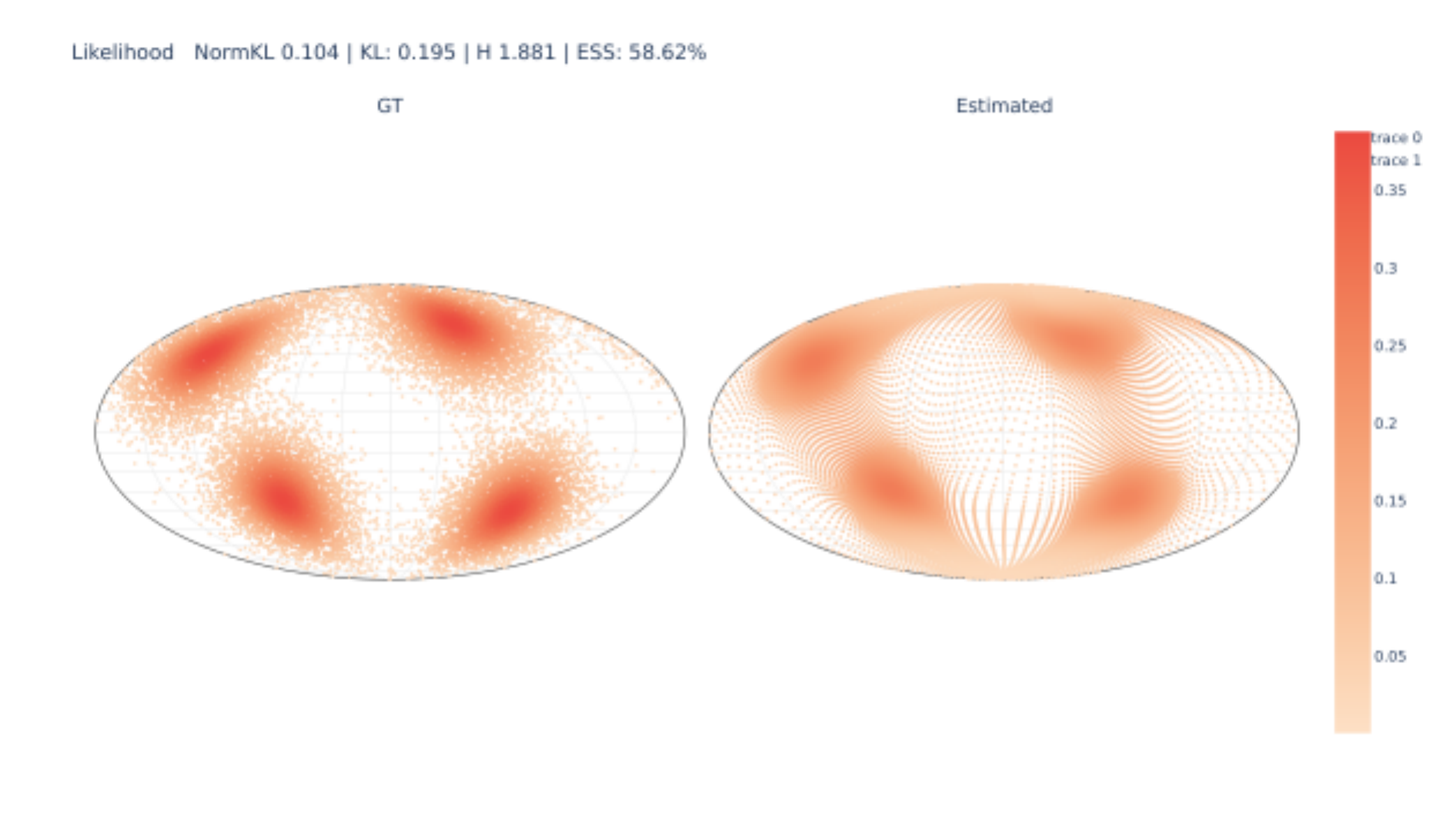}
            \end{overpic}
            \caption{\label{fig:rcpmvsvqe_gt_samp} Ground Truth}
        \end{subfigure}
        \hfill
        \begin{subfigure}[b]{0.24\linewidth}
            \begin{overpic}
                [trim=15.35cm 5.5cm 3cm 6cm,clip,width=\linewidth, grid=False]{./figs/rezende/rcpm_samples.pdf}
                \put(79,-2){ 6.67}
            \end{overpic}
            \caption{\label{fig:rcpmvsvqe_rcpm_samp} RCPM-KL}
        \end{subfigure}
        \hfill
        \begin{subfigure}[b]{0.24\linewidth}
            \begin{overpic}
                [trim=15.35cm 5.5cm 3cm 6cm,clip,width=\linewidth, grid=False]{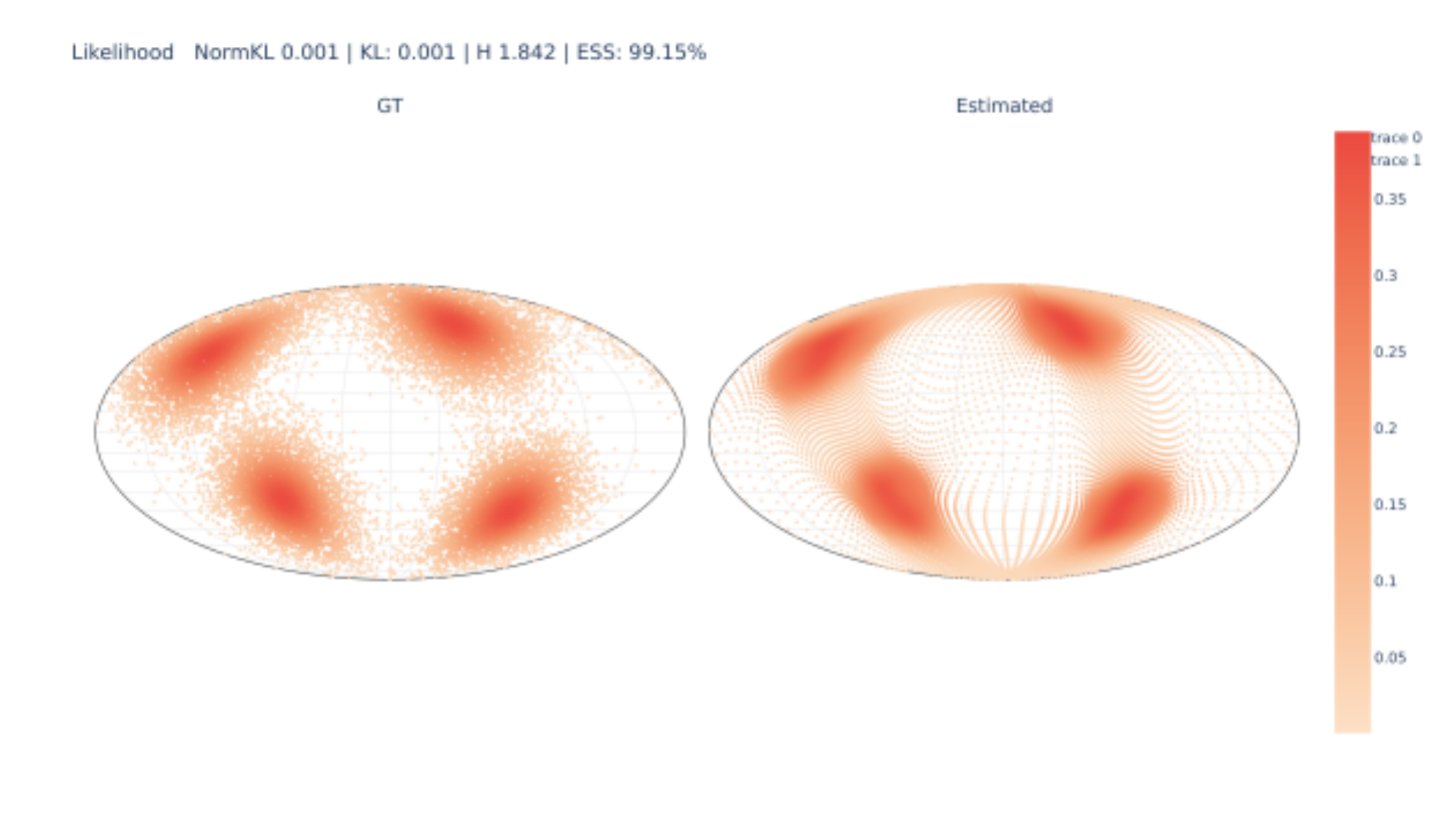}
                \put(79,-2){  6.30}
            \end{overpic}
            \caption{\label{fig:rcpmvsvqe_rcpm_samp} RCPM-LH}
        \end{subfigure}
        \hfill
        \begin{subfigure}[b]{0.24\linewidth}
            \begin{overpic}
                [trim=15.35cm 5.5cm 3cm 6cm,clip,width=\linewidth, grid=False]{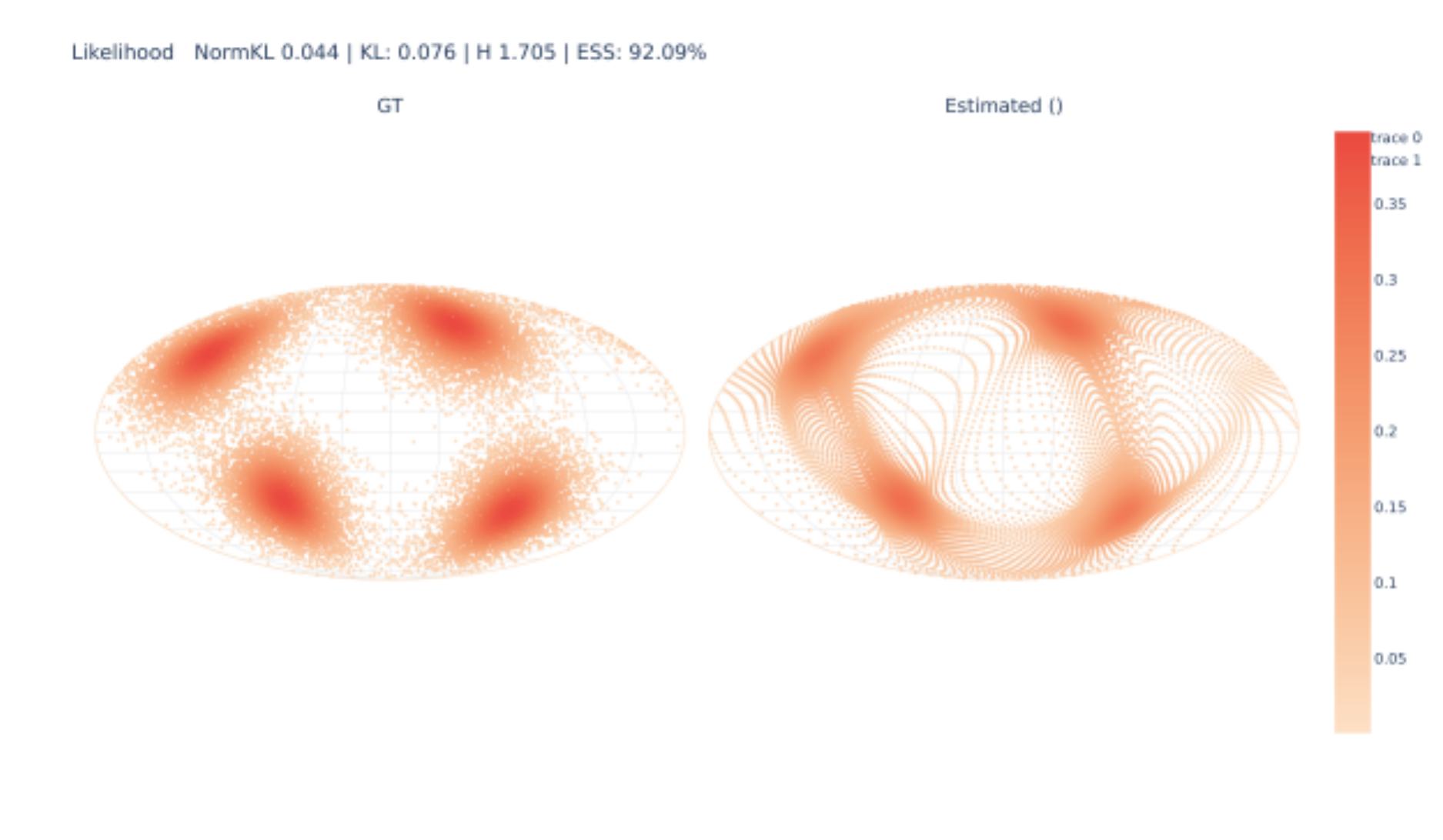}
                \put(79,-2){ 3.84}
            \end{overpic}
            \caption{\label{fig:rcpmvsvqe_vqe_samp} M-VQE}
        \end{subfigure}
    \caption{\label{fig:rcpmvsvqe_vqe} Sampling (KDE-$L_1$ $\times 10^{-5}$)}
    \end{subfigure}
    \caption{\label{fig:rcpmvsvqe} \textbf{Likelihood and Sampling on the `Multimodal von-Mises' distribution.} We compare the results obtain with our M-VQE and the RCPM method \citep{cohen2021rcpm} trained with KL-divergence (RCPM-KL) and Maximum log-likelihood (RCPM-LH). We also report the ESS$_\%$ and KDE-$L_1$ ($\times 10^{-5}$) values.}
\end{figure}

\subsection{Comparison to \cite{cohen2021rcpm}}
While we adopt the parametrization strategy for $c$-concave function introduced by \cite{cohen2021rcpm}, our work differs from \cite{cohen2021rcpm} in two crucial ways: (i) we employ a different training strategy; (ii) we retrieve a family of maps conditioned on covariates $\rvec{X}$, whereas \cite{cohen2021rcpm} retrieve a single map.

\cite{cohen2021rcpm} propose two alternative strategies to train a normalizing flow between the base and target distributions: (i) training only the forward map by minimizing the KL-divergence w.r.to the groundtruth target samples, (ii) training only the inverse map by performing maximum likelihood in with respect to the base distribution.
In contrast to these both training strategies, we retrieve both forward and inverse maps \textit{simultaneously} by solving the optimal transport problem.

In Figure \ref{fig:rcpmvsvqe}, we compare these three training strategies, both visually and quantitatively, by evaluating the likelihood and sampling quality on the `Multimodal von-Mises' distribution. 
Across all strategies, to allow for a fair comparison, we train a single $c$-concave potential discretized at 200 points ($\{\vec{z}_i, \alpha_i\}_{i=1}^200$). 
Likelihoods are computed by leveraging the Jacobian of the inverse of the quantile function $Q^{-1}_{\rvec{Y}}$. 
For RCPM with KL-divergence (RCPM-KL), we train the forward potential $\varphi$ and then calculate the backward potential using the $c$-transform to compute the likelihood. 
Conversely, for RCPM with maximum likelihood (RCPM-LH), we sample points using the $c$-transform of the trained backward potential.
The results highlight that our method attains a comparable ESS$_\%$ and the smallest KDE-$L_1$ error. Notably, RCPM-LH achieves a nearly perfect likelihood, possibly attributed to the chosen loss function. However, in terms of sampling quality, our model, M-VQE, remains superior.

\section{\label{app:exp}\uppercase{Experimental details}}

\subsection{Hyperparameters}
We train all models for $5 \times 10^{4}$ iterations, using Adam optimizer with a learning rate set at $10^{-3}$. We discretize the c-concave potentials using 200 points ($\alpha$'s) and approximate the potentials with 4-6 layered c-concave functions.

\textbf{Multimodal von-Mises.} For the Multimodal von-Mises dataset, we train an M-CVQF with a learning rate of 0.001, incorporating a single c-concave $\beta$ potential.

\textbf{Synthetic datasets.} In the case of the synthetic datasets, our approach involves training an M-CVQF with a learning rate of 0.001, implementing a c-concave $\beta$ potential comprising 4 layers. The c-concave function is further augmented with inner sizes of $[2, 4, 8]$.

\textbf{Continental drift.} For the Continental Drift dataset, we train an M-CVQF using a learning rate of 0.0005. The c-concave $\beta$ potential is composed of 6 layers, while the c-concave function encompasses inner sizes of $[4, 8, 12, 16, 20, 24]$. The dataset contains 3.8M samples.

\textbf{Dihedral angles.} In the context of the Dihedral Angles dataset, our approach involves training an M-CVQF with a learning rate of 0.0005. Similar to the Continental Drift dataset, the c-concave $\beta$ potential consists of 6 layers, and the c-concave function's inner sizes are specified as $[4, 8, 12, 16, 20]$. The dataset contains 350k samples.

\textbf{Spherical uniform distribution $\mathcal{U}_{\Sphere{n}}$.}   
We compute the manifold uniform distribution on the $n$-sphere sampling each component of the vector $\vec{v}\in \R^n$ from a random uniform distribution and then normalizing $\vec{v}$ such that $\parallel \vec{v} \parallel = 1$.

\textbf{Torus uniform distribution $\mathcal{U}_{\mathcal{T}^2}$.} 
We compute the manifold uniform distribution on the torus sampling from the spherical uniform distributions $\mathcal{U}_{\Sphere{1}}$ on the $\Sphere{1}$ and then concatenating the results:
$\vec{u} = [\vec{u}_1,\vec{u}_2] \sim \mathcal{U}_{\mathcal{T}^2}$ with $\vec{u}_1,\vec{u}_2 \sim \mathcal{U}_{\Sphere{1}}$.

\begin{figure}
    \centering
    \begin{subfigure}[b]{0.24\linewidth}
        \begin{overpic}
            [trim=0.2cm 0.2cm 0.2cm 0.2cm,clip,width=\linewidth]{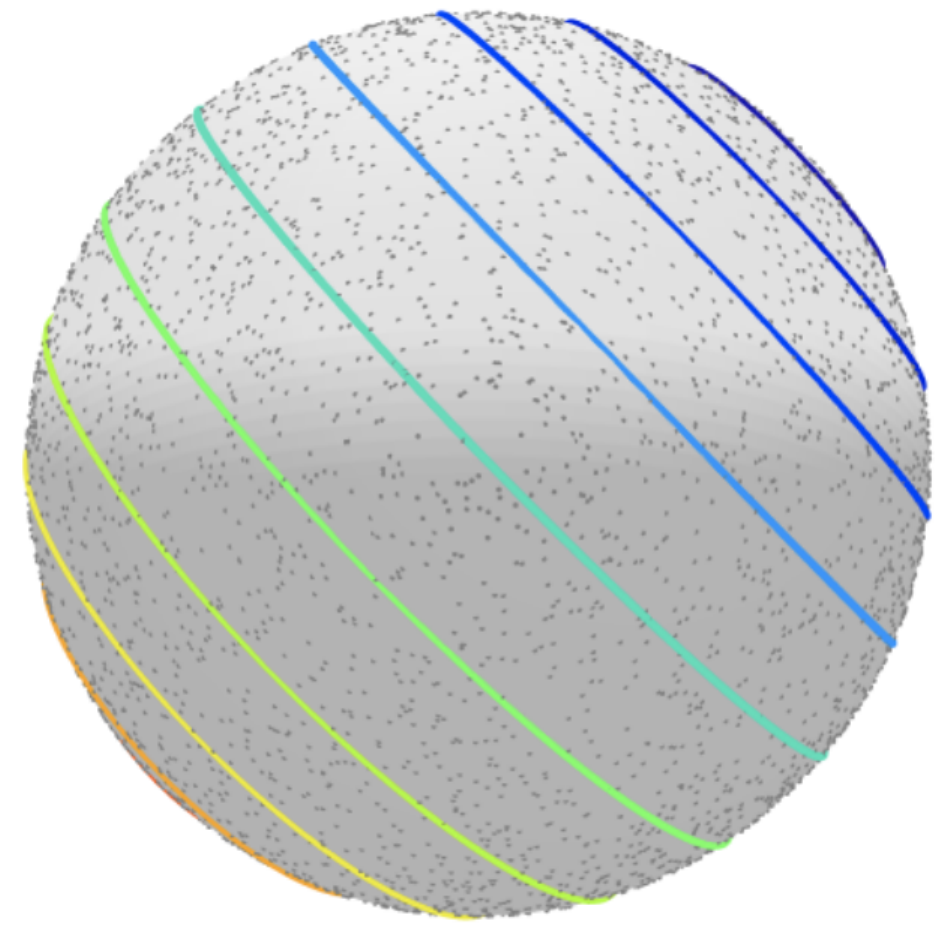}
        \end{overpic}
        \caption{$\Sphere{n}$}
    \end{subfigure}
    \hspace{1cm}
    \begin{subfigure}[b]{0.34\linewidth}
        \begin{overpic}
            [width=0.88\linewidth]{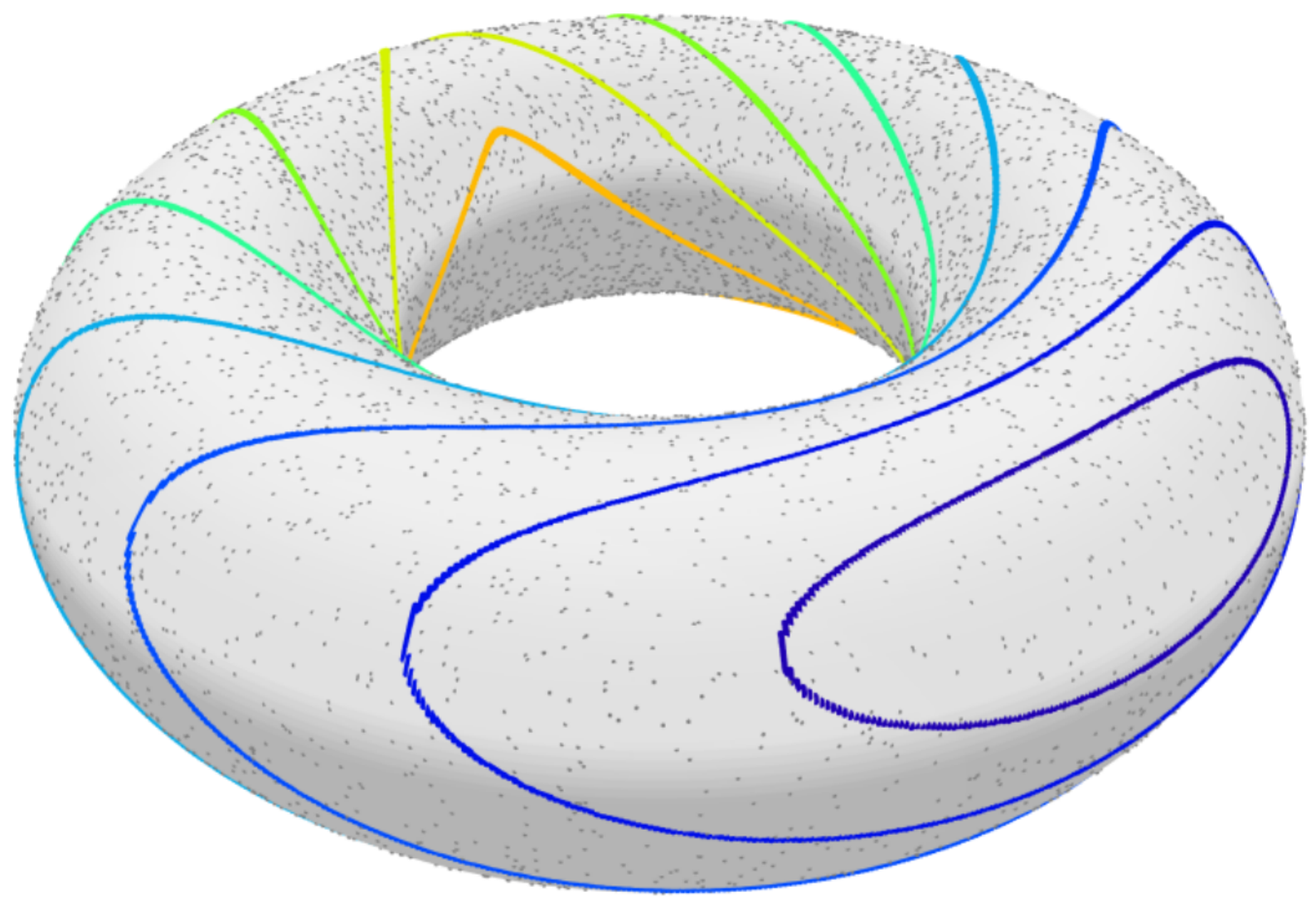}
        \end{overpic}
        \begin{overpic}
            [trim=0cm -3.2cm 0cm 0cm,clip,width=0.1\linewidth]{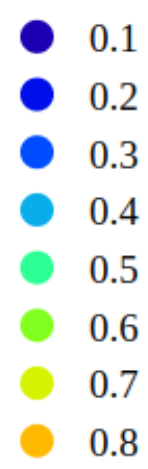}
            \put(10,100){\footnotesize $\tau$}
        \end{overpic}
        \caption{$\mathcal{T}^2$}
    \end{subfigure}
    \caption{\textbf{$\tau$-contours of the Manifold Uniform Distributions $\mathcal{U}_\Manifold$}}
    \label{fig:U_contours}
\end{figure}

\subsection{\label{app:conf_sets} Creating confidence sets}
Given a point $\vec{\omega} \in \Manifold$, we define a confidence set with a level of confidence $(1- \tau)$ on a manifold uniform distribution as the set of points contained in a $\tau$-contour with pole $\vec{\omega}$: 
\begin{equation} \label{eq:confset_base}
    \mathbb{C}_{(1-\tau)}^{\rvec{U}} = \{\vec{u} \in \Manifold : 
    C^*_{\vec{\omega}}(\vec{u}) \leq \tau \}
\end{equation}
$C^*_{\vec{\omega}}$ is a function that maps distances $d_\Manifold(\vec{\omega}, \vec{u})$, with $\vec{u} \in \Manifold$, to the probabilities $\tau \in [0,1]$. We build this function empirically for each pole $\vec{\omega}$. 
We compute the set of points $ \mathcal{C}_{\kappa}^* = \left\{\vec{u} \in \Manifold : d_\Manifold(\vec{\omega}, \vec{u}) = \kappa \right\}$ and the amount of probability  $\tau_\kappa$ contained in $\mathcal{C}_{\kappa}^*$  as the percentage of points $\vec{u} \in \Manifold$ with $d_\Manifold(\vec{\omega}, \vec{u}) \leq \kappa$.
The function $C^*_{\vec{\omega}}$ is then the interpolation of the pairs $\left\{ (\kappa, \tau_\kappa)\right\}_{i=1}^{N_\kappa}$.
Figure \ref{fig:confset} shows the amount of probability contained in the contours as their distance from the pole increases with and without using $C^*_{\vec{\omega}}$.

\subsubsection{$\tau$-contours}

The $\tau$-contours are computed using the vector quantile function $\hat{Q}_{\rvec{Y}}$. First of all, we compute the $\tau$-contours $\mathcal{C}_{\tau}^{\rvec{U}}$ on the base distribution $\mu$ (Figure \ref{fig:U_contours}).
Then we map $\mathcal{C}_{\tau}^{\rvec{U}}$ to the target distribution using the learned vector quantile function $\hat{Q}_{\rvec{Y}}$: 
$\mathcal{C}_{\tau}^{\rvec{Y}}:=\hat{Q}_{\rvec{Y}}(\mathcal{C}_{\tau}^{\rvec{U}})$.
\section{\label{app:add_exp} \uppercase{Additional experiments results}}
In this section, we present additional experimental results and accompanying plots that extend upon the findings reported in the main paper.

\begin{table}
    \centering
    \caption{ \textbf{KDE-$L_1$ ($\times 10^{-4}$) computed over samples of M-VQR and GT}. We report the values for all the conditional distributions considered in the paper.}
    \label{tab:kde_gt}
    \begin{tabular}{cccc}
    $\Manifold$  & ${\rvec{Y}|\rvec{X}}$  & M-VQR & GT \\ \toprule
  \multicolumn{1}{l|}{\multirow{3}{*}{ $\Sphere{2}$ }}  &     Cond. Multimodal& $ 14.2\pm  2.09$ &  $ 8.21 \pm 2.69 $\\ 
\multicolumn{1}{l|}{}     &Scaled Star  & $ 2.25\pm  0.99$ &  $ 2.95 \pm 1.25 $\\
\multicolumn{1}{l|}{}     & Scaled Heart & $ 4.46 \pm 1.18 $ &  $ 3.59\pm  1.21$\\
\multicolumn{1}{l|}{}     & Continental Drift &  $ 17.2\pm 2.6 $ &   $ 9.45\pm 1.06 $\\ \midrule
\multicolumn{1}{l|}{\multirow{3}{*}{ $\mathcal{T}^2$ }} & Cond. Multimodal& $ 17.5\pm 2.92 $ &  $ 11.4\pm  4.87$\\
\multicolumn{1}{l|}{}     & Scaled Star  & $ 5.85 \pm 1.92 $ &  $ 3.44\pm  2.07$\\
\multicolumn{1}{l|}{}     & Scaled Heart  & $7.93 \pm 2.14$ &  $ 4.17 \pm  2.1$\\
\multicolumn{1}{l|}{}     & Dihedral angles  & $ 26.1\pm 5.6 $ &   $ 7.34\pm 2.88 $\\

    \end{tabular}
    
\end{table}

\paragraph{KDE-$L_1$ scores.}
Table \ref{tab:kde_gt} provides a detailed breakdown of the KDE-$L_1$ scores for all the conditional distributions featured in the primary paper. Specifically, we furnish both the mean and standard deviation of the KDE-$L_1$ values, calculated for two distinct scenarios: (1) between ground truth samples and samples generated by our method (M-VQR), and (2) solely among samples drawn from the ground truth distribution (GT). The latter value aids us in determining the extent to which any discrepancies are attributed to finite sampling, thus allowing us to measure the effectiveness of our method in accurately capturing the ground truth distribution through sampling.

\begin{figure*}[h!]
\centering
\hfill
\begin{subfigure}[b]{0.3\linewidth}
\centering
\begin{overpic}
[trim=1cm 3.5cm 0cm 2cm,clip,width=\linewidth, grid=false]{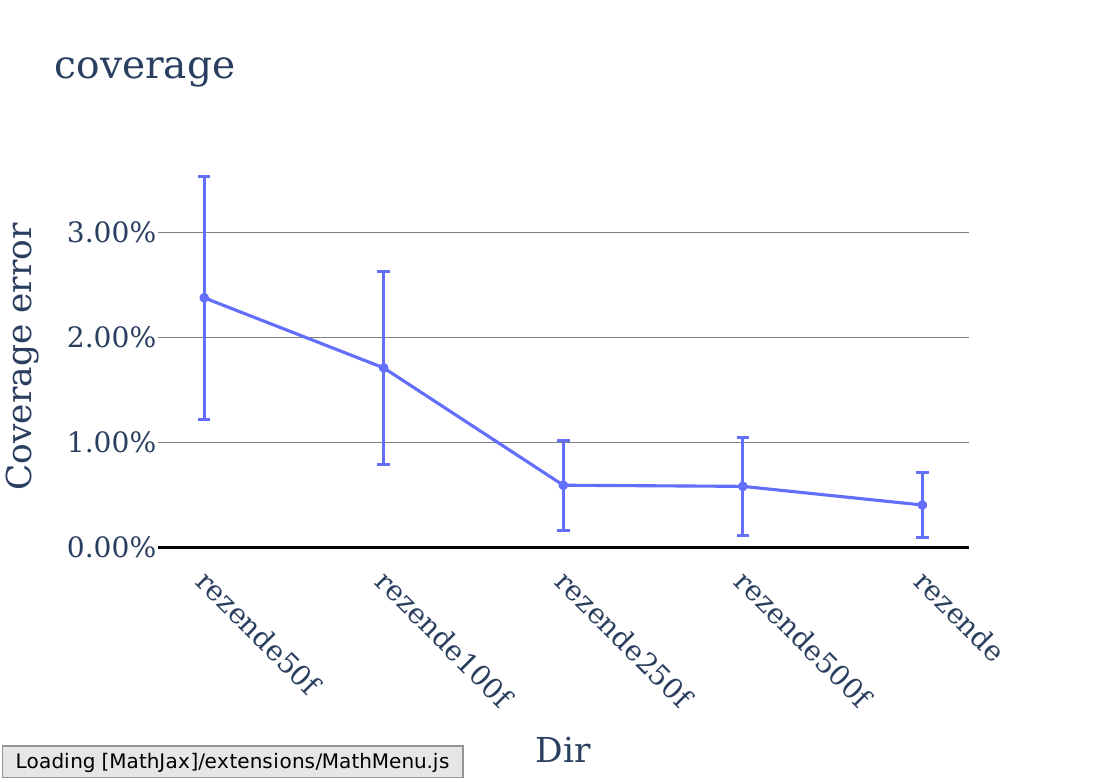}
\put(-4,7){\rotatebox{90}{\tiny Coverage error}}
\put(11,-2){\tiny 2500}
\put(28,-2){\tiny 5000}
\put(44,-2){\tiny 12500}
\put(60,-2){\tiny 25000}
\put(79,-2){\tiny 50000}
\put(20,-6){\tiny Number of train samples}
\end{overpic}
\end{subfigure}
\hfill
\begin{subfigure}[b]{0.3\linewidth}
\centering
\begin{overpic}
[trim=1cm 3.5cm 0cm 2cm,clip,width=\linewidth, grid=false]{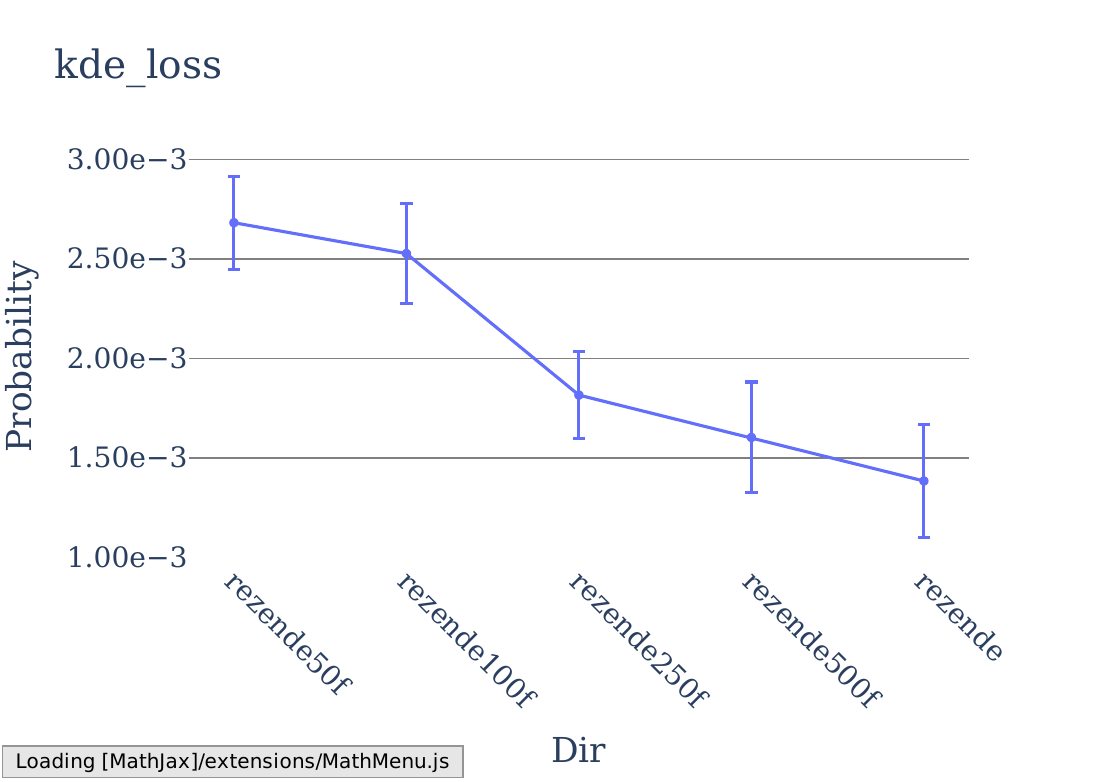}
\put(-4,14){\rotatebox{90}{\tiny KDE-$L_1$}}
\put(13,-2){\tiny 2500}
\put(29,-2){\tiny 5000}
\put(45,-2){\tiny 12500}
\put(62,-2){\tiny 25000}
\put(79,-2){\tiny 50000}
\put(20,-6){\tiny Number of train samples}
\end{overpic}
\end{subfigure}
\hfill
\begin{subfigure}[b]{0.3\linewidth}
\centering
\begin{overpic}
[trim=1cm 3.5cm 0cm 2cm,clip,width=\linewidth, grid=false]{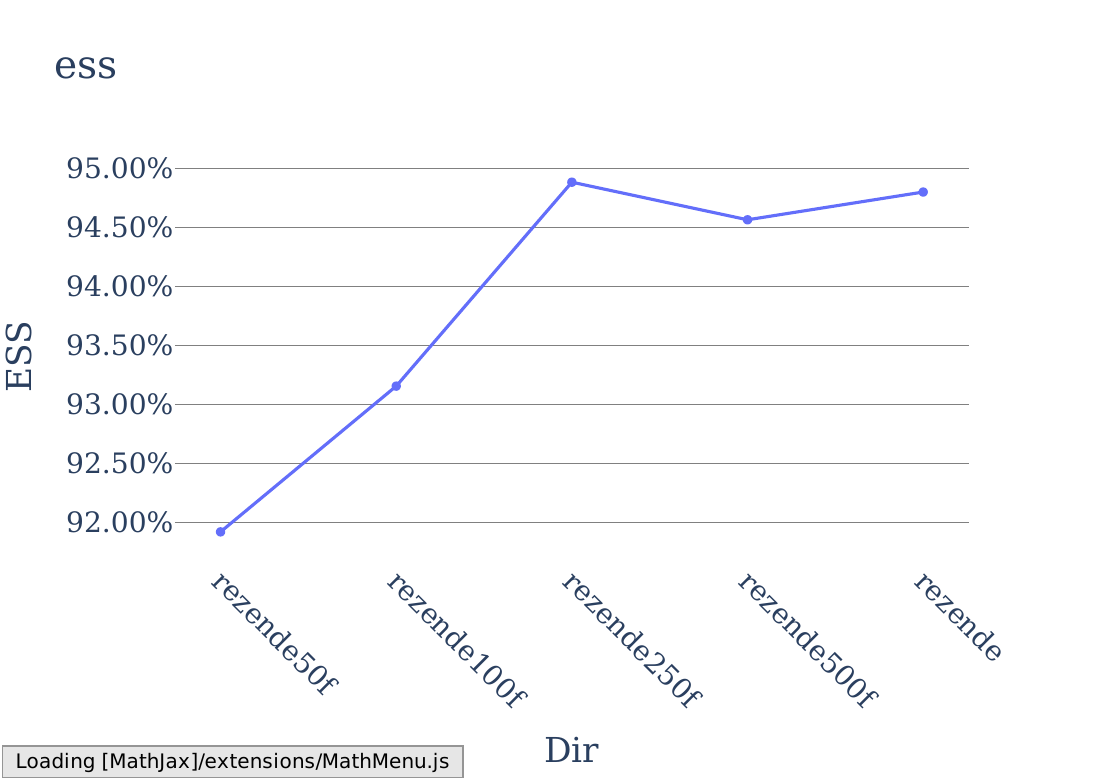}
\put(-4,18){\rotatebox{90}{\tiny ESS}}
\put(13,-2){\tiny 2500}
\put(29,-2){\tiny 5000}
\put(45,-2){\tiny 12500}
\put(62,-2){\tiny 25000}
\put(79,-2){\tiny 50000}
\put(20,-6){\tiny Number of train samples}
\end{overpic}
\end{subfigure}
\vspace{0.1cm}
\caption{\label{fig:samples_abl} \textbf{Performance variation as the number of training samples increases.} We report the Coverage error,  KDE-$L_1$ and ESS(\%) for the M-VQR trained on the Conditional Multimodal distribution on a sphere.}
\end{figure*}

\paragraph{Sample complexity.} 
For M-VQE, we experimented with ablating the number of samples, and observed that the estimation accuracy (ESS, KDE-L1) saturates at $N=1k$. In contrast, M-VQR has much higher sample complexity. This is intuitive because modeling conditional distributions is significantly harder than estimating a single distribution (as in M-VQE). To demonstrate this, we performed an ablation study of M-VQR on conditional multi-modal distributions with increasing number of samples, as presented in Figure \ref{fig:samples_abl}.
The results suggest that the Coverage error and ESS saturate at $N=12.5k$, while KDE-L1 error drops as N increases.

\begin{figure*}[h!]
\centering
\begin{subfigure}[b]{0.3\linewidth}
\centering
\begin{overpic}
[trim=2.07cm 1.9cm 3cm 2.5cm,clip,width=\linewidth, grid=False]{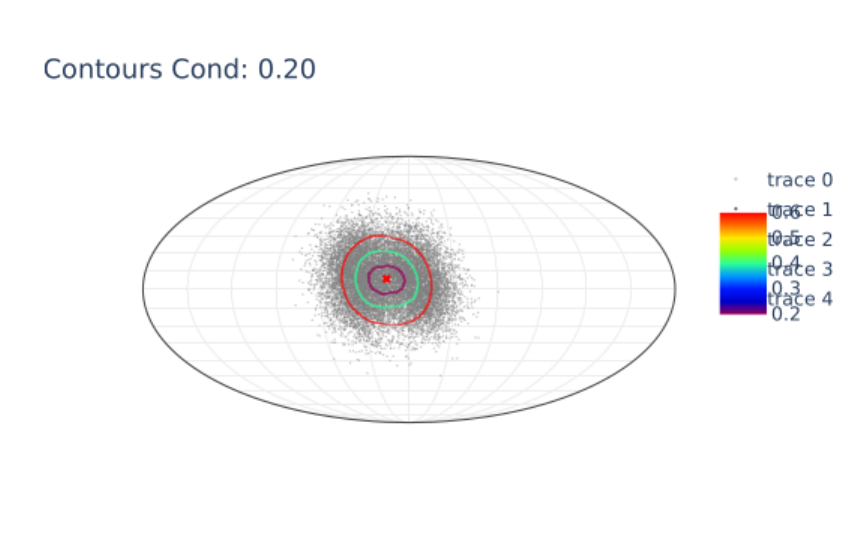}
\end{overpic}
\caption{\label{fig:cont_duec} \rvec{Y}|\rvec{X}=0.2}
\end{subfigure} 
\hspace{0.2cm}
\begin{subfigure}[b]{0.3\linewidth}
\centering
\begin{overpic}
[trim=2.07cm 1.9cm 3cm 2.5cm,clip,width=\linewidth, grid=false]{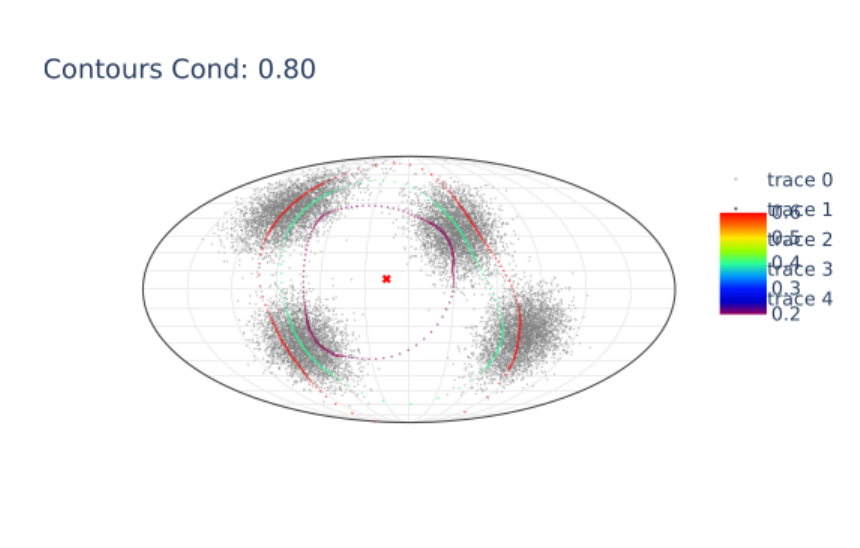}
\end{overpic}
\caption{\label{fig:cont_pleist} \rvec{Y}|\rvec{X}=0.8}
\end{subfigure} 
\begin{subfigure}[b]{0.04\linewidth}
\centering
\begin{overpic}
[trim=4.52cm -3.5cm 3.45cm 2.79cm,clip,width=\linewidth, grid=false]{./figs/synth/legend_synth}
\put(10,100){\footnotesize $\tau$}
\end{overpic}
\end{subfigure} 
\hfill
\begin{subfigure}[b]{0.3\linewidth}
\centering
\begin{overpic}
[trim=0cm 0.5cm 0cm 2cm,clip,width=\linewidth, grid=false]{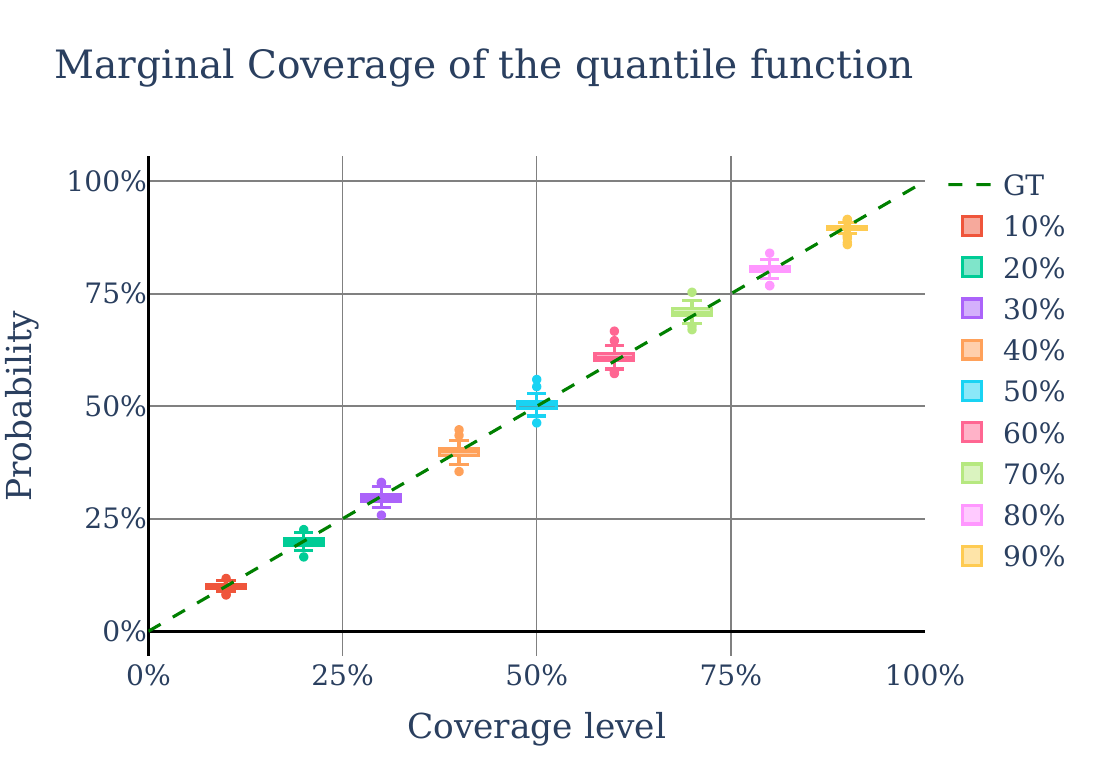}
\end{overpic}
\caption{\label{fig:cov_contdrift} Marginal Coverage}
\end{subfigure} 

    \begin{subfigure}[b]{0.49\linewidth}
        \begin{overpic}
        [trim=1.5cm 5.5cm 3cm 4cm,clip,width=\linewidth, grid=false]{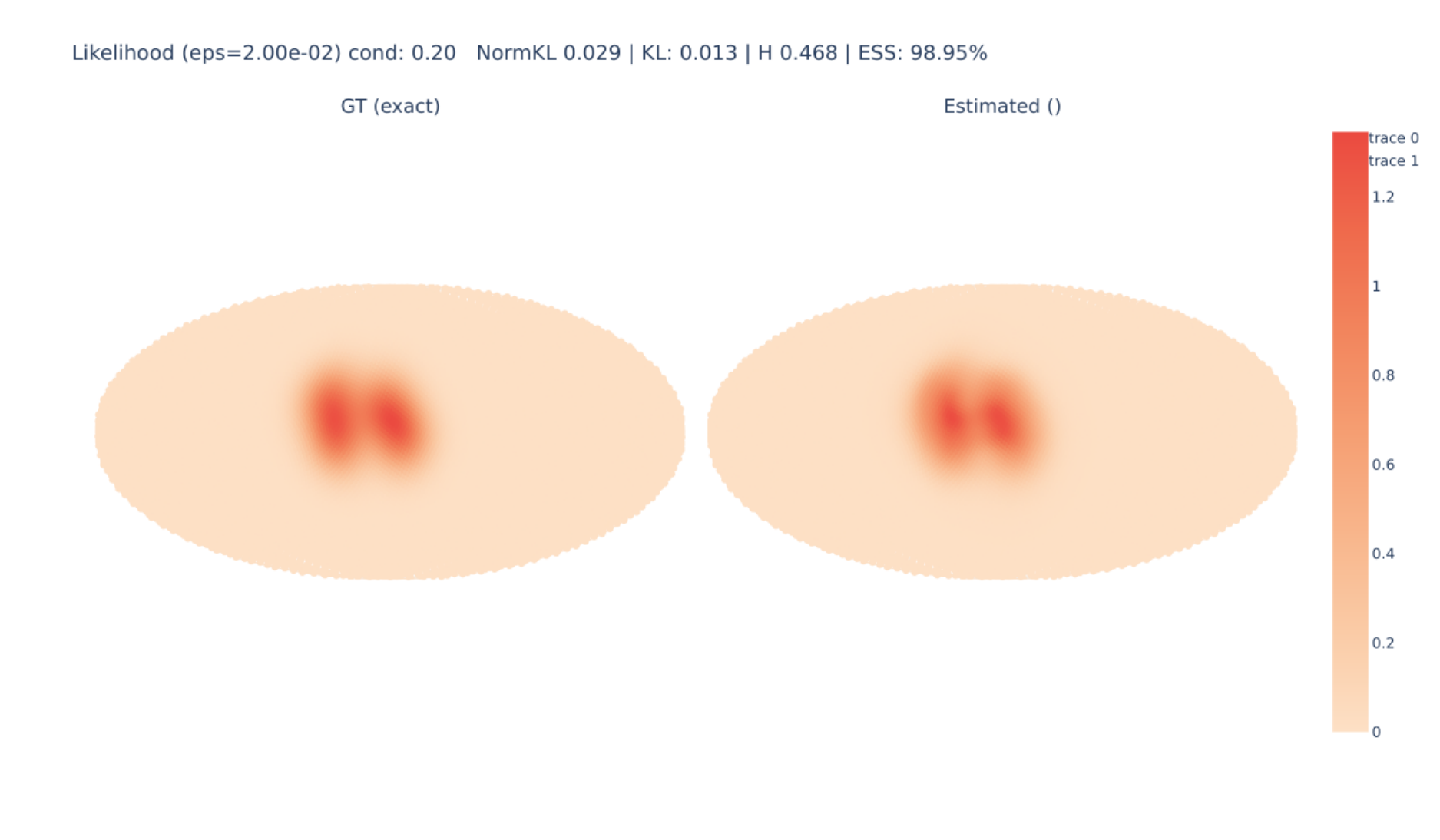}
        \put(12,26){ Ground Truth}
        \put(64,26){ Estimated}
        \put(85,0){\tiny $ESS_\% = 99\%$}
        \end{overpic}
    \caption{\label{fig:contdrift_200_kde_lh}{\rvec{Y}|\rvec{X}=0.2}}
    \end{subfigure}
    \hfill
    \begin{subfigure}[b]{0.49\linewidth}
        \begin{overpic}
        [trim=1.5cm 5.5cm 3cm 4cm,clip,width=\linewidth, grid=false]{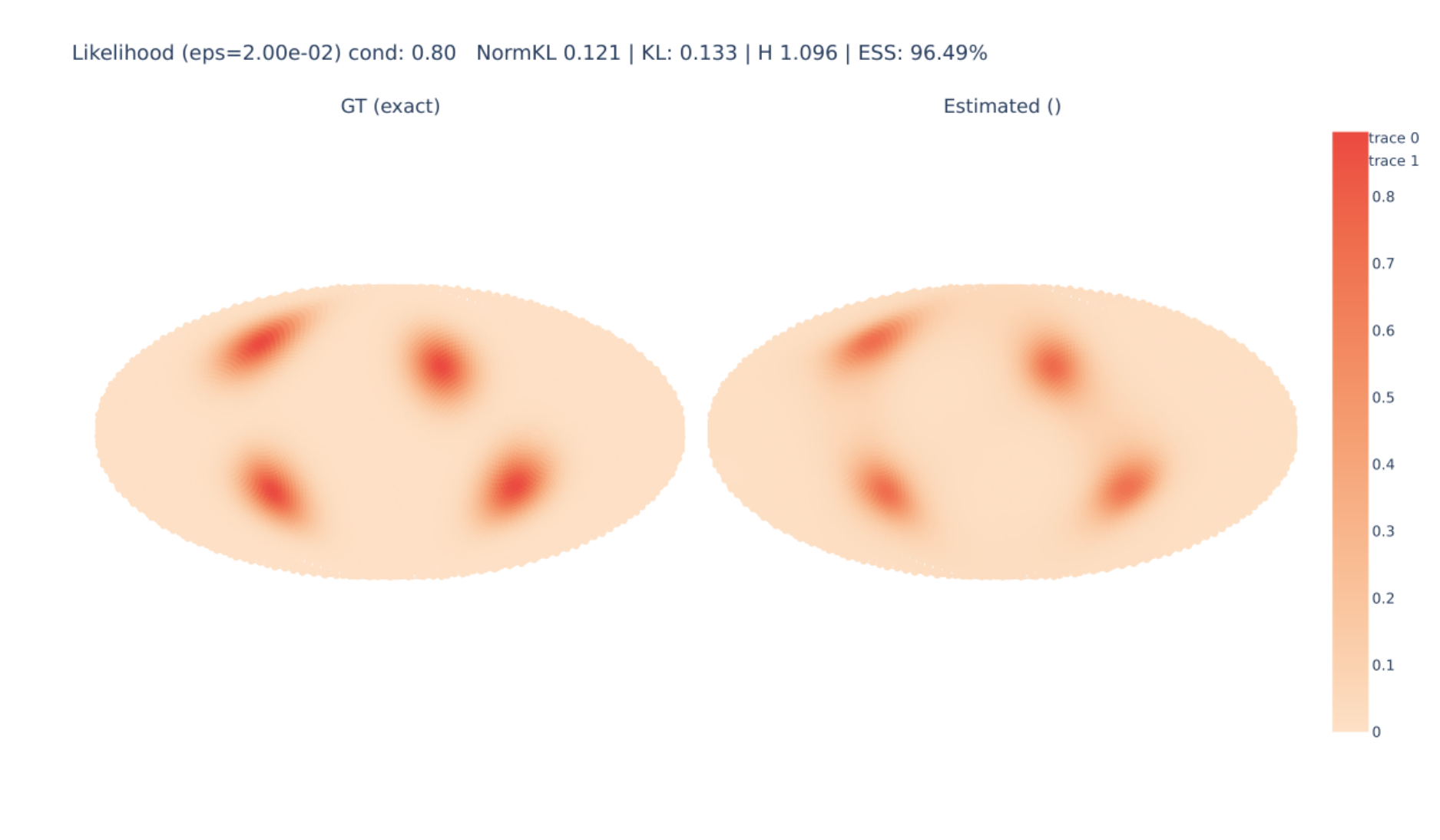}
        \put(12,26){ Ground Truth}
        \put(64,26){ Estimated}
        \put(85,0){\tiny $ESS_\% = 96\%$}
        \end{overpic}
    \caption{\label{fig:contdrift_200_kde_lh}{\rvec{Y}|\rvec{X}=0.8}}
    \end{subfigure}

\caption{\label{fig:cont_multitr}
\textbf{$\tau$-confidence sets and Likelihood computed with M-VQR on the `Translated Multimodal' dataset.} Subfigures (a) and (b) report $\tau$-contours overlayed on the ground truth samples, for different values of $\tau$.
Mollweide projection is used to visualize the whole sphere.
Graph (c) shows the coverage achieved by the model as a function of the requested coverage level, averaged over the different conditionings with relative confidence bars.
Subfigures (d) and (e) show the likelihood and the ESS$_\%$. 
}
\end{figure*}

\paragraph{Translated Multimodal.}
In the main paper, we considered synthetic datasets where conditioning governed the scale of the distribution. Here, we consider a more intricate scenario: a mixture of four von-Mises distributions defined on the manifold $\Manifold$, where the conditioning controls the positions of the modes.
Figure \ref{fig:cont_multitr} showcases the results on the $\Sphere{2}$ sphere. We train an M-VQR model employing a $c$-concave $\beta$ potential with 4 layers. The $c$-concave function is further enriched with inner sizes of $[2, 4, 8]$. Remarkably, our method consistently generates contours with all the desired characteristics: they are nested, smooth, and constitute valid contours, with a mean coverage error of $1.01\% \pm 0.97\%$. Additionally, we present the likelihood, which achieves a notable ESS$_\%$ of 93.72\% and a mean KDE-$L_1$ value of $(14.38 \pm 5.03) \times 10^{-4}$. The ground truth KDE-$L_1$ stands at $(7.759\pm 1.52) \times 10^{-4}$. These results further demonstrate our approach accurately captures the conditional distribution.

\begin{figure}[t]
    \centering
    \begin{overpic}
        [trim= 1.9cm 3cm 3cm 3cm, clip, grid=false, width=\linewidth]{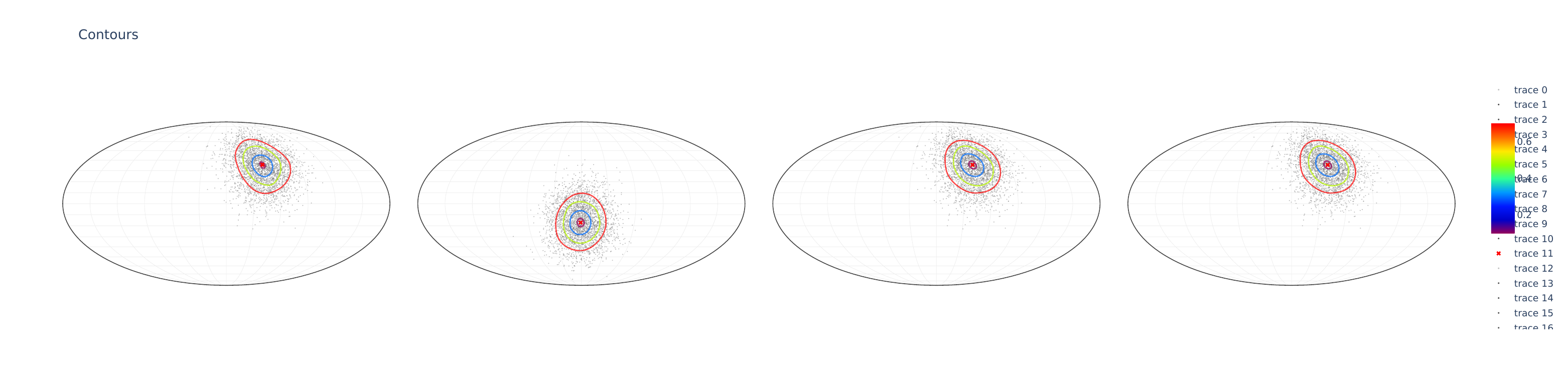}
        \put(11,14){$\rvec{Y}_1$}
        \put(18,0){1.30}
        \put(36,14){$\rvec{Y}_2$}
        \put(43,0){1.03}
        \put(60,14){$T^{-1} \rvec{Y}_2$}
        \put(67,0){1.30}
        \put(85,14){$T^{-1} z_2$}
        \put(92,0){1.21}
    \end{overpic}
    \caption{\textbf{Impact of Transformation $T$ on Estimated Contours.} In this visualization, the first two plots depict the contours derived from two distinct M-VQE models, each trained on different distributions. The second distribution is obtained by applying a transformation $T$ to the first, i.e., $\rvec{Y}_2 = T \rvec{Y}_1$. The third plot showcases the contours produced using the second model and subsequently applying the inverse transformation to these contours, yielding $T^{-1} \rvec{Y}_2$. The fourth plot illustrates the contours generated by applying the inverse transformation to the support points $z_2$ learned by the second model, resulting in $T^{-1} z_2$. Under each plot, we report the KDE-L$_1$ ($\times 10^{-3}$) error.}
    \label{fig:tr_contour}
\end{figure}
\paragraph{Transformation of samples vs learned points of the $c$-concave function.}
We explore the impact of a transformation denoted as $T$ on the learned $c$-concave function, focusing on the $\Sphere{2}$ sphere and two von-Mises distributions, $\rvec{Y}_1$ and $\rvec{Y}_2$. The second distribution, $\rvec{Y}_2$, is derived by applying a 3D rotation to the first, signifying the effect of $T$.
We train two separate models, M-VQE$_1$ and M-VQE$_2$, on these distributions and investigate the consequences of applying the inverse transformation $T^{-1}$ to the second model. Specifically, we apply $T^{-1}$ to both the contours estimated by M-VQE$_2$ ($T^{-1} \rvec{Y}_2$) and the learned support points, denoted as $z_2$, from M-VQE$_2$ ($T^{-1} z_2$). The final two plots in Figure \ref{fig:tr_contour} reveal that the results are remarkably close to the contours estimated by M-VQE$_1$.
To quantify this closeness, we compare the coverage error of M-VQE$_1$ with $T^{-1} \rvec{Y}_2$ and $T^{-1} z_2$, resulting in mean differences of $0.60\% \pm 0.43\%$ and $0.93\% \pm 0.65\%$, respectively. Furthermore, we compute the KDE-L$_1$ error for all four cases and observe that the error remains consistent across these scenarios. 
This analysis highlights the learned points where the $c$-concave function is discretized roughly corresponds to the location of the density that is being modeled.

\begin{figure*}[t]
\centering
\begin{subfigure}[b]{\linewidth}
    \begin{subfigure}[b]{0.44\linewidth}
        \begin{subfigure}[b]{\linewidth}
        \centering
        \begin{overpic}
        [trim=2cm 7cm 5cm 6.5cm,clip,width=\linewidth, grid=False]{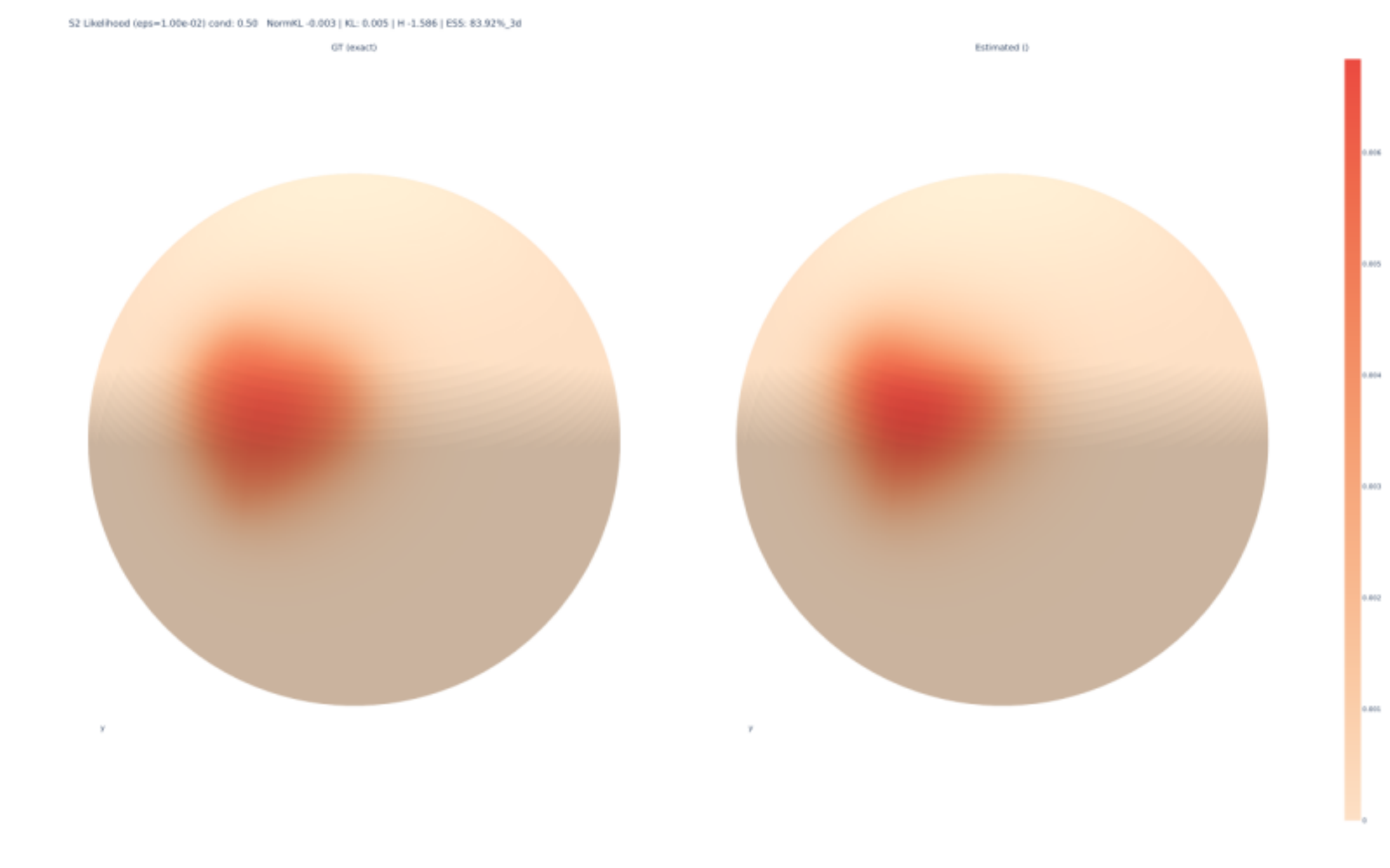}
        \put(9,46){Ground Truth}
        \put(64,46){Estimated}
        \put(85,0){\tiny $ESS_\% = 84\%$} 
        \put(-2,22){\rotatebox[origin=c]{90}{$x=0.5$}}
        \end{overpic}
        \end{subfigure} 
        \begin{subfigure}[b]{\linewidth}
        \centering
        \begin{overpic}
        [trim=2cm 7cm 5cm 6.5cm,clip,width=\linewidth, grid=false]{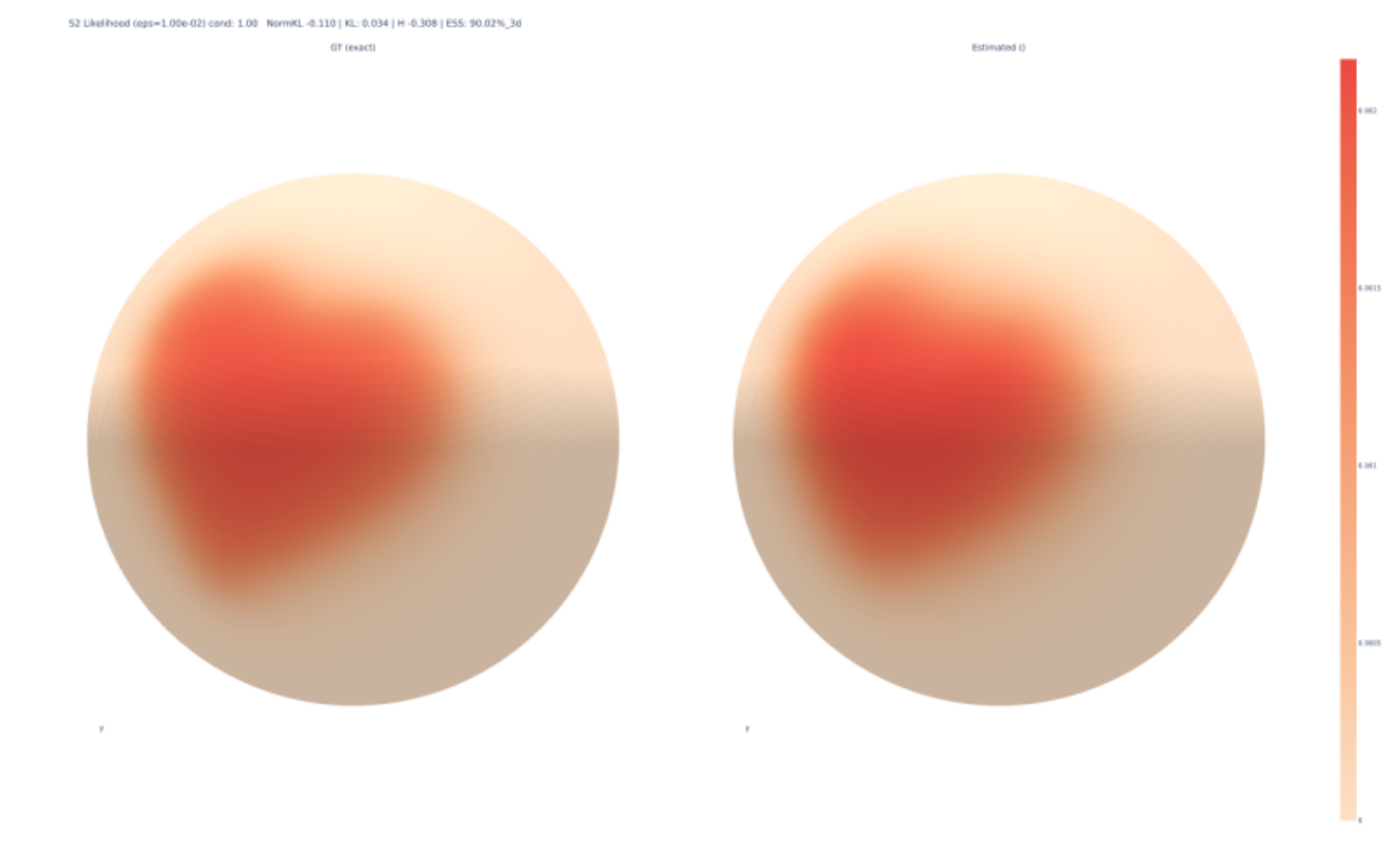}
        \put(85,0){\tiny $ESS_\% = 90\%$} 
        \put(-2,23){\rotatebox[origin=c]{90}{$x=1$}}
        \end{overpic}
        \end{subfigure} 
        \caption{\label{fig:s2_heart_lh} Scaled Heart}
    \end{subfigure}
    \hspace{0.7cm}
    \begin{subfigure}[b]{0.44\linewidth}
        \begin{subfigure}[b]{\linewidth}
        \centering
        \vspace{0.3cm}
        \begin{overpic}
        [trim=2cm 7cm 5cm 6.5cm,clip,width=\linewidth, grid=false]{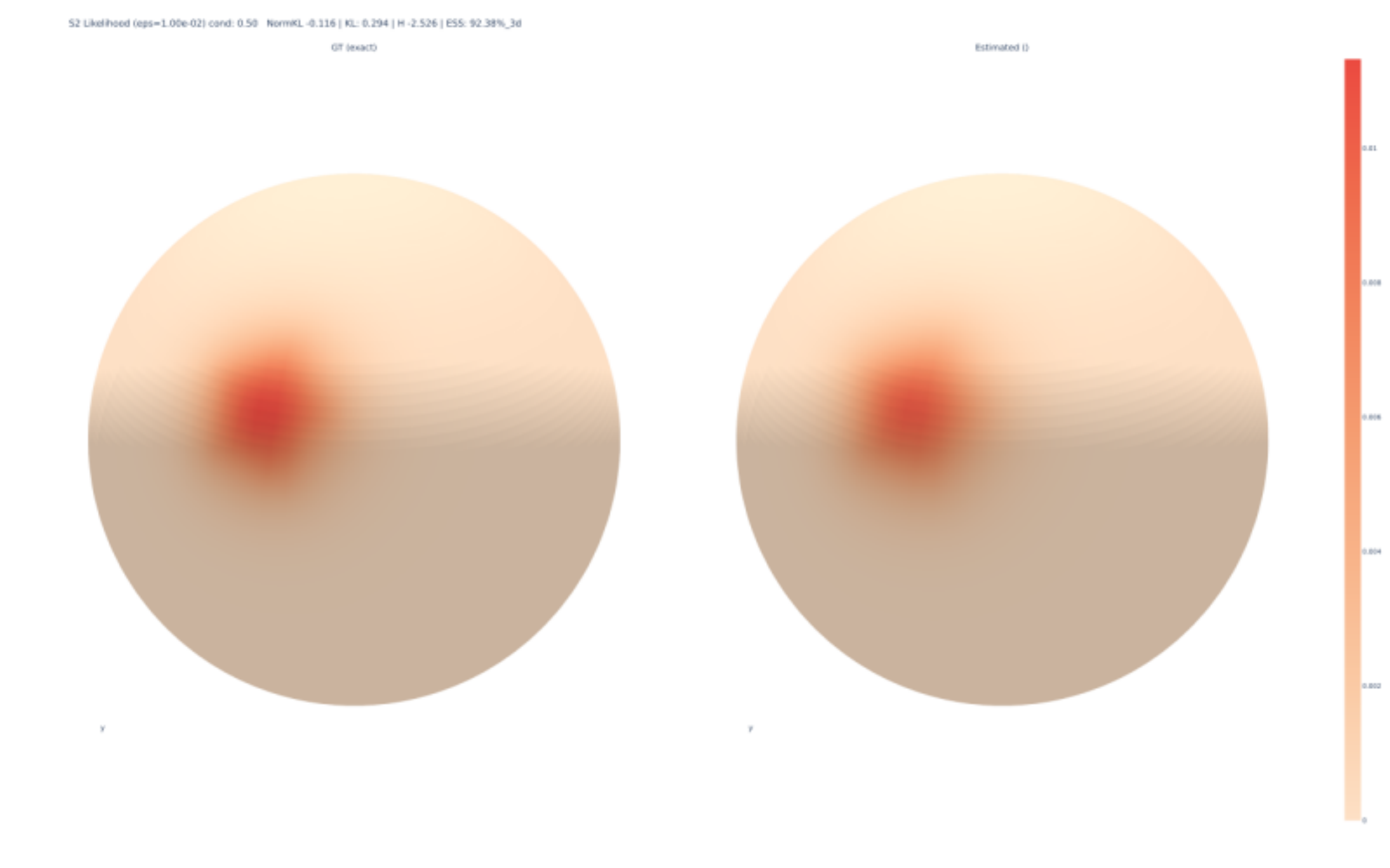}
        \put(9,46){Ground Truth}
        \put(64,46){Estimated}
        \put(85,0){\tiny $ESS_\% = 92\%$} 
        \put(-2,22){\rotatebox[origin=c]{90}{$x=0.5$}}
        \end{overpic}
        \end{subfigure} 
        \begin{subfigure}[b]{\linewidth}
        \centering
        \begin{overpic}
        [trim=2cm 7cm 5cm 6.5cm,clip,width=\linewidth, grid=false]{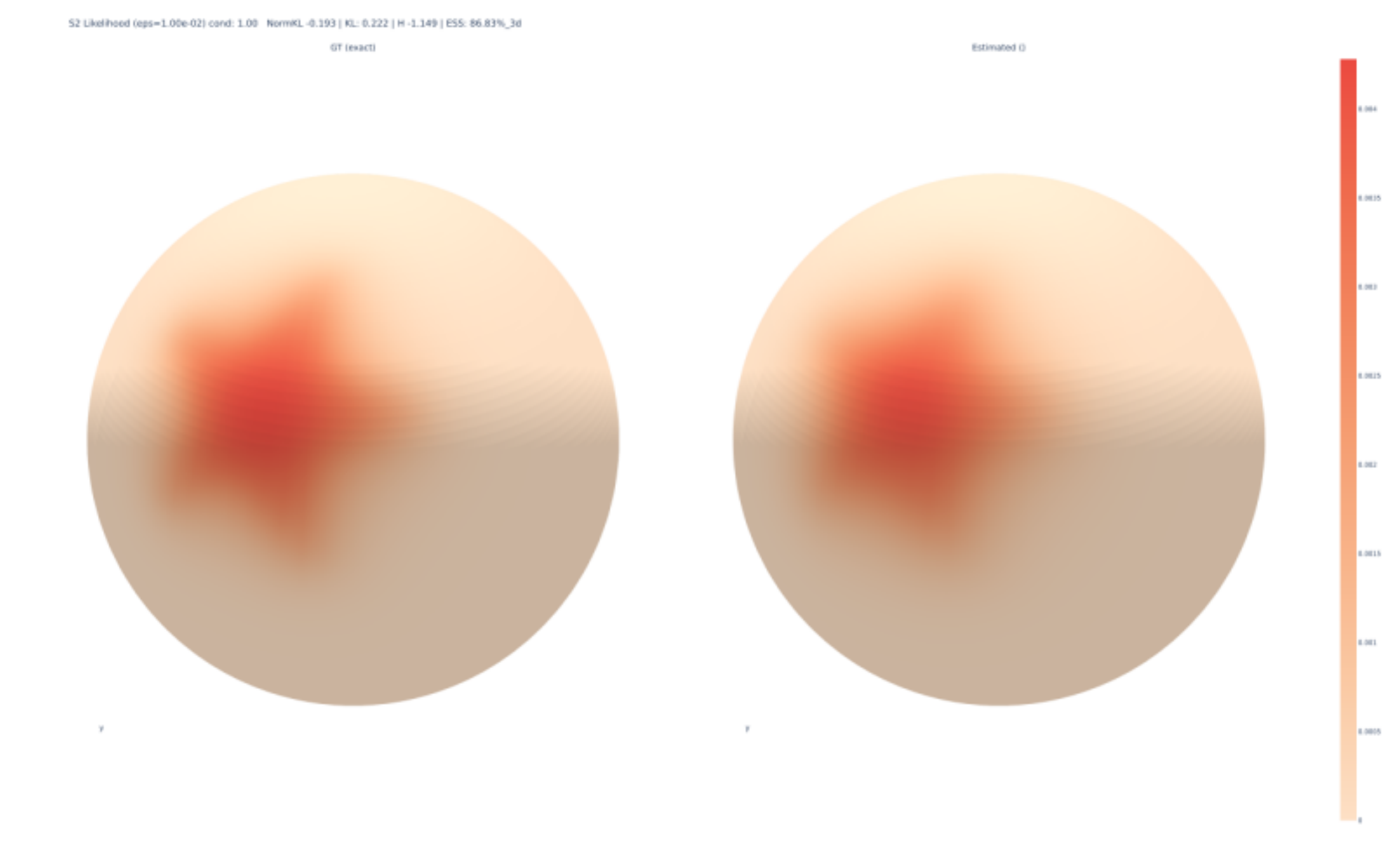}
        \put(85,0){\tiny $ESS_\% = 87\%$} 
        \put(-2,23){\rotatebox[origin=c]{90}{$x=1$}}
        \end{overpic}
        \end{subfigure} 
\caption{\label{fig:s2_star_lh} Scaled Star}
\end{subfigure}
\caption{\label{fig:s2_lh} $\Sphere{2}$ }
\end{subfigure}

\begin{subfigure}[b]{\linewidth}
    \begin{subfigure}[b]{0.44\linewidth}
        \begin{subfigure}[b]{\linewidth}
        \centering
        \begin{overpic}
        [trim=2.6cm 8cm 4.5cm 11cm,clip,width=0.95\linewidth, grid=False]{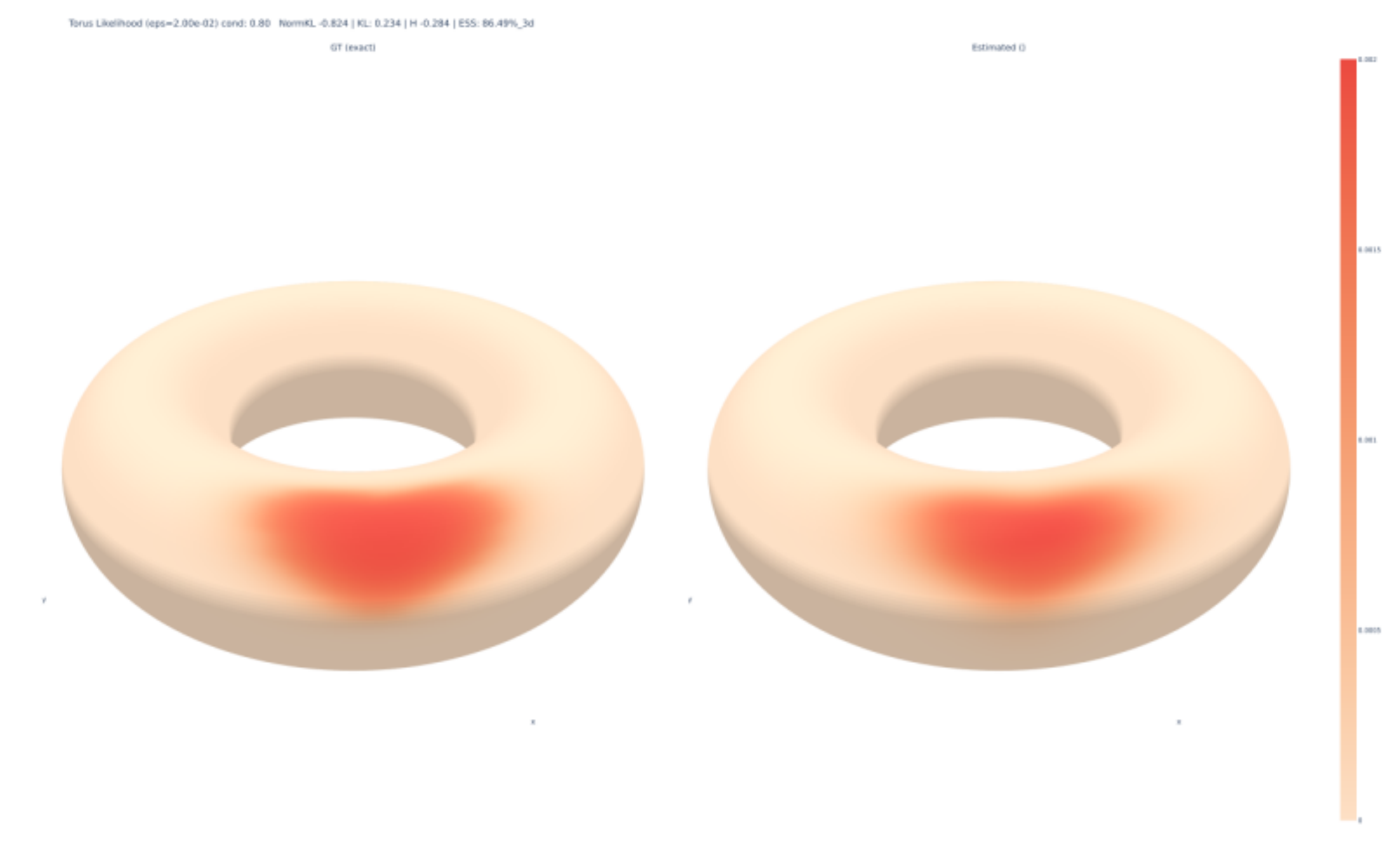}
        \put(9,35){Ground Truth}
        \put(64,35){Estimated}
        \put(85,0){\tiny $ESS_\% = 87\%$} 
        \put(-4,15){\rotatebox[origin=c]{90}{$x=0.8$}}
        \end{overpic}
        \end{subfigure} 
        \begin{subfigure}[b]{\linewidth}
        \centering
        \begin{overpic}
        [trim=2.6cm 8cm 4.5cm 11cm,clip,width=0.95\linewidth, grid=false]{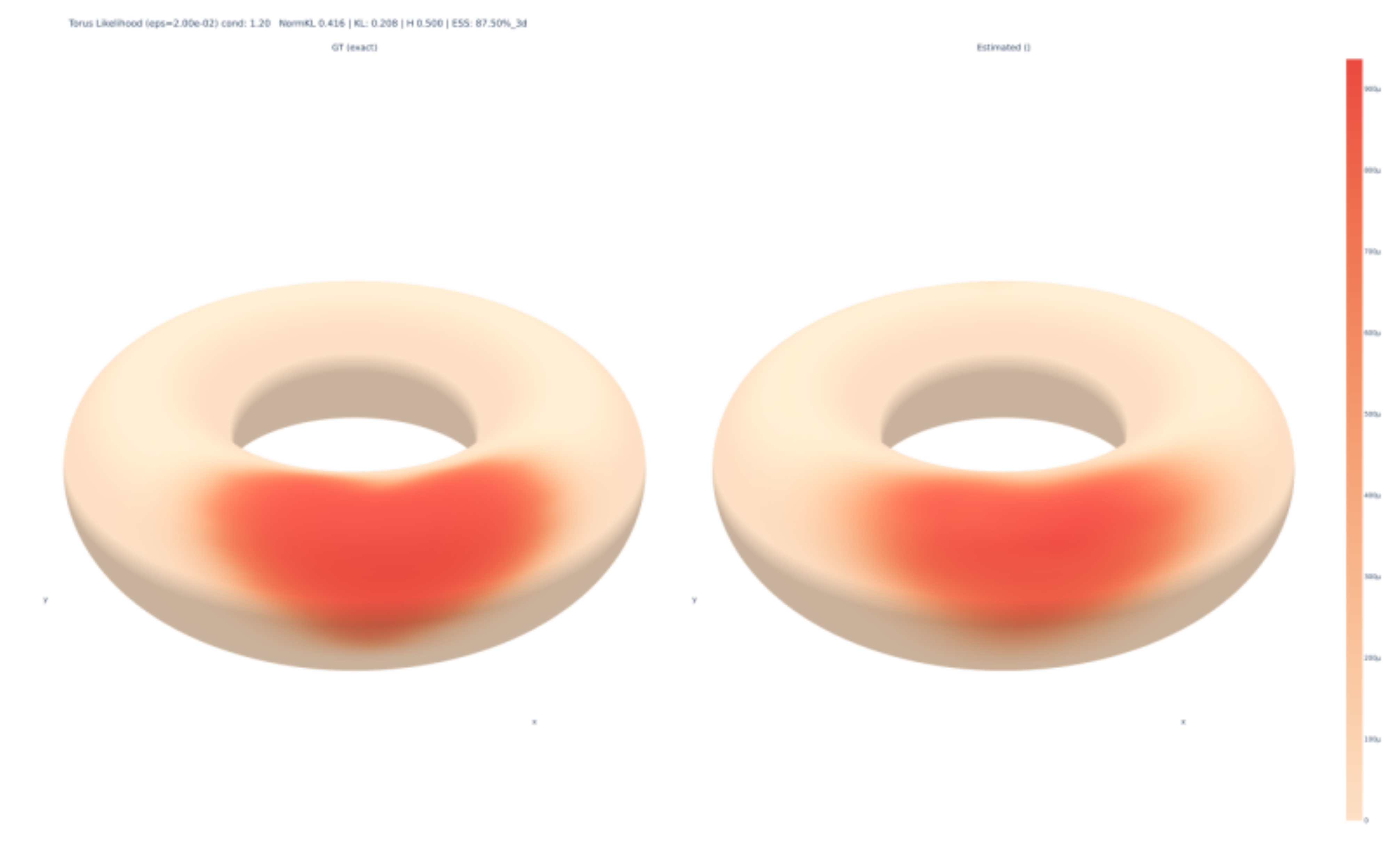}
        \put(85,0){\tiny $ESS_\% = 86\%$} 
        \put(-4,15){\rotatebox[origin=c]{90}{$x=1.2$}}
        \end{overpic}
        \end{subfigure} 
        \caption{\label{fig:t2_heart_lh} Scaled Heart}
    \end{subfigure}
    \hspace{0.7cm}
    \begin{subfigure}[b]{0.44\linewidth}
        \begin{subfigure}[b]{\linewidth}
        \centering
        \vspace{0.3cm}
        \begin{overpic}
        [trim=2.6cm 8cm 4.5cm 11cm,clip,width=0.95\linewidth, grid=false]{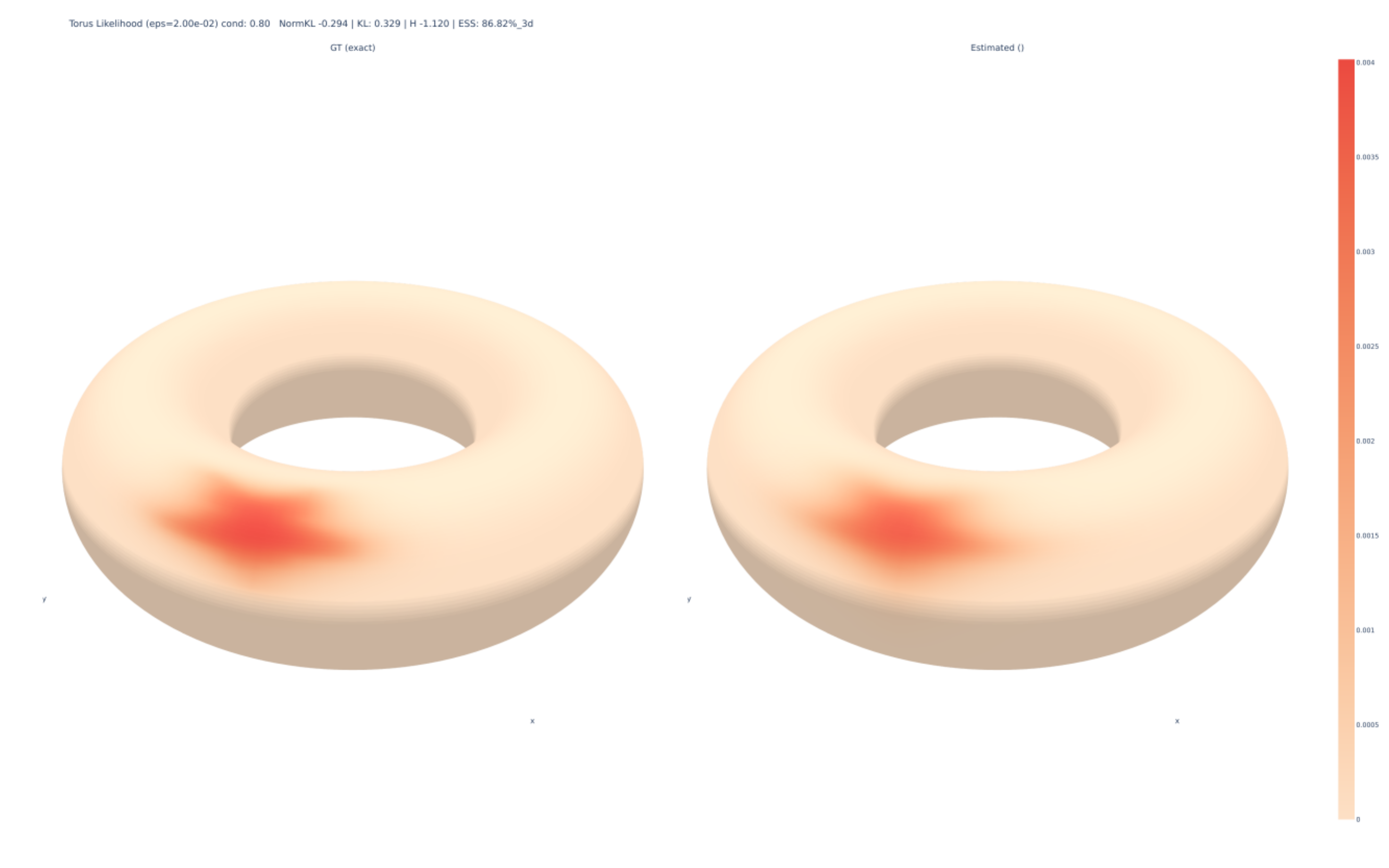}
        \put(9,35){Ground Truth}
        \put(64,35){Estimated}
        \put(85,0){\tiny $ESS_\% = 87\%$} 
        \put(-4,15){\rotatebox[origin=c]{90}{$x=0.8$}}
        \end{overpic}
        \end{subfigure} 
        \begin{subfigure}[b]{\linewidth}
        \centering
        \begin{overpic}
        [trim=2.6cm 8cm 4.5cm 11cm,clip,width=0.95\linewidth, grid=false]{./figs/synth/tos_lh_0.80.pdf}
        \put(85,0){\tiny $ESS_\% = 89\%$} 
         \put(-4,15){\rotatebox[origin=c]{90}{$x=1.2$}}
        \end{overpic}
        \end{subfigure} 
        \caption{\label{fig:t2_star_lh} Scaled Star}
\end{subfigure}
\caption{\label{fig:t2_lh} $\mathcal{T}^2$}
\end{subfigure}

\caption{\label{fig:synth_lh}
\textbf{Likelihood function $p_{\rvec{Y}|\rvec{X}}$ for the Synthetic distributions.}
The covariate $\rvar{X}$ controls the distribution scale. $ESS_{\%}$ values are also reported.}
\end{figure*}

\paragraph{M-VQR: synthetic data experiments.}
Figure \ref{fig:synth_lh} showcases the likelihood alongside the ESS$_\%$ (Effective Sample Size as a percentage) for two distinct conditioning examples within each synthetic distribution featured in the main paper. Notably, in all cases presented, the ESS$_\%$ exceeds 84\%, demonstrating that our method efficiently computes the likelihood of the target conditional distribution. This effectiveness is further illustrated in the accompanying plots.

\paragraph{M-VQR: Real data experiments.}

\begin{figure}[t]
    \centering
    \begin{subfigure}[b]{\linewidth}
        \begin{overpic}
        [trim=1.75cm 6.5cm 3cm 7cm,clip,width=0.62\linewidth, grid=false]{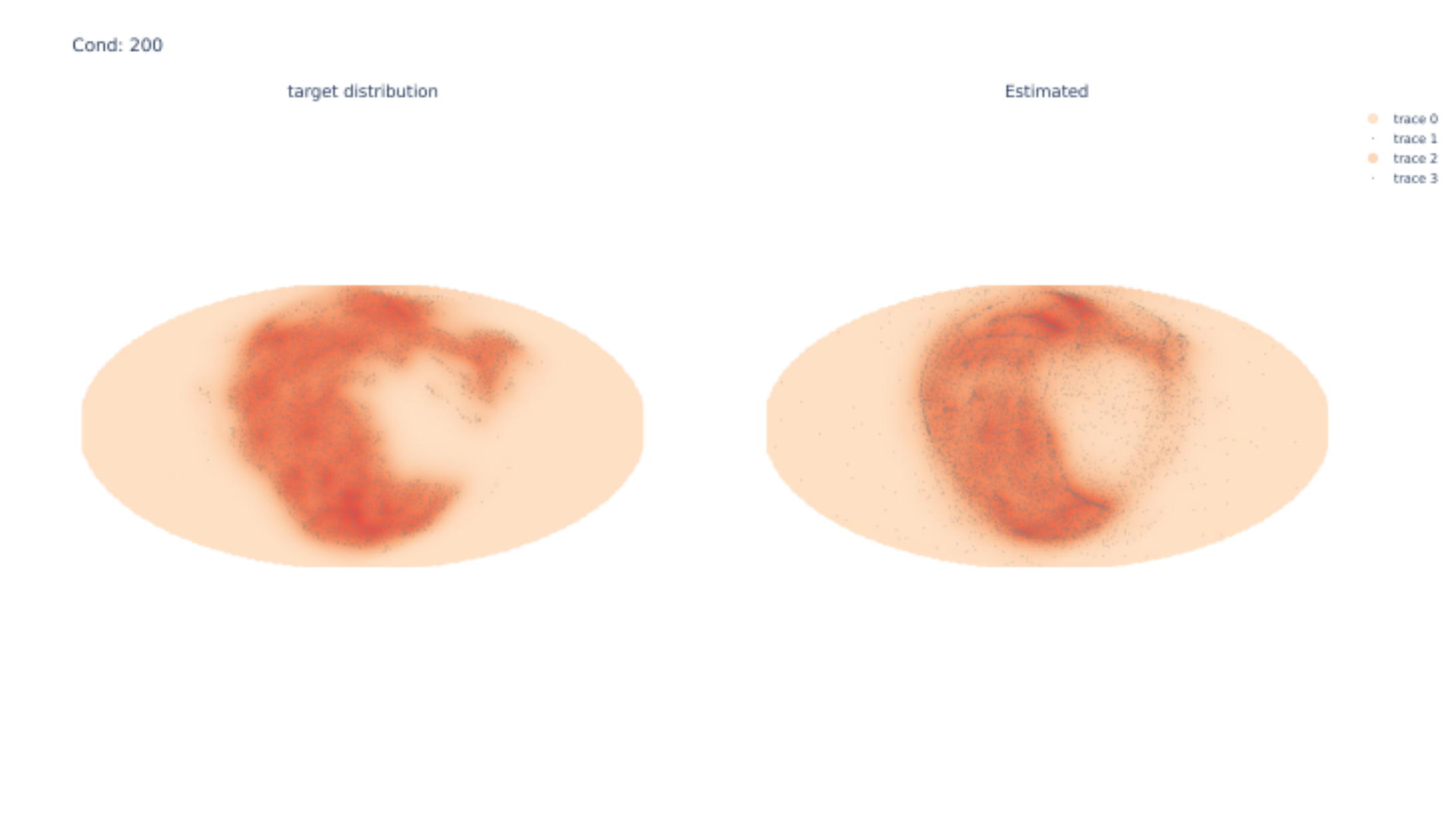}
        \put(14,24){ Sampling GT}
        \put(62,24){ Sampling M-VQR}
        \end{overpic}
        \hspace{0.5cm}
        \begin{overpic}
        [trim=15.4cm 5.55cm 3cm 5.8cm,clip,width=0.28\linewidth, grid=false]{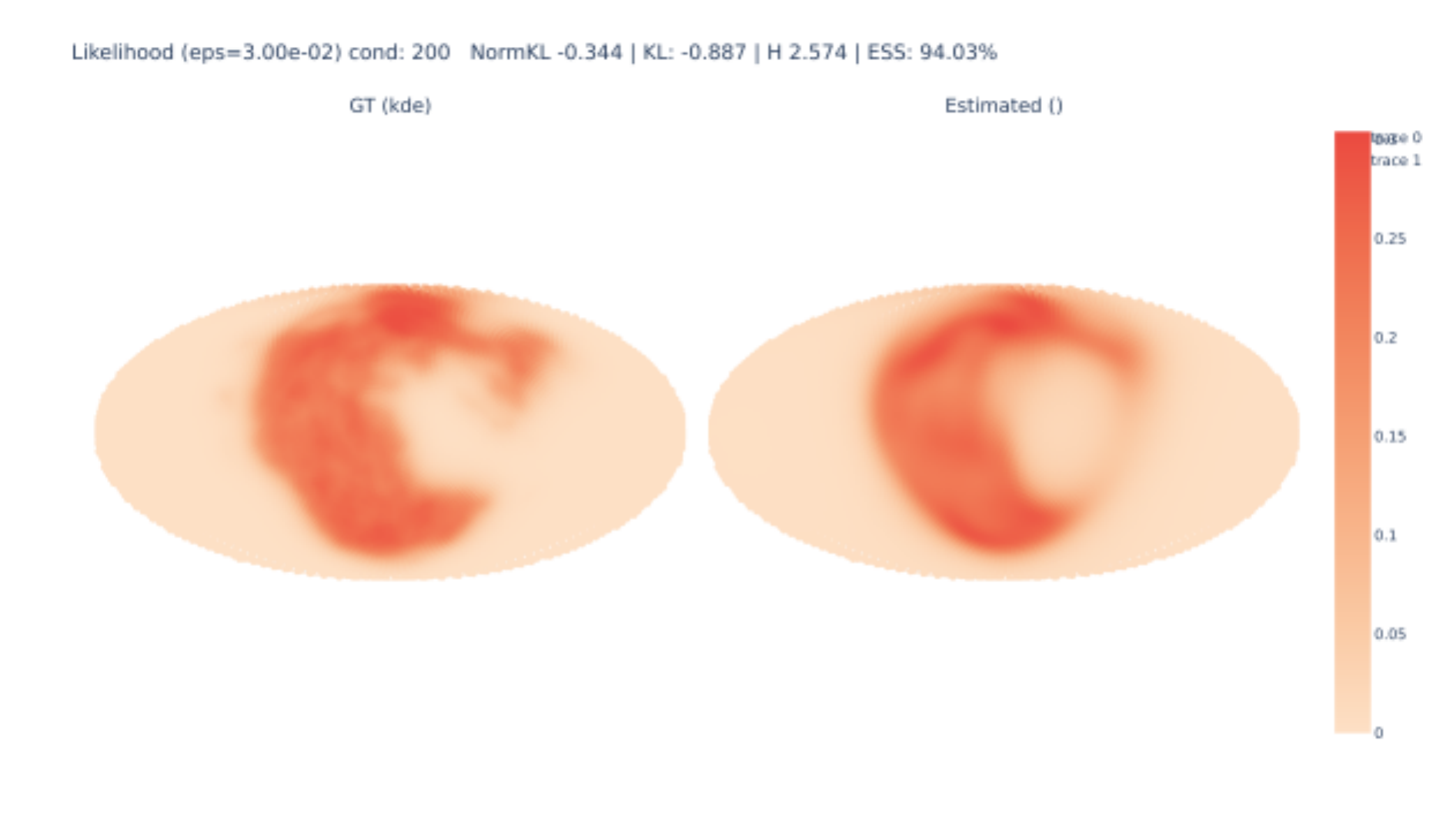}
        \put(32,52){ Likelihood}
        \end{overpic}
    \caption{\label{fig:contdrift_200_kde_lh}{\rvec{Y}|\rvec{X}=200}}
    \end{subfigure}

    \vspace{0.2cm}
    \begin{subfigure}[b]{\linewidth}
        \begin{overpic}
        [trim=1.75cm 6.5cm 3cm 7cm,clip,width=0.62\linewidth, grid=false]{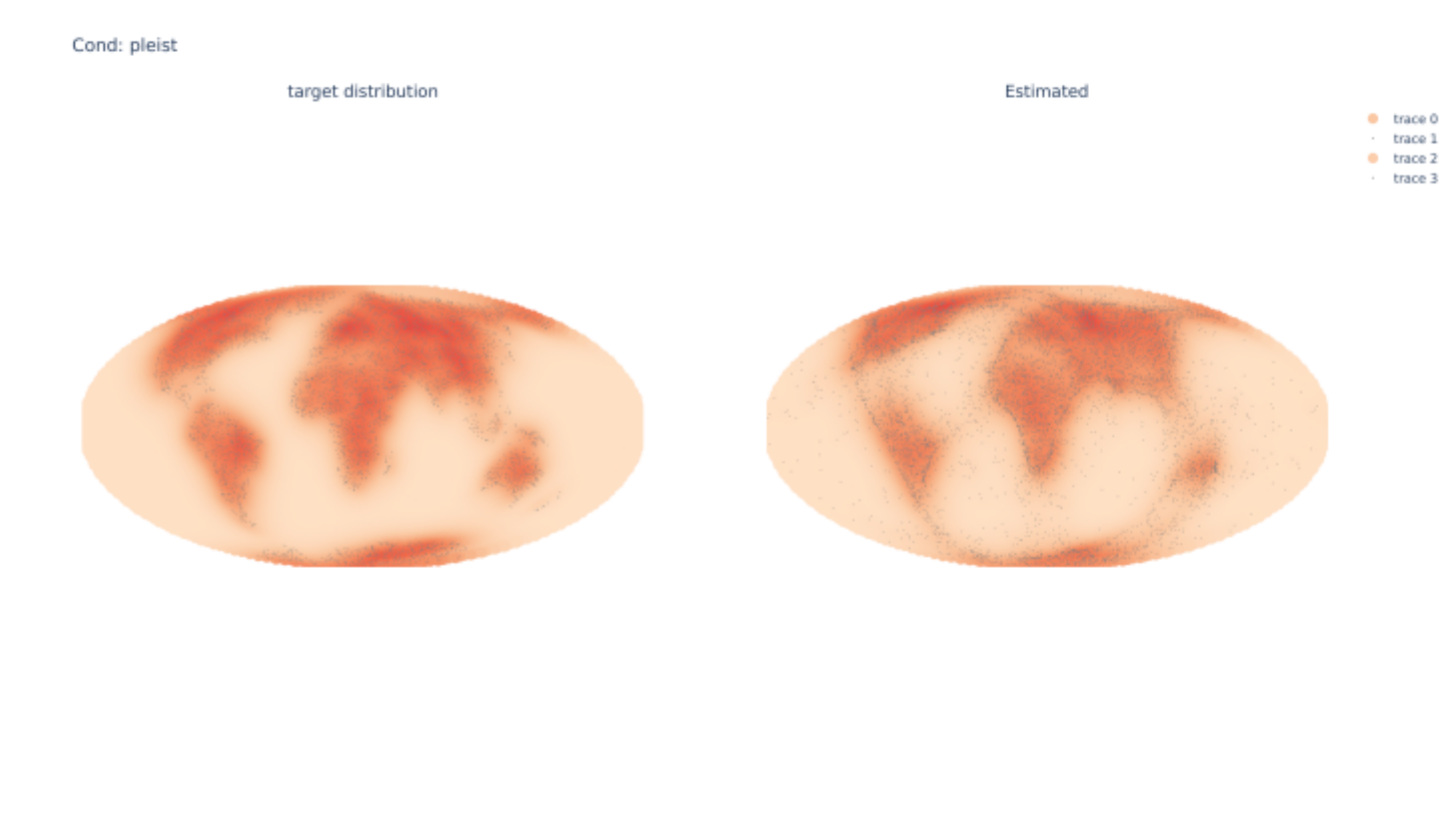}
        \put(14,24){ Sampling GT}
        \put(62,24){ Sampling M-VQR}
        \end{overpic}
        \hspace{0.5cm}
        \begin{overpic}
        [trim=15.4cm 5.55cm 3cm 5.8cm,clip,width=0.28\linewidth, grid=false]{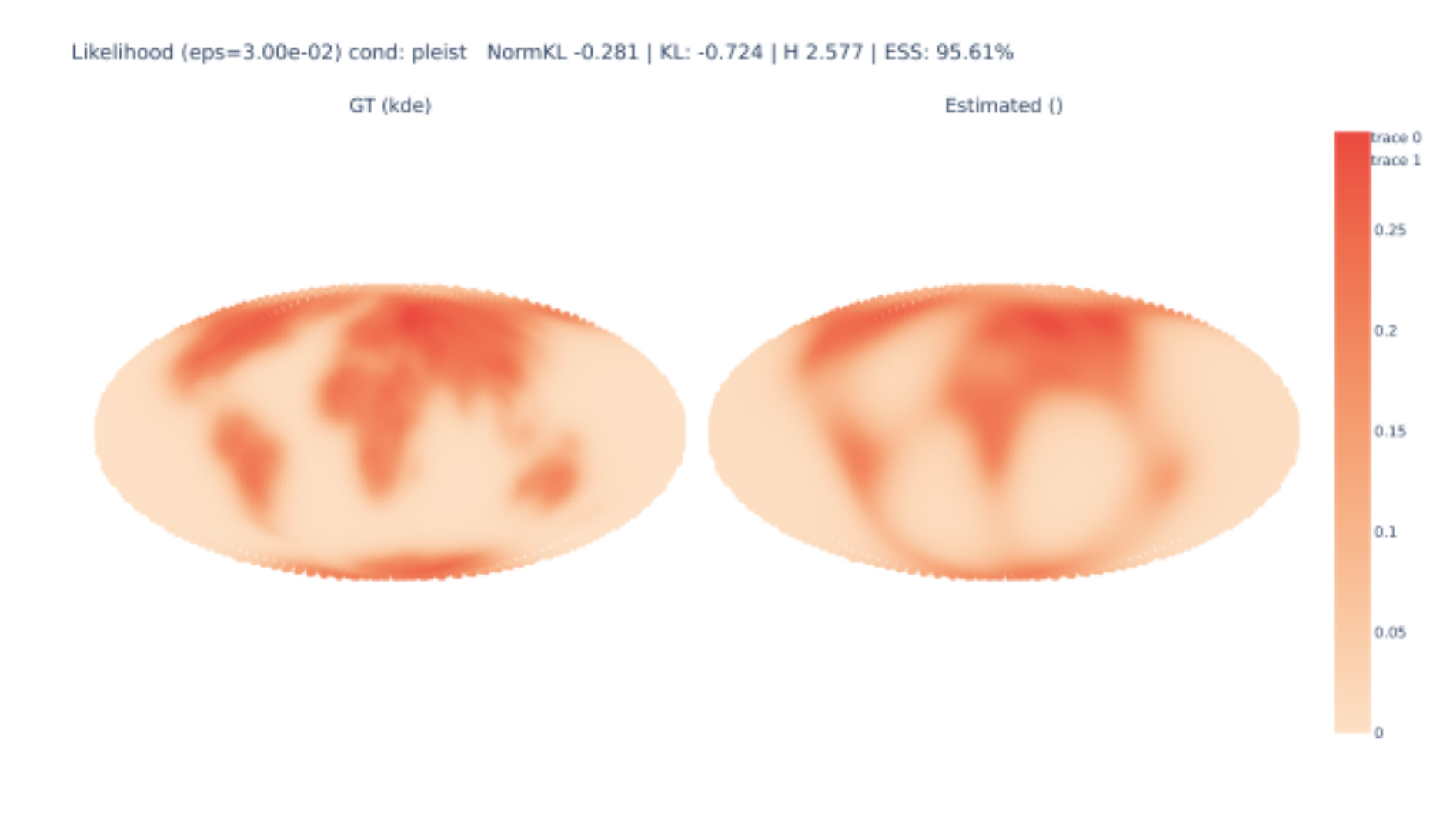}
        \put(32,52){ Likelihood}
        \end{overpic}
    \caption{\label{fig:contdrift_200_kde_lh}{\rvec{Y}|\rvec{X}=pleist}}
    \end{subfigure}
    \caption{\textbf{Sampling and Likelihood $p_{\rvec{Y}|\rvec{X}}$ from M-VQR on the `Continental Drift' dataset.}}
    \label{fig:contdrift_kde_lh}
\end{figure}

\begin{figure}[t]
    \centering
    \begin{subfigure}[b]{\linewidth}
    \hspace{0.5cm}
        \begin{overpic}
        [trim=1.25cm 1.5cm 2.5cm 2.54cm,clip,width=0.61\linewidth, grid=false]{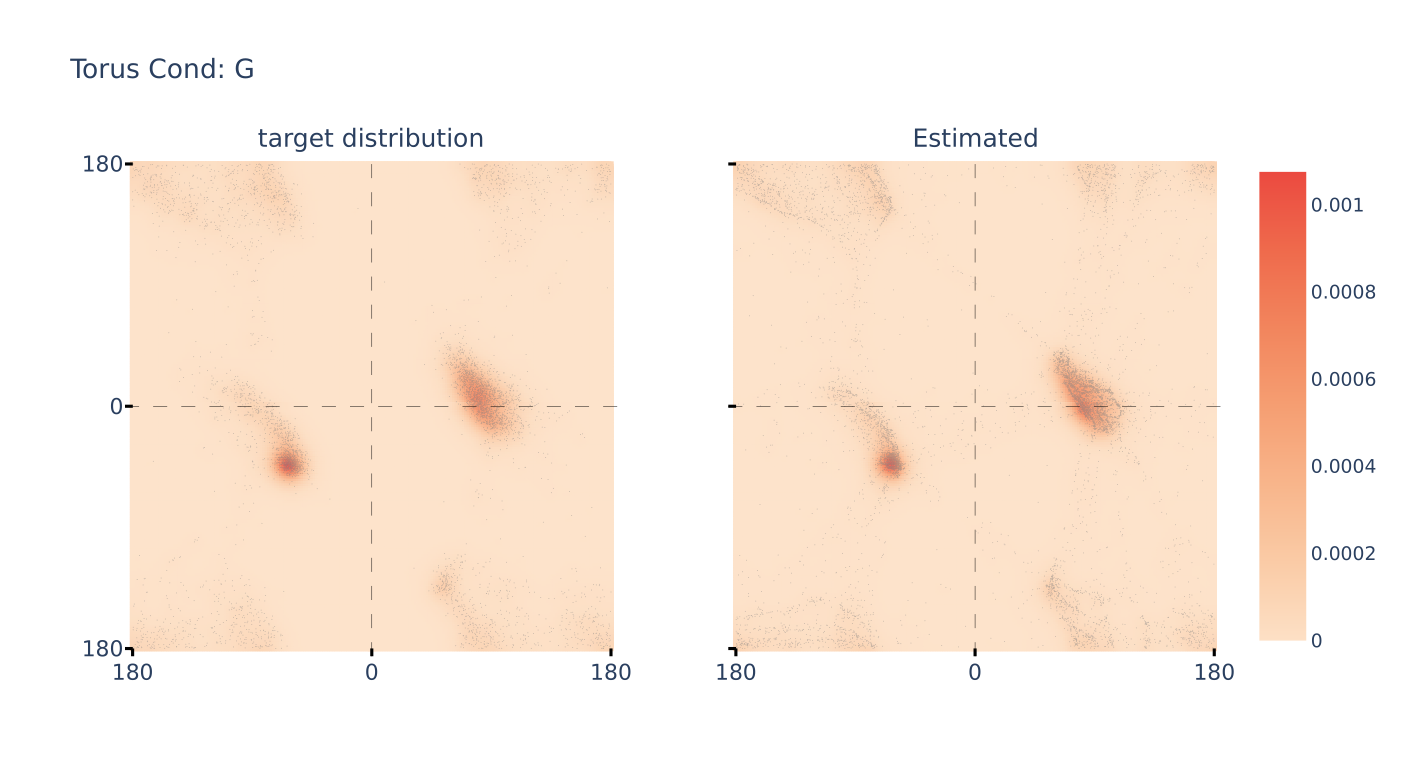}
        \put(15,46){ Sampling GT}
        \put(62,46){ Sampling M-VQR}
        \put(24,-1){\tiny {$\phi (^{\circ})$}}
        \put(74,-1){\tiny {$\phi (^{\circ})$}}
        \put(0,24){\tiny \rotatebox[origin=c]{90}{$\psi (^{\circ})$}}
        \end{overpic}
        \hspace{0.2cm}
        \begin{overpic}
        [trim=12cm 1.5cm 2.2cm 2.52cm,clip,width=0.29\linewidth, grid=false]{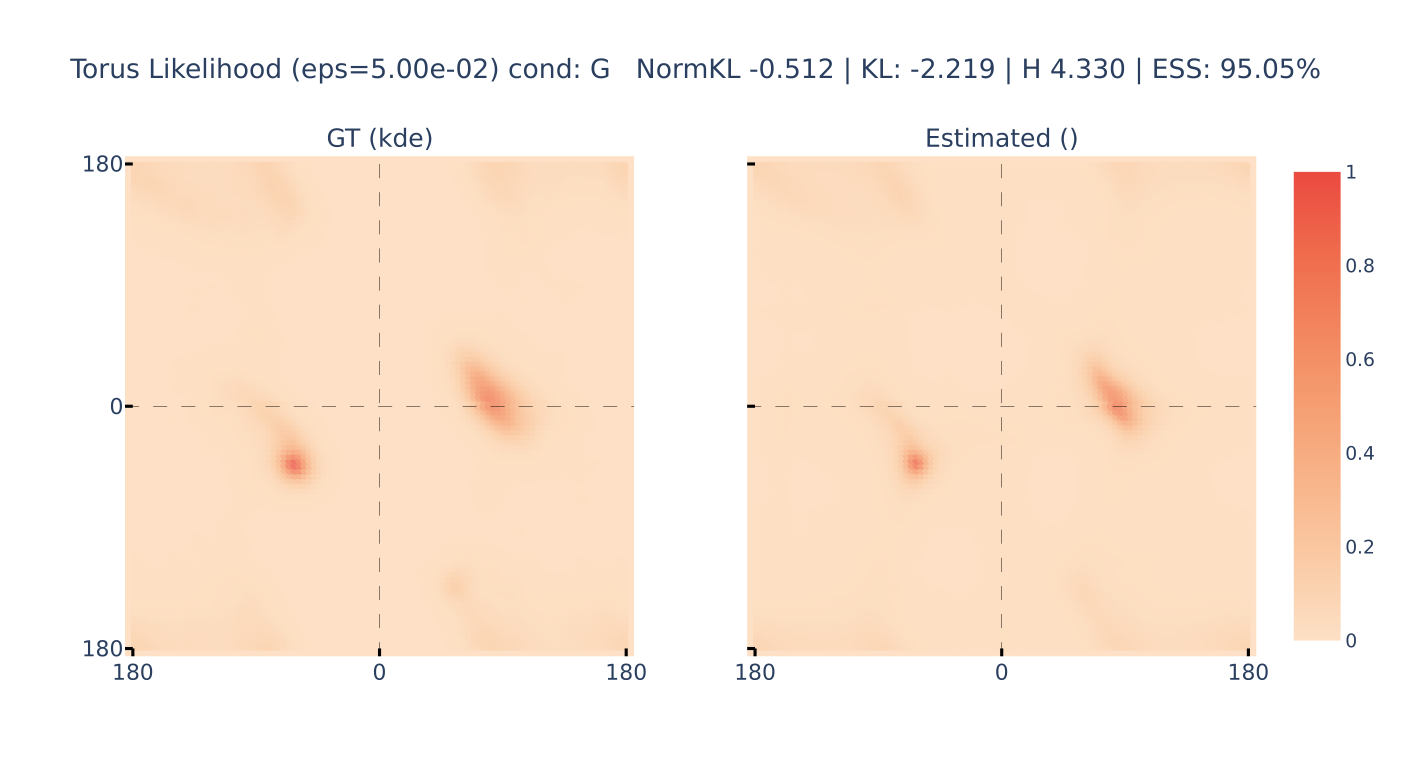}
        \put(32,96){ Likelihood}
        \put(48,-1){\tiny {$\phi (^{\circ})$}}
        \end{overpic}
    \caption{\label{fig:codonN_kde_lh} {\rvec{Y}|\rvec{X}=\text{G}}}
    \end{subfigure}
    
    \vspace{0.5cm}
    \begin{subfigure}[b]{\linewidth}
    \hspace{0.5cm}
        \begin{overpic}
        [trim=1.25cm 1.5cm 2.5cm 2.54cm,clip,width=0.61\linewidth, grid=false]{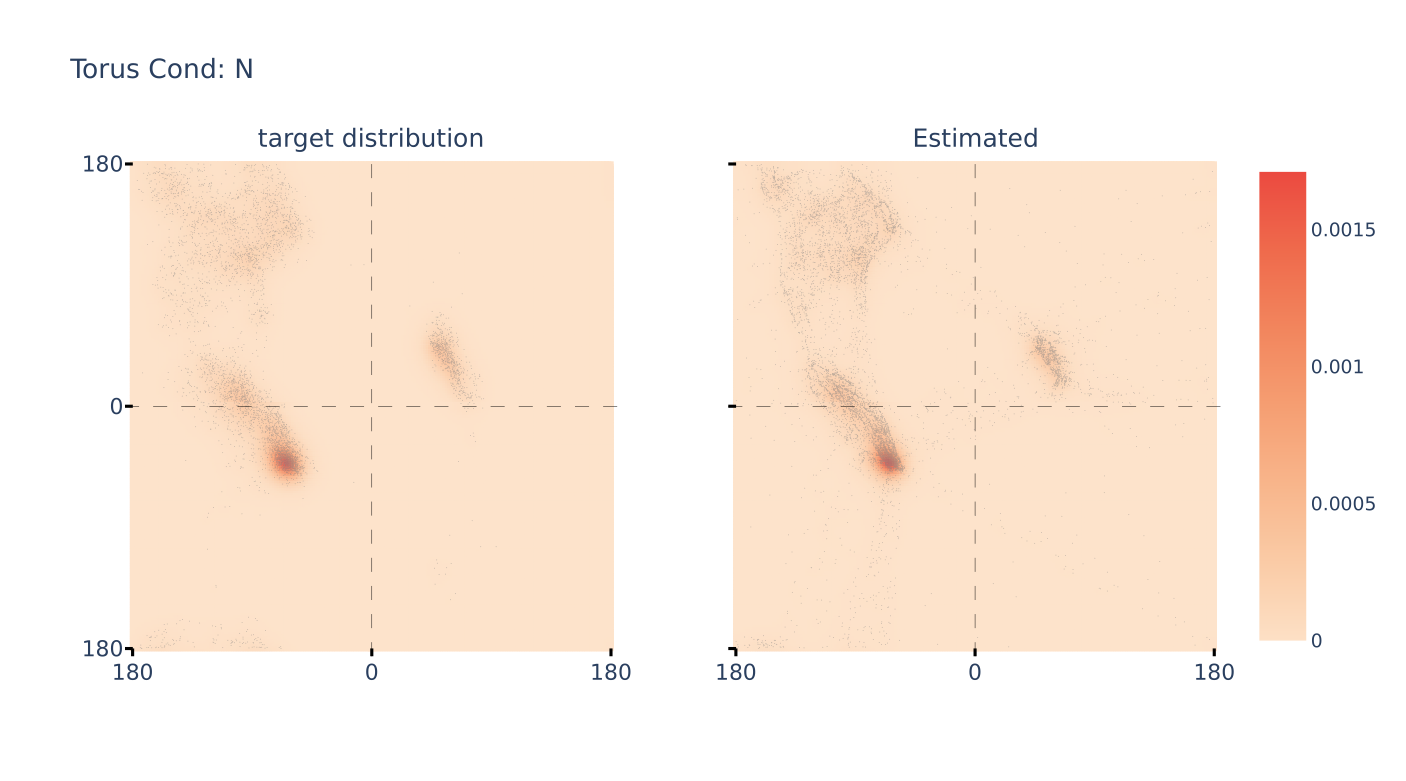}
        \put(15,46){ Sampling GT}
        \put(62,46){ Sampling M-VQR}
        \put(24,-1){\tiny {$\phi (^{\circ})$}}
        \put(74,-1){\tiny {$\phi (^{\circ})$}}
        \put(0,24){\tiny \rotatebox[origin=c]{90}{$\psi (^{\circ})$}}
        \end{overpic}
        \hspace{0.2cm}
        \begin{overpic}
        [trim=12cm 1.5cm 2.2cm 2.52cm,clip,width=0.29\linewidth, grid=false]{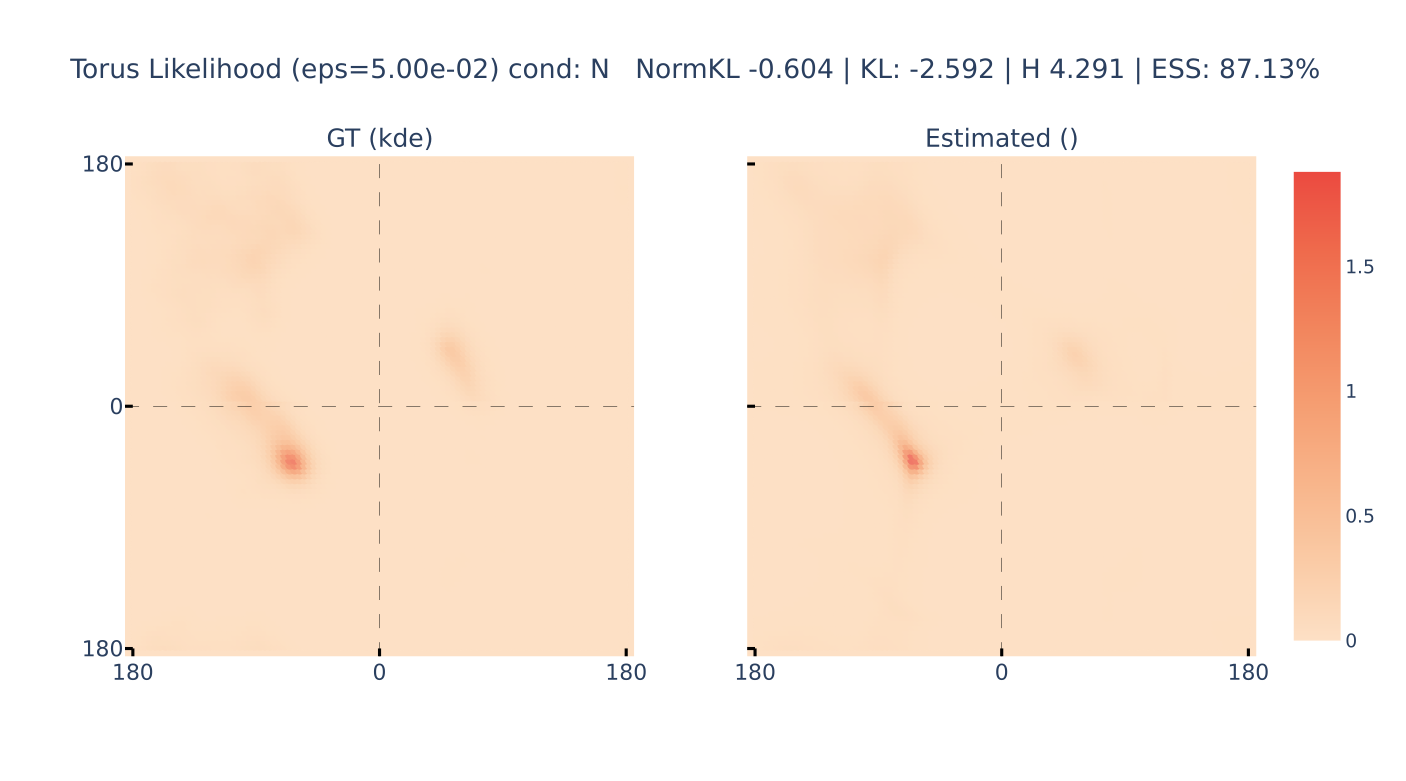}
        \put(32,96){ Likelihood}
        \put(48,-1){\tiny {$\phi (^{\circ})$}}
        \end{overpic}
    \caption{\label{fig:codonNN_kde_lh}{\rvec{Y}|\rvec{X}=\text{N}}}
    \end{subfigure}

    \vspace{0.5cm}
    \begin{subfigure}[b]{\linewidth}
    \hspace{0.5cm}
        \begin{overpic}
        [trim=1.25cm 1.5cm 2.5cm 2.54cm,clip,width=0.61\linewidth, grid=false]{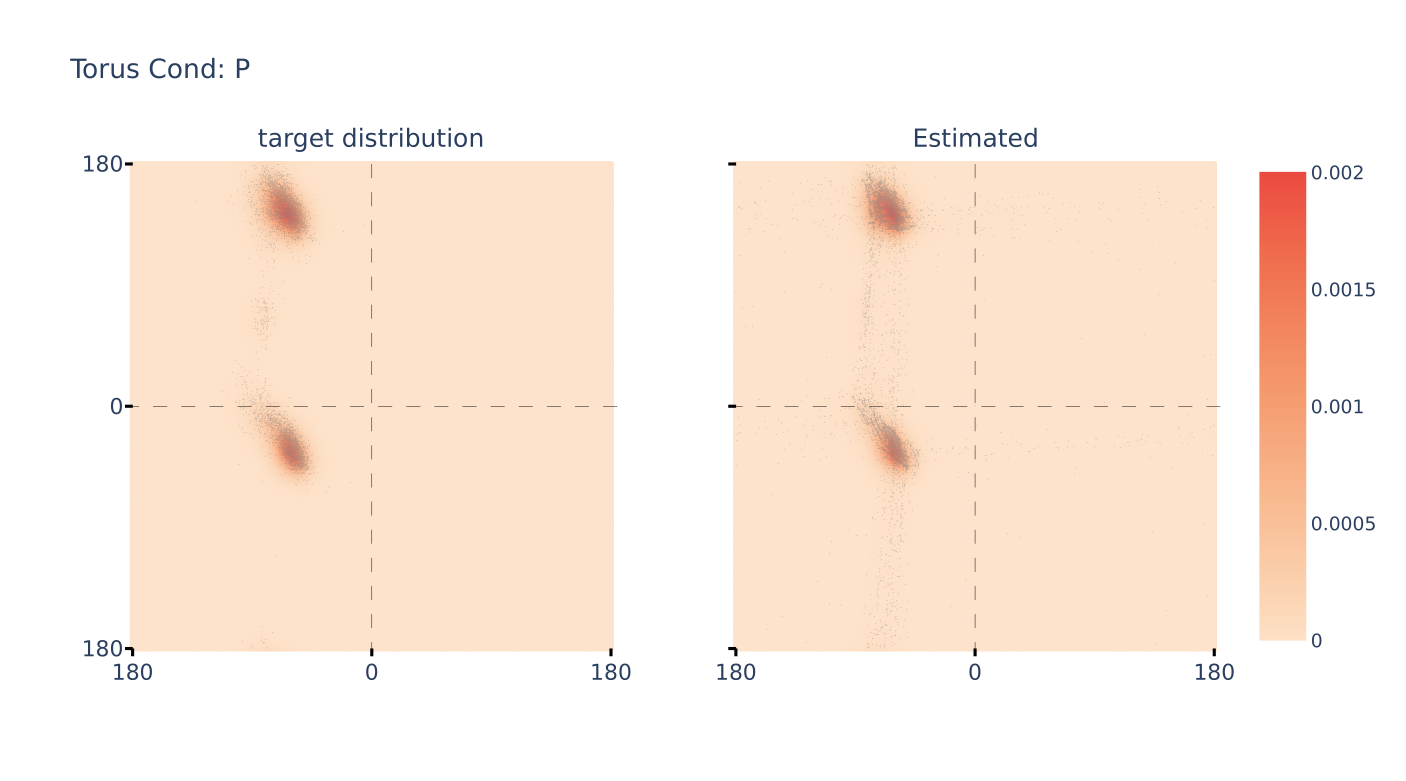}
        \put(15,46){ Sampling GT}
        \put(62,46){ Sampling M-VQR}
        \put(24,-1){\tiny {$\phi (^{\circ})$}}
        \put(74,-1){\tiny {$\phi (^{\circ})$}}
        \put(0,24){\tiny \rotatebox[origin=c]{90}{$\psi (^{\circ})$}}
        \end{overpic}
        \hspace{0.3cm}
        \begin{overpic}
        [trim=12cm 1.5cm 2.2cm 2.52cm,clip,width=0.29\linewidth, grid=false]{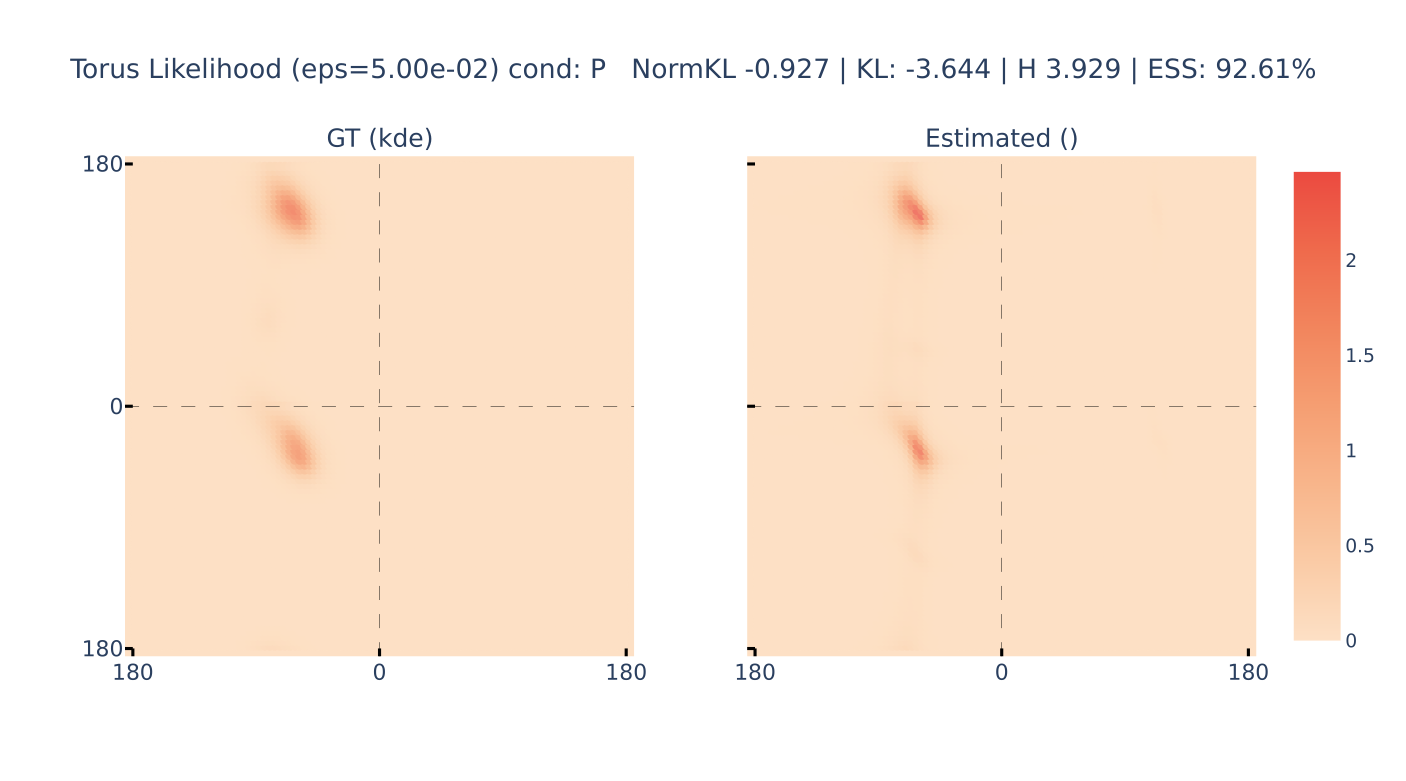}
        \put(32,96){ Likelihood}
        \put(48,-1){\tiny {$\phi (^{\circ})$}}
        \end{overpic}
    \caption{\label{fig:codonN_kde_lh} {\rvec{Y}|\rvec{X}=\text{P}}}
    \end{subfigure}
    \caption{\textbf{Sampling and Likelihood $p_{\rvec{Y}|\rvec{X}}$ from M-VQR on the `Dihedral Angles' dataset.}}
    \label{fig:codon_kde_lh}
\end{figure}

Figure \ref{fig:contdrift_kde_lh} and \ref{fig:codon_kde_lh} present the results of likelihood and sampling calculations by our method, M-VQR, applied to the real data distributions examined in the primary paper. It is evident that, in all cases, M-VQR's sampled data closely aligns with the ground truth, with a few outliers that may be attributed to discrete formulation approximations.
Moreover, we illustrate the likelihoods $p_{\rvec{Y}|\rvec{X}}$ computed using M-VQR. In this context, we lack a ground truth likelihood for direct comparison, as these distributions are derived from finite sets of samples. However, by comparing the computed likelihoods with the ground truth sampling, we observe that higher values of $p_{\rvec{Y}|\rvec{X}}$ tend to correspond to regions with a denser concentration of samples.

\end{document}